\documentclass[lettersize,journal]{IEEEtran}
\usepackage{amsmath,amsfonts}
\usepackage{algorithmic}
\usepackage{algorithm}
\usepackage{array}
\usepackage[caption=false,font=normalsize,labelfont=sf,textfont=sf]{subfig}
\usepackage{textcomp}
\usepackage{stfloats}
\usepackage{url}
\usepackage{verbatim}
\usepackage{graphicx}
\usepackage{cite}
\usepackage{booktabs} 
\usepackage{hyperref}
\usepackage{multirow}
\usepackage{tabularx}
\usepackage{amssymb}

\begin{document}

\title{Deploying Foundation Model Powered Agent Services: A Survey}

\author{Wenchao Xu, Jinyu Chen, Peirong Zheng, Xiaoquan Yi, Tianyi Tian, Wenhui Zhu, Quan Wan, Haozhao Wang, Yunfeng Fan, Qinliang Su, Xuemin Shen,~\IEEEmembership{Fellow,~IEEE,}
\thanks{W. Xu, J. Chen, P. Zheng, and Y. Fan are with the Department of Computing, The Hong Kong Polytechnic University, Hong Kong SAR, China (e-mail:\{ jinyu.chen, peirong.zheng, yunfeng.fan\}@connect.polyu.hk, wenchao.xu@polyu.edu.hk).
X. Yi, H. Wang, and Q. Wan are with The School of Computer Science and Technology, Huazhong University of Science and Technology, Wuhan, China (e-mail:\{yixiaoquan, hz\_wang\}@hust.edu.cn, muyou@bupt.edu.cn). T. Tian and W. Zhu are with the School of Information and Communication Engineering, Beijing University of Posts and Telecommunications, Beijing, China (e-mail: \{tian\_tianyi,wenhui\_zhu \}@bupt.edu.cn). 
Q. Su is with the School of Computer Science and Engineering, Sun Yat-sen University, Guangzhou, China (suqliang@mail.sysu.edu.cn)
X. Shen is with the Department of Electrical and Computer Engineering, University of Waterloo, Waterloo, Canada (e-mail: sshen@uwaterloo.ca).}
}

\markboth{Manuscript}%
{Shell \MakeLowercase{\textit{et al.}}: A Sample Article Using IEEEtran.cls for IEEE Journals}


\maketitle

\begin{abstract}
Foundation model (FM) powered agent services are regarded as a promising solution to develop intelligent and personalized applications for advancing toward Artificial General Intelligence (AGI). To achieve high reliability and scalability in deploying these agent services, it is essential to collaboratively optimize computational and communication resources, thereby ensuring effective resource allocation and seamless service delivery. In pursuit of this vision, this paper proposes a unified framework aimed at providing a comprehensive survey on deploying FM-based agent services across heterogeneous devices, with the emphasis on the integration of model and resource optimization to establish a robust infrastructure for these services. Particularly, this paper begins with exploring various low-level optimization strategies during inference and studies approaches that enhance system scalability, such as parallelism techniques and resource scaling methods. The paper then discusses several prominent FMs and investigates research efforts focused on inference acceleration, including techniques such as model compression and token reduction. Moreover, the paper also investigates critical components for constructing agent services and highlights notable intelligent applications. Finally, the paper presents potential research directions for developing real-time agent services with high Quality of Service (QoS).
\end{abstract}


\section{Introduction}
\begin{figure}
    \centering
    \includegraphics[width=1\linewidth]{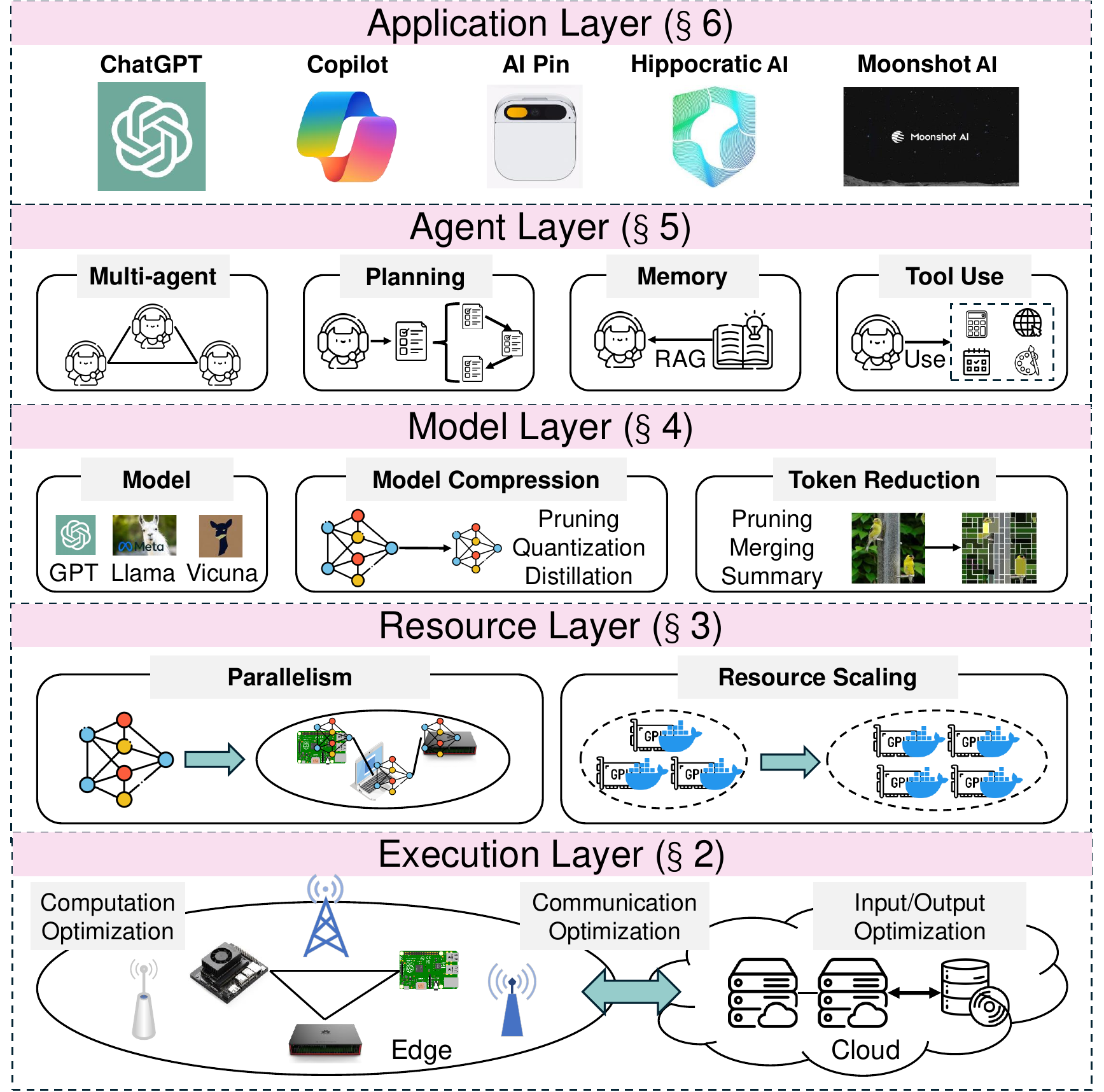}
    \caption{The framework of FM-powered agent services. The execution layer runs model inference with low-level optimizations. The resource layer focuses on designing strategies for parallelism and resource scaling. The model and agent layers work on optimizing FMs and various agent components. The application layer constructs different intelligent applications.}
    \label{fig:survey_all}
\end{figure}

\begin{figure*}[t]
  \centering
  \includegraphics[width=\linewidth]{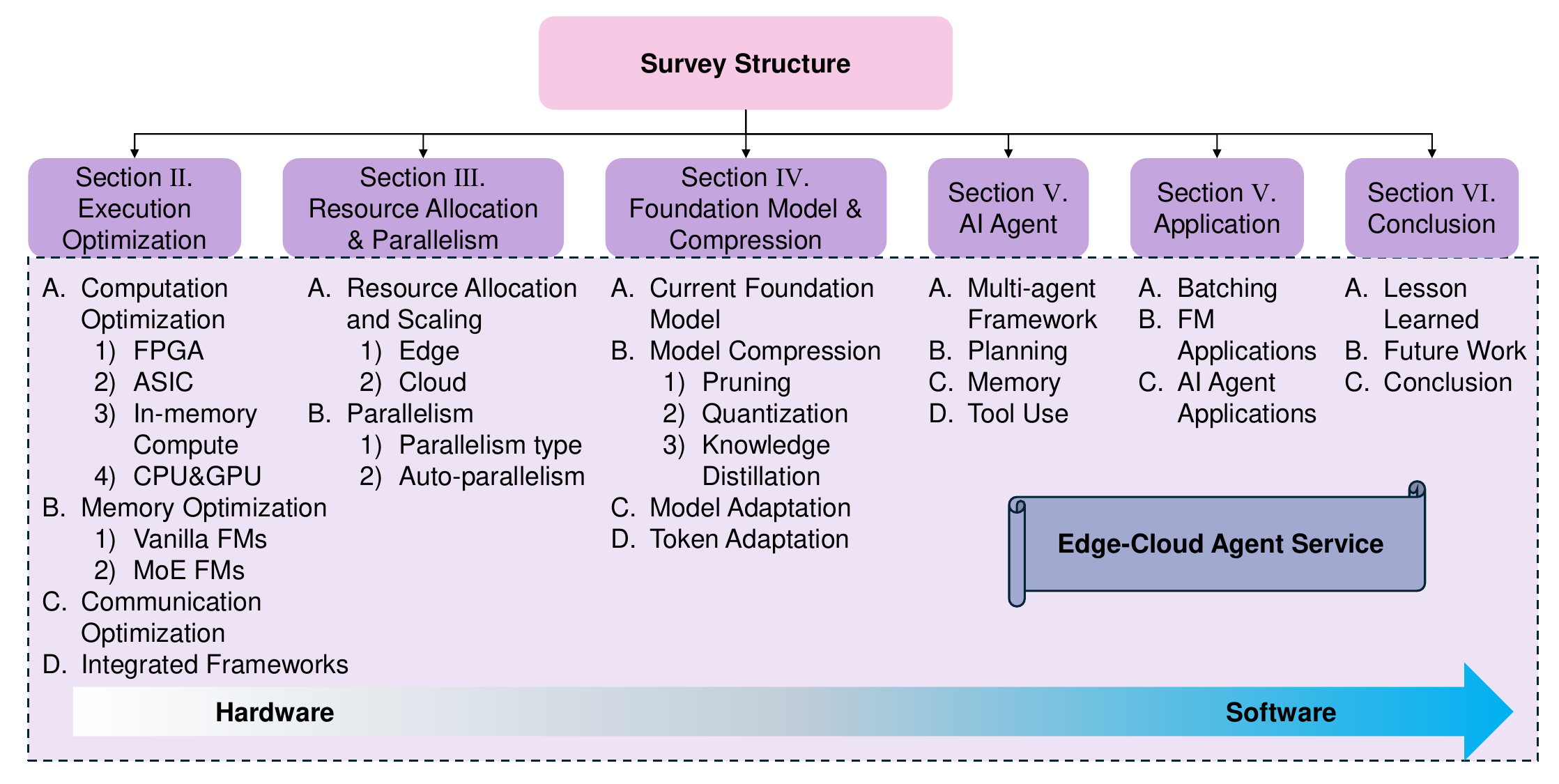}
  \caption{The framework of our survey. Each technical session corresponds to a layer in Figure~\ref{fig:survey_all}.}
  \label{fig:framework}
\end{figure*}

\IEEEPARstart{T}{he} rapid advancement of artificial intelligence (AI) has positioned foundation models (FMs) as a cornerstone of innovation,
driving progress in various fields such as natural language processing, computer vision, and autonomous systems. These models, characterized by their vast parameter spaces and extensive training on broad datasets, incubate numerous applications from automated text generation to advanced multi-modal question answering and autonomous robot services~\cite{bommasani2022opportunities}. Some popular FMs, such as GPT, Llama, ViT, and CLIP, are pivotal in pushing the boundaries of AI capabilities, offering sophisticated solutions for processing and analyzing large volumes of data across different formats and modalities.
The continuous advancement of FMs significantly enhances AI's ability to comprehend and interact with the world in a manner akin to human cognition.

However, traditional FMs are typically confined to providing question-and-answer services and generating responses based on pre-existing knowledge, often lacking the ability to incorporate the latest information or employ advanced tools. 
FM-powered agents are designed to enhance the capability of FM.
These agents are incorporated with dynamic memory management, long-term task planning, advanced computational tools, and interactions with the external environment~\cite{xi2023rise}.
For example, FM-powered agents can call different external APIs to access real-time data, perform complex calculations, and generate updated responses based on the most current information available. 
This approach improves the reliability and accuracy of the responses and enables more personalized interactions with users.


Developing a serving system with low latency, high reliability, high elasticity, and minimal resource consumption is crucial for delivering high-quality agent services to users.
Such a system can efficiently manage varying query loads while maintaining swift response and reducing resource costs. 
Moreover, constructing a serving system on heterogeneous edge-cloud devices is a promising solution to leverage the idle computational resources at the edge and the abundant computational clusters available in the cloud.
The collaborative inference of edge-cloud devices can enhance overall system efficiency by dynamically allocating tasks to various edge-cloud machines based on computational load and real-time network conditions.

\vspace{-3pt}
Although many research works investigate edge-cloud collaborative inference for small models, deploying FMs under this paradigm for diverse agent services still faces several severe challenges.  
First, the fluctuating query load severely challenges the model serving. 
The rapidly growing number of users want to experience intelligent agent services with FMs. For example, as of April 2024, ChatGPT has approximately 180.5 million users, with around 100 million of users being active weekly~\cite{nerdynav_chatgpt_stats}. 
These users access the service at different times, resulting in varying request rates.
An elastic serving system should dynamically scale the system capacity according to the current system characteristics.
Secondly, the parameter space of an FM is particularly large, reaching the scale of several hundred billion, which is a significant challenge to the storage system. However, the storage capacity of edge devices and the consumer GPU is limited, making it unable to accommodate an entire model. 
The large number of parameters results in significant inference overhead and long execution latency. Therefore, it is necessary to design model compression methods and employ different parallelism approaches in diverse execution environments.
In addition, users have different service requirements and inputs in different applications. For example, some applications prioritize low latency, while others prioritize high accuracy. This necessitates dynamic resource allocation and adjustment of the inference process. 
Moreover, AI agents need to deal with lots of hard tasks under complex environments, which require effective management of large-scale memory, real-time processing of updated rules, and specific domain knowledge. Additionally, agents possess distinct personalities and roles, necessitating the design of an efficient multi-agent collaboration framework.

\vspace{-3pt}
To address the aforementioned challenges and promote the development of the real-time FM-powered agent service, this survey proposes a unified framework and investigates various research works from different optimization aspects.
This framework is shown in Figure~\ref{fig:survey_all}.
The bottom layer is the execution layer, where edge or cloud devices execute an inference with FMs. 
Joint computation optimization, I/O optimization, and communication optimization are applied to accelerate inference and promote the building of a powerful infrastructure for FMs.
The resource layer, comprised of two components, facilitates the deployment of the model on various devices. Parallelism methods design different model splitting and placement strategies to utilize the available resources and improve throughput collaboratively.  Resource scaling dynamically adjusts the hardware resources during runtime based on query load and resource utilization, thereby improving overall scalability.
The model layer focuses on the optimization of FMs. Two lightweight methods, including model compression and token reduction, are specifically designed to promote the widespread adoption of FMs.
Based on these FMs, many AI agents are constructed to accomplish various tasks. Numerous methods have been proposed to enhance the four key components of agents, which encompass the multi-agent framework, planning capabilities, memory storage, and tool utilization.
Ultimately, leveraging the aforementioned techniques, all kinds of applications can be developed to deliver intelligent and low-latency agent services to users.

\vspace{-6pt}
\subsection{Previous Works}

Many research works focus on system optimization to deploy machine learning models in edge-cloud environments.
KACHRIS reviews some hardware accelerators for Large Language Models (LLMs) to address the computational challenges~\cite{kachris2024survey}.
Tang et al. summarize scheduling methods designed for optimizing both network and computing resources~\cite{10304187}.
Miao et al. present some acceleration methods to improve the efficiency of LLMs~\cite{miao2023efficient}. 
This survey includes system optimizations, such as memory management and kernel optimization, as well as algorithm optimizations, such as architectural design and compression algorithms, to accelerate model inference.
Xu et al. focus on the deployment of Artificial Intelligence-Generated Content (AIGC), and they provide an overview of mobile network optimization for AIGC, covering the processes of dataset collection, AIGC pre-training, AIGC fine-tuning, and AIGC inference~\cite{10398474}.
Djigal et al. investigate the application of machine learning and deep learning techniques in resource allocation for Multi-access Edge Computing (MEC) systems~\cite{9858872}. The survey includes resource offloading, resource scheduling, and collaborative allocation.
Many research works propose different algorithms to optimize the design of FMs and agents.
~\cite{bommasani2022opportunities}, ~\cite{zhao2023survey} and ~\cite{naveed2024comprehensive} present popular FMs, especially LLMs.
~\cite{zhu2023survey}, ~\cite{wang2024model} and ~\cite{xu2024survey} summarizes model compression and inference acceleration methods for LLM.
~\cite{xi2023rise}, ~\cite{wang2024survey}, and ~\cite{guo2024large} review the challenges and progress for the development of agents.

In summary, the above studies either optimize edge-cloud resource allocation and scheduling for small models or design acceleration or efficiency methods for large FMs. 
To the best of our knowledge, this paper is the first comprehensive survey to review and discuss the deployment of real-time FM-powered agent services in heterogeneous devices, a research direction that has gained significant importance in recent years.
We design a unified framework to fill this research gap and review current research works from different perspectives.
This framework not only delineates essential techniques for the deployment of FMs but also identifies key components of FM-based agents and corresponding system optimizations specifically tailored for agent services.

\vspace{-6pt}
\subsection{Contribution}

This paper presents a comprehensive survey on the deployment of FM-powered agent services in edge-cloud environments, covering optimization approaches spanning from hardware to software layers.
For the convenience of readers, we provide an outline of the survey in Figure~\ref{fig:framework}.
The contributions of this survey are summarized in the following aspects:
\begin{itemize}
    \item This survey proposes the first comprehensive framework to provide a deep understanding of the deployment of FM-powered agent services within the edge-cloud environment. Such a framework holds the potential to foster the advancement of AGI greatly.
    \item From a low-level hardware perspective, we present research on various runtime optimization methods and resource allocation and scheduling methods. These techniques are designed to establish a reliable and flexible infrastructure for FMs.
    \item From a high-level software perspective, we elucidate research efforts focused on model optimization and agent optimization, thereby offering diverse opportunities for building intelligent and lightweight agent applications.
\end{itemize}

The remainder of this article is organized as follows: Section~\ref{sec:execution} presents some low-level execution optimization methods. Section~\ref{sec:resource} describes resource allocation and parallelism mechanisms. Section~\ref{sec:model} discusses current FMs, as well as techniques for model compression and token reduction. Section~\ref{sec:agent} illustrates key components for agents. Section~\ref{sec:app} presents batching methods and some related applications. Finally, Section~\ref{sec:conclusion} discusses the future works and draws conclusions.

\section{Execution Optimization}
\label{sec:execution}

\subsection{Computation Optimization}

\begin{figure}
    \centering
    \includegraphics[width=0.9\linewidth]{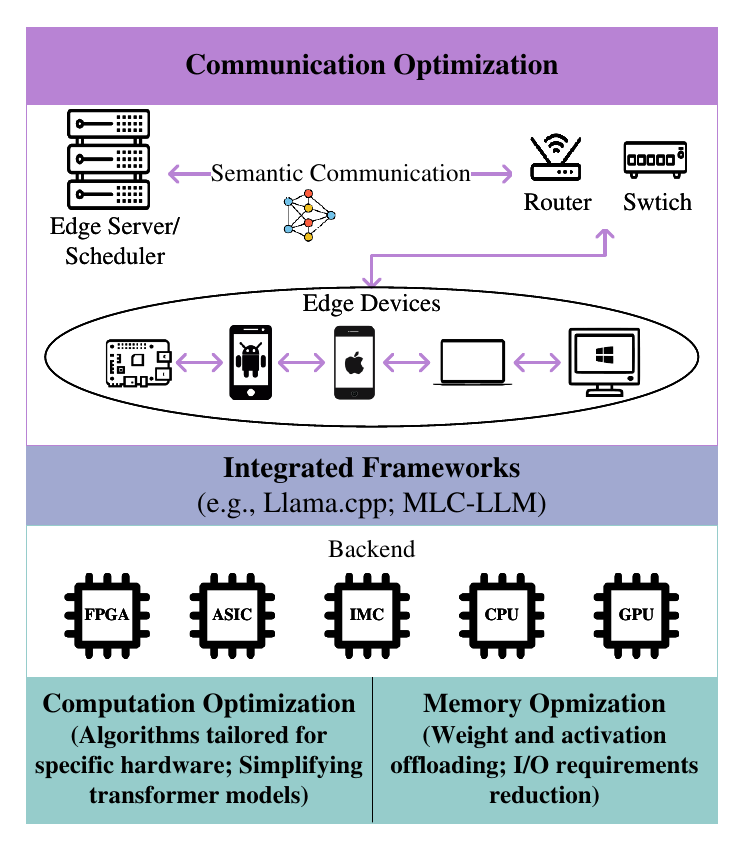}
    \caption{A multi-layer optimization framework for edge computing systems serving FMs. At the hardware level, heterogeneous resources such as FPGAs, ASICs, IMCs, CPUs, and GPUs are utilized and optimized in terms of computation and memory. The integrated frameworks can support heterogeneous hardware. The network level focuses on semantic communication.}
    \label{fig:computing_node}
\end{figure}

Edge deployment of FMs poses severe challenges due to the heterogeneity of edge devices in terms of computing ability, hardware architecture, and communication bandwidth~\cite{wang2024feddse}. 
It is important to design computation optimization methods for different devices to accelerate model inference.
\autoref{fig:computing_node} provides an overview of this chapter. This diagram provides a global perspective for algorithm design (e.g., computation optimization). It also depicts the components of an edge computing system, including heterogeneous backends and the network. 
\autoref{tab-computing-resource} shows some commonly used hardware devices at the edge. Traditionally, devices with CPUs have been deployed at the edge environment. However, due to their limited parallel computing capabilities and memory constraints, various specialized accelerators are designed for Deep Learning (DL) tasks.
FPGAs have gained widespread attention for their programmable and parallel computing capabilities.
ASICs, while non-programmable after manufacture, offer high speed and low power consumption. 
Many edge devices, such as personal computers and smartphones, have different computational resources, such as GPU and CPU. Consequently, many approaches focus on optimizing computational efficiency by jointly utilizing these resources for model inference. 
In-memory computing, with non-Von-Neumann architecture, has emerged as a prominent research area because its intrinsic parallelism can significantly reduce the I/O latency and improve computational efficiency. 
To enhance the understanding of different devices, we provide some detailed information about different hardware devices as below.

\begin{table}[t]
    \centering
    \caption{Computing Resource Types and Features}
    \label{tab-computing-resource}
    \begin{tabularx}{0.45\textwidth}{@{}p{3cm}X@{}}
    \toprule
    \textbf{Resource Type} & \textbf{Features and Functions} \\
    \midrule
    {Field Programmable Gate Arrays \newline (FPGAs)} & FPGAs are integrated circuits that can be reconfigured at the hardware level after manufacturing. It can have higher performance and lower latency than GPU accelerators.  \\
    {Application-Specific Integrated Circuit \newline(ASIC)} & ASIC is a customized accelerator for a specific use scenario. It is not programmable after manufacturing and offers computing comparable to FPGAs. The producer decides the specifications. \\
    {In-memory Compute \newline(IMC)} & IMC is a non-von-Neumann method, able to conduct computation in the memory. For instance, IMC utilizes physical processes to execute addition and multiplication, thus accelerating matrix-vector computing.\\ 
    {Central Processing Unit \newline(CPU)} & CPU is a General-purpose processor; It executes basic operations and instructions and can run lowly parallel computing. \\
    {Graphics Processing Unit \newline(GPU)} & GPU excels in parallel computing and is dominantly used as accelerators for DL.\\
    \bottomrule
    \end{tabularx}
\end{table}

\subsubsection{FPGAs}
FPGAs are widely deployed in edge computing applications and are re-configurable hardware devices that are efficient in power consumption.
FMs are built upon Transformers, which contain non-linear computation components, such as layer normalization, SoftMax, and non-ReLU activation functions. Serving these models requires specific accelerator designs. 
On the other hand, matrix multiplications are conducted during the inference of FMs. Their substantial computational complexity poses challenges for optimizing FPGA-based accelerators. To solve these problems, a specialized hardware accelerator is designed for Multi-Head Attention (MHA) and feed-forward networks (FFN) in Transformers \cite{lu2020hardware}. It incorporates a matrix partitioning strategy to optimize resource sharing between Transformer blocks, a computation flow that maximizes systolic array utilization, and optimizations of nonlinear functions to diminish complexity and latency.
MnnFast provides a scalable architecture specifically for Memory-augmented Neural Networks~\cite{jang2019mnnfast}. 
It incorporates a column-based streaming algorithm, zero-skipping optimization, and a dedicated embedding cache to tackle the challenges posed by large-scale memory networks. 
NPE offers software-like programmability to developers~\cite{khan2021npe}. Unlike previous FPGA designs that implement specialized accelerators for each nonlinear function, NPE can be easily upgraded for new models without requiring extensive reconfiguration, making it a more cost-effective solution.
DFX is a multi-FPGA system that optimizes the latency and throughput for text generation. It leverages model parallelism and an optimized, model-and-hardware-aware dataflow to handle the sequential nature of text generation. It demonstrates the superior speed and cost-effectiveness of FPGA compared to GPU implementations\cite{hong2022dfx}.

Additionally, the development of Transformer-OPU (Overlay Processor)\cite{bai2023Transformer} provides a flexible and efficient FPGA-based processor that expedites the computation of Transformer networks.
Moreover, implementing a tiny Transformer model through a Neural-ODE (Neural Ordinary Differential Equation) approach \cite{okubo2024cost} leads to a substantial reduction in model size and power usage, making it ideal for edge computing devices in Internet-of-Things (IoT) applications.
FlightLLM \cite{zeng2024flightllm} addresses the computational efficiency, memory bandwidth utilization, and compilation overheads on FPGAs for LLM inference. It includes a configurable Digital Signal Processor (DSP) chain optimized for varying sparsity patterns in LLMs and always-on-chip activations during decoding and a length adaptive compilation method for dynamic sparsity patterns and input lengths.

In summary, the above designs for FPGA-based FM inference focus on several key aspects: enhancing computational efficiency through specialized architectures such as systolic arrays and DSP chains, optimizing memory usage via techniques like matrix partitioning and dedicated caches, reducing latency through efficient dataflows and model-and-hardware-aware optimizations, and improving adaptability and programmability to accommodate diverse and evolving model structures and non-linear functions. 

\subsubsection{ASIC}
An ASIC is an integrated circuit chip customized for a specific application. For example, router ASICs can handle packet processing and signal modulation. It often includes microprocessors, memory, and other components as a System-on-Chip (SoC).
Recent advancements in ASIC design significantly enhance the performance and efficiency of attention mechanisms, which is crucial for applications across NLP and CV. The $A^3$ \cite{ham20203} accelerator employs algorithmic approximations and a prototype chip to significantly enhance energy efficiency and processing speed. Essentially, the attention mechanism is a content-based search to evaluate the correlation between the currently processed token and previous tokens. $A^3$ addresses the inefficiency of matrix-vector multiplication in the self-attention mechanism, which is sub-optimal for the content-based search, by implementing an efficient greedy candidate search method. Similarly, ELSA (Efficient, Lightweight Self-Attention) \cite{ham2021elsa} tackles the quadratic complexity of self-attention by selectively filtering out less important relations in self-attention and implements an ASIC to achieve high energy efficiency.

SpAtten\cite{wang2021spatten} prunes Transformer models at both the token and head levels and reduces the model size with quantization to minimize the computational and memory demands of Transformers.
Sanger\cite{lu2021sanger} framework enables sparse attention mechanisms through a reconfigurable architecture that supports dynamic software pruning and efficient sparse operations. Additionally, the Energon\cite{zhou2022energon} co-processor, working together with other FM accelerators, introduces a dynamic sparse attention mechanism that uses a mix-precision multi-round filtering algorithm to optimize query-key pair evaluations.
The above ASIC solutions demonstrate significant advancements in speed and energy efficiency compared to traditional devices while still preserving high accuracy. They pave the way for real-time, resource-efficient implementations for complex FMs.

\subsubsection{In-memory Compute}
In-memory computing (IMC) is an emerging computational paradigm performing computational tasks directly within memory, eliminating frequent I/O requirements between processors and memory units. IMC is a scalable and energy-efficient solution to handle long sequence data in Transformers.
\cite{guo2020att} introduces ATT, a fault-tolerant Resistive Random-access Memory (ReRAM) accelerator specifically designed for attention-based neural networks. This accelerator capitalizes on the high-density storage capabilities and low leakage power of ReRAM to address the compatibility issues between traditional neural network architectures and the complex data flow of attention mechanisms. 
ReTransformer\cite{laguna2021memory} is a ReRAM-based IMC architecture for Transformers. It accelerates the scaled dot-product attention mechanism, utilizes a matrix decomposition technique to avoid storing intermediate results, and designs the sub-matrix pipelines of MHA.

The iMCAT utilizes a combination of crossbar arrays to store the matrix in the memory array and uses Content Addressable Memories (CAM) to overcome the significant memory and computational bottlenecks in processing long sequences with MHA~\cite{laguna2021memory}. 
Recent works design different optimization methods to deploy computationally intensive Transformer models on edge AI accelerators~\cite{reidy2023work}. 
The Google TPU, categorized as ASIC and IMC, is widely used in edge and cloud computing. 
Large Transformer models can be executed efficiently on the Coral Edge TPU by optimizing the computational graph and employing quantization techniques, ensuring real-time inference with minimal energy consumption. 
The techniques employed in IMC, including matrix decomposition and quantization, reduce the computational resources required to serve FMs, thereby facilitating the deployment of FMs at the edge.

\subsubsection{CPU \& GPU}

Current optimization algorithms, such as CPU-GPU hybrid computation, are hardware-independent and can be applied to various backends. 
Different methods are designed to accelerate Transformer inference by optimizing the MHA, FFN, skip connections, and normalization modules, which are identified as critical bottlenecks in Transformer architectures \cite{he2023simplifying}.
\cite{choi2022accelerating} aims to optimize the execution time of the SoftMax layer by decomposing it into multiple sub-layers and then fusing them with adjacent layers. LLMA \cite{yang2023inference} leverages the overlap between the text generated by LLMs and existing reference texts to enhance computational parallelism and speed up the inference process without compromising output quality. These algorithms are hardware-agnostic and can be applied to various backends besides CPU and GPU. UltraFastBERT optimized inference by activating only a fraction of the available neurons, thereby maintaining competitive performance levels \cite{belcak2023exponentially}. Intel CPU clusters can serve LLMs by adopting the algorithms \cite{shen2023efficient}, such as quantization and optimized kernels, specifically designed to accelerate computations on CPUs.

Many optimizations have been proposed by coordinately utilizing GPUs and CPUs to improve the efficiency and speed of Transformer models.
Similarly, \cite{song2023powerinfer} introduces PowerInfer, a high-speed LLM inference engine that leverages the high locality and power-law distribution in neuron activation to reduce GPU memory demands and CPU-GPU data transfers. \cite{zhao2024hetegen} aims to reduce latency by coordinating CPU and GPU computing while mitigating the I/O bottlenecks by overlapping the data processing and I/O time. \cite{sheng2023flexgen} utilizes a single GPU to serve LLMs, prioritizing throughput at the expense of latency. 
It designs a scheduling algorithm to schedule model parameters, KV caches, and activations between CPU, GPU, and disk.
These advancements focus on hardware-independent algorithms and CPU-GPU collaborative inference strategies, signifying a robust movement towards more efficient FMs.

\vspace{-6pt}
\subsection{Memory Optimization}
\label{subsec:memory_optimization}
\subsubsection{Vanilla FMs}

Besides computational costs, memory overhead is another major challenge in deploying LLMs at resource-constrained edge devices. 
The inference process of the LLM involves two key stages: the prefill and decode phases. The prefill phase is responsible for creating the KV cache based on the prompts, while the decode phase focuses on generating the subsequent token autoregressively.
The first stage is compute-intensive, and the primary challenge of the decoding stage lies in the memory wall.
This barrier prevents the loading of the vast parameters associated with LLMs and the maintenance of the KV cache throughout the inference process.
Consequently, considerable research effort has focused on memory scheduling, utilizing multiple storage levels to execute a single model. 
For example, jointly optimizing CPU and GPU memory usage makes it possible to deploy LLMs on personal computers, smartphones, and other weak devices.

Contextual sparsity, which refers to the input-dependent activation of a small subset of neurons in LLMs, has been exploited to reduce computational and memory costs. \cite{liu2023deja} proposes a method to predict contextual sparsity on the fly and an asynchronous, hardware-aware implementation to accelerate LLM inference. \cite{alizadeh2023llm} explores storing model parameters in flash memory and loading only the required subset during inference, considering bandwidth, energy constraints, and read throughput.

Several studies have focused on optimizing the attention mechanism in LLMs to reduce I/O costs. \cite{shazeer2019fast} introduces the Multi-Query Attention to accelerate decoding by reducing memory bandwidth requirements during incremental inference. \cite{ainslie2023gqa} proposes Grouped-Query Attention (GQA), an intermediate between MHA and Multi-Query Attention (MQA), achieving a balance between quality and speed. \cite{kwon2023efficient} introduces PagedAttention, which is inspired by virtual memory and paging techniques, to efficiently manage key-value cache memory and integrate it into vLLM, a high-throughput LLM serving system. \cite{dao2022flashattention} proposes FlashAttention, an I/O-aware attention algorithm that reduces memory accesses between High Bandwidth Memory (HBM) and on-chip Static Random-Access Memory (SRAM) in GPU, which avoids fully loading the large attention matrix, reducing I/O overhead significantly. \cite{dao2023flashattention} further improves FlashAttention to enhance efficiency and scale Transformers to longer sequence lengths by strategically reducing the number of non-matrix-multiplication operations and paralleling the sequence dimension. This adjustment enables improved resource utilization across the GPU thread block. FlashDecoding++ \cite{hong2023flashdecoding} introduces asynchronized partial parallel SoftMax. SoftMax requires subtracting a value from all input numbers before exponential computing to avoid overflow of the denominator. The subtracted value is usually the maximum input value, so synchronization is required to obtain this maximum value when computing softmax in parallel. However, FlashDecoding++ can avoid this synchronization by utilizing a unified max value pre-decided according to the statistical LLM-specified input distribution to the SoftMax.

BMInf\cite{han2022bminf} introduces the Big Model Inference and Tuning toolkit, which employs model quantization, parameter-efficient tuning, and CPU-GPU scheduling optimization to reduce computational and memory costs. \cite{patel2023splitwise} proposes Splitwise, a technique that schedules the prompt computation and token generation phases to different machines, optimizing hardware utilization and overall system efficiency.  
FastServe is a distributed serving system that optimizes job completion time for LLMs in interactive AI applications~\cite{wu2023fast}. 
\cite{miao2023specinfer} introduces SpecInfer, a system that accelerates generative LLM serving through tree-based speculative inference and verification. 
LLMCad is an innovative on-device inference engine specifically designed for efficient generative NLP tasks~\cite{xu2023llmcad}. It executes LLM on weak devices with limited memory capacity by utilizing a compact LLM that resides in memory to generate tokens and a high-precision LLM for validation.

\subsubsection{MoE FMs}

Mixture-of-Experts (MoE), replacing an FFN with a router and multiple FFNs, is a promising approach to enhance the efficiency and scalability of LLMs. However, deploying MoE-LLMs presents challenges in memory scheduling because of their huge number of parameters and the uncertain choices of experts, particularly in resource-constrained environments. Recent research has addressed these challenges through novel architecture designs, model compression techniques, and efficient inference engines. \cite{rajbhandari2022deepspeed} proposes DeepSpeed-MoE, an end-to-end solution for training and deploying large-scale MoE models. \cite{yi2023edgemoe} introduces EdgeMoE, an on-device inference engine designed for MoE-LLMs that loads popular experts to GPU to prioritize memory and computational efficiency. 
\cite{eliseev2023fast} proposes a novel expert offloading strategy utilizing the intrinsic properties of MoE-LLMs. It uses Least Recently Used (LRU) caching to store experts since certain experts are reused across consecutive tokens. It also accelerates the loading process by predicting the selection of experts for future layers based on the hidden states of earlier layers.

Several studies have focused on developing efficient serving systems and inference engines for MoE-LLMs. \cite{xue2024moe} introduces MOE-INFINITY, an efficient MoE-LLMs serving system that implements cost-efficient expert offloading through activation-aware techniques, significantly reducing latency overhead and deployment costs. \cite{kamahori2024fiddler} proposes Fiddler, a resource-efficient inference engine that orchestrates CPU and GPU collaboration to accelerate the inference of MoE-LLMs in resource-constrained settings. Furthermore, \cite{hwang2023pre} introduces Pre-gated MoE, an algorithm-system co-design approach that employs a novel pre-gating function to enhance the inference capabilities of MoE-LLMs by enabling more efficient activation of sparse experts and reducing the memory footprint. These advancements in MoE-LLMs deployment and inference engines demonstrate the ongoing efforts to make these powerful models more accessible and efficient across various computing environments.

\vspace{-6pt}
\subsection{Communication Optimization}

In addition to computation and memory optimization, communication overhead is another major challenge in deploying large models within the edge-cloud environment. The heterogeneity of edge devices and communication channels necessitates a scheduler for computation and network resources when executing FM tasks across different devices. 

\cite{hong2023intelligence} introduces an ``Intelligence-Endogenous Management Platform" for Computing and Network Convergence (CNC), which efficiently matches the supply and demand within a highly heterogeneous CNC environment. The CNC brain is prototyped using a deep reinforcement learning model. It theoretically comprises four key components: perception, scheduling, adaptation, and governance, collectively supporting the entire CNC lifecycle. Specifically, Perception perceives the incoming service request and real-time computing resources. Scheduling assigns the workload to heterogeneous computing nodes in a heterogeneous network. Adaptation is adapting to dynamic resources by ensuring the continuity of services through backup measures, and Governance is the self-governed decentralized computing nodes.

\cite{liu2024resource} discusses the integration of LLMs into 6G vehicular networks, focusing on the challenges and solutions related to computational demands and energy consumption. It proposes a framework where vehicles handle initial LLM computations locally and offloads more intensive tasks to Roadside Units (RSUs), leveraging edge computing and 6G networks' capabilities. The authors formulate a multi-objective optimization framework that aims to minimize the cumulative cost incurred by vehicles and RSUs in processing computational tasks, including the cost associated with communication.
LinguaLinked\cite{zhao2023lingualinked} is a system to deploy LLMs on distributed mobile devices. The high memory and computational demands of these models typically exceed the capabilities of a single mobile device. It utilized load balancing and ring topology to optimize the delay of computation and communication while maintaining the original model structure. The results demonstrate improvements in inference throughput on mobile devices.
\cite{jiang2024megascale} presents MegaScale, a production system designed to train and deploy LLMs across over 10,000 GPUs. The synchronization of model parameters and gradients across GPUs can become a bottleneck as the number of GPUs increases. MegaScale addresses communication challenges through a combination of parallelism strategies, overlapping techniques, and network optimizations, resulting in improved training efficiency and fault tolerance~\cite{wang2021losp}.

Semantic communication enables much more efficient utilization of limited network resources at the edge by only transmitting the essential semantic information needed for communication purposes. \cite{mu2022heterogeneous} investigates multiple access designs to facilitate the coexistence of semantic and bit-based transmissions in future networks. The authors propose a heterogeneous semantic and bit communication framework where an access point simultaneously sends semantic and bit streams to a semantics-interested user and a bit-interested user. \cite{qin2024computing} presents a framework for semantic communications enabled by computing networks, aiming to provide sufficient computational resources for semantic processing and transmission. This framework leverages computing networks to support semantic communication. It introduces key techniques to optimize the network, such as semantic sampling and reconstruction, semantic-channel coding, and semantic-aware resource allocation and optimization based on cloud-edge-end computing coordination. Two use cases, an end-cloud computing-enabled video transmission system, and a semantic-aware task offloading system, are provided to demonstrate the advantages of the proposed framework.

\vspace{-9pt}
\subsection{Integrated Frameworks}

With the emergence of computational and memory optimization methods, numerous integrated frameworks have integrated these techniques, supporting various LLMs and lowering the barriers to LLM deployment.
These frameworks leverage CPU and GPU backends, enabling the widespread local execution of LLMs.

The most popular framework to deploy LLMs at the edge is the llama.cpp\cite{llamacpp}, which pioneered the open-source implementation of LLM execution using only C++. 
We provide an overview of various open-source frameworks and their respective features in \autoref{integrated-frameworks}. 
The frameworks can be categorized into heterogeneous and GPU-only backends, ranging from mobile and embedded devices to high-end computing systems. 
Different suppliers offer specialized development platforms and APIs for their hardware products (i.e., CPU and GPU). CUDA (Compute Unified Device Architecture) is a parallel computing platform developed for NVIDIA GPU. ROCm (Radeon Open Compute platform) is a software platform for high-performance computing using AMD GPU. Metal is a low-overhead hardware-accelerated 3D graphic and compute shader API developed by Apple for its own GPUs. Some frameworks like Vulkan and OpenCL (Open Computing Language) facilitate development across different hardware and operating systems. Vulkan is an API for cross-platform access to GPUs for graphics and computing. OpenCL is a framework for writing programs that execute across heterogeneous backends, including CPU, GPU, FPGA, and other hardware. 



\begin{table*}[htbp]
    \centering
    \caption{Integrated Frameworks}
    \label{integrated-frameworks}
    \begin{tabularx}{\textwidth}{@{}p{2.5cm}XX@{}}
    \toprule
    \textbf{Framework} & \textbf{Features and Functions} &\textbf{Backend}\\
    \midrule
    LLaMA.cpp\cite{llamacpp} & High-performance inference of LLaMA and other LLMs &CPU(x64, ARM), GPU(CUDA, ROCm, Metal), OpenCL \\
    MLC-LLM\cite{mlc-llm} & Accelerating LLMs inference using Machine Learning Compilation &GPU(CUDA, ROCm, Metal, Vulkan, WebGPU), OpenCL\\
    MNN-LLM\cite{mnn-llm} & Inference for Mobile Neural Network LLMs &CPU(x64, ARM), GPU(CUDA), OpenCL\\
    FastChat\cite{zheng2023judging} & Platform for both training and inference of LLM-based chatbots & CPU(x64), GPU(CUDA, ROCm, Metal)\\
    DeepSpeed\cite{aminabadi2022deepspeed} & DeepSpeed-Inference integrates multiple parallelisms and custom kernels, communication, and memory optimizations. &CPU(x64, ARM), GPU(CUDA, ROCm)\\
    OpenVINO\cite{gorbachev2019openvino} & Convert FMs and deploy them on Intel hardware & CPU(x64), GPU(OpenCL)\\
    MLLM\cite{mllm} & Mobile LLMs Inference &CPU(x86, ARM)\\
    FP6\cite{xia2024fp6} & Mixed precision for LLMs &GPU(CUDA) \\
    Colossal-AI\cite{li2023colossal} &  Distributed training and inference for LLMs using multiple parallelisms&GPU(CUDA)\\
    Megatron-LM\cite{narayanan2021efficient} & Efficient LLMs inference on GPUs with system-level optimizations&GPU(CUDA)\\ 
    TensorRT-LLM\cite{tensorrtllm} & NVIDIA TensorRT for LLMs inference &GPU(CUDA)\\
    \bottomrule
    \end{tabularx}
\end{table*}

Recently, several novel frameworks have been specifically designed for the deployment of LLM-based applications/agents.
These frameworks provide abstract interfaces that facilitate the design of complex LLM-based applications, such as chain-of-thought reasoning and retrieval-augmented LLMs.
LangChain is a powerful framework aimed at simplifying the development of applications that interact with LLMs~\cite{Langchain}. It offers a set of tools and abstractions that enable developers to construct complex, dynamic workflows by chaining various operations, such as querying a language model, retrieving documents, or processing data.
Parrot is a serving system for LLM applications that optimizes performance across multiple requests~\cite{lin2024parrot}. It introduces the concept of a semantic variable to define the input or output information of LLM requests.
To enhance system performance, Parrot schedules requests based on a predefined execution graph and latency sensitivity, and it shares the key-value cache among requests to accelerate the prefilling process.
SGLang is another LLM agent framework designed to efficiently execute complex language model programs~\cite{zheng2024sglangefficientexecutionstructured}. It simplifies the programming of LLM applications by providing primitives for generation and parallelism control. Additionally, it accelerates the execution of these applications by reusing key-value caches, enabling faster constrained decoding and designing API speculative execution.

\section{Resource Allocation and Parallelism}
\label{sec:resource}

\subsection{Resource Allocation and Scaling}

\begin{figure}
    \centering
    \includegraphics[width=0.75\linewidth]{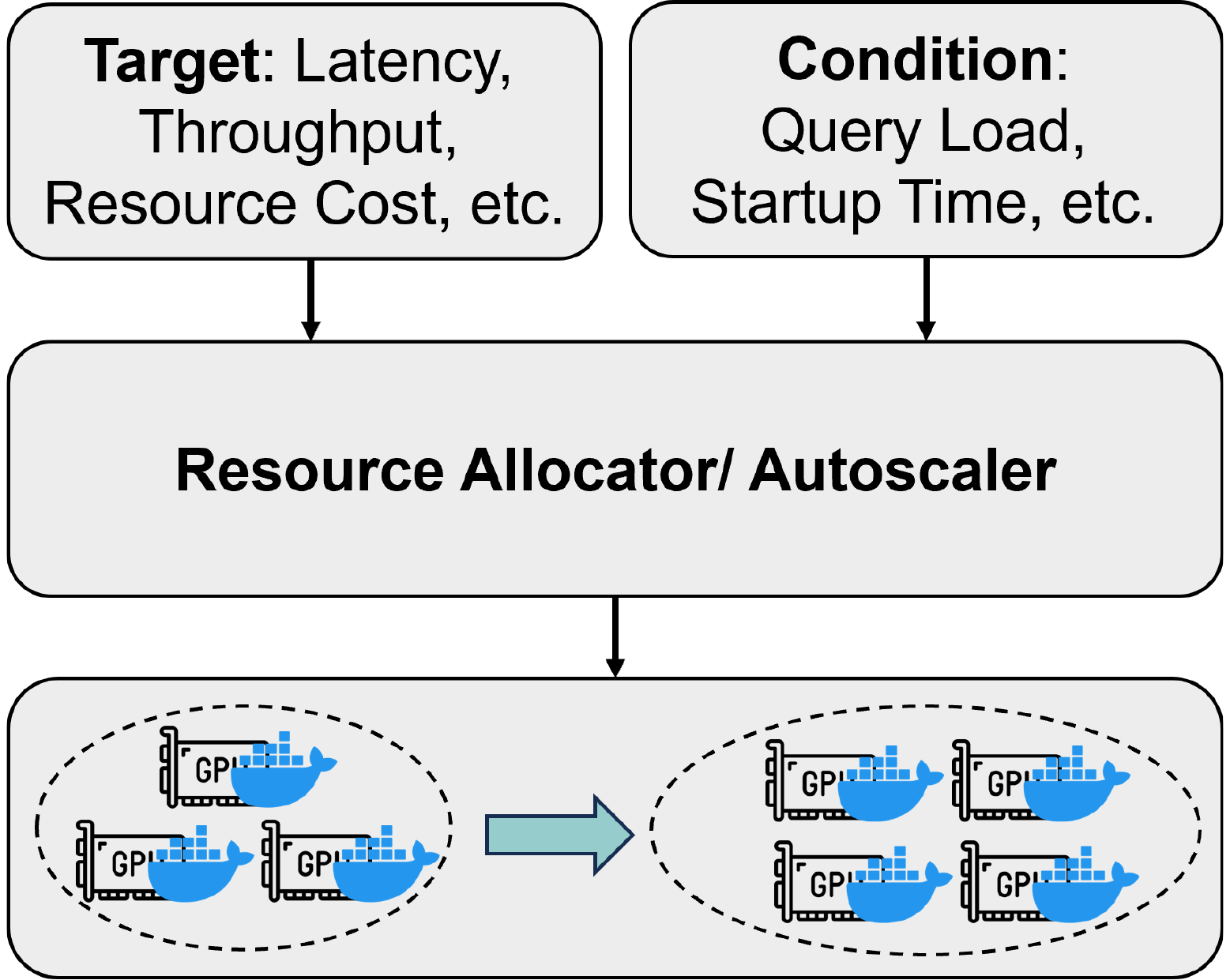}
    \caption{The illustration of resource allocation. Resource allocation in a serving framework primarily involves dynamically adjusting the resource allocation strategy based on real-time resource conditions and query load.}
    \label{fig:cloud}
\end{figure}

Edge-cloud computing leverages strong cloud servers and distributed edge devices to handle tasks near the data source. 
As shown in Figure~\ref{fig:cloud}, adaptive resource allocation is crucial for achieving an optimal balance between system performance and cost in edge-cloud environments. Resource management facilitates the optimal utilization of system resources, including processing power, storage space, and network bandwidth. 
Adaptive algorithms are designed to automatically adjust resource configurations based on real-time workloads and environmental changes, ensuring optimal performance under varying conditions. 

Despite their numerous advantages, several challenges also arise in edge-cloud environments.
1) In edge-cloud environments, resources need to be allocated and managed between the central cloud and multiple edge nodes. Distributed resource management may increase the system's complexity and require more fine-grained scheduling and coordination mechanisms.
2) The load in edge computing scenarios is highly dynamic. Accurately predicting this query load and dynamically adjusting resource allocation demands is difficult. 
3) To meet the real-time processing requirements, the system must quickly adapt to changes in execution environments and adjust processing strategies and resource allocations in time. 
4) Modern computing environments provide a variety of computing resources (CPU, GPU, TPU, etc.). Optimizing the allocation of these heterogeneous resources for various models based on their computational requirements and current workloads is a complex task.

\vspace{-3pt}
We have summarized previous research works on resource allocation and adaptive optimization in Table~\ref{tab:resource allocation and adaptation}. The collected research works are divided into two categories: optimization for model services in cloud environments and optimization for IoT applications in edge computing environments.
\emph{1) Cloud Environments.} This task is to efficiently deploy, manage, and scale machine learning models in the cloud. The objective is optimizing resource usage, reducing computing costs, and achieving performance goals in the face of dynamic and unpredictable query loads. This includes resource provisioning algorithms and strategies for utilizing heterogeneous computing resources.
\emph{2) Edge Environments.} Because of the unique architectures and application scenarios in edge environments, data needs to be processed closer to the user or the geographic location of the data source. This requires well-designed scheduling algorithms and fault tolerance and resilience mechanisms to minimize latency and mitigate network overhead.

\begin{table*}
\centering
\caption{Summary of resource allocation and adaptation methods.}
\label{tab:resource allocation and adaptation}
\begin{tabular}{@{}p{1cm}llp{4cm}p{8cm}@{}}       
\toprule
\textbf{Scenario} & \textbf{Ref.} & \textbf{Year} & \textbf{Target} & \textbf{Method}  \\ 
\midrule
Cloud & Clipper~\cite{crankshaw2017clipper} & 17 & Accuracy, Latency and Throughput & Model Containers.\\
& MArk~\cite{zhang2019mark} & 19 & Latency and Resource Cost & Predictive scaling.
\\
& Nexus~\cite{shen2019nexus} & 19 & Latency and Throughput
& Squishy bin packing.\\
& InferLine~\cite{crankshaw2020inferline} & 20 & Latency and Resource Cost & 1. The low-frequency planner. 2. The high-frequency tuner.\\
& Clockwork~\cite{gujarati2020serving} & 20 & Latency & DNN Workers.\\
& INFaaS ~\cite{romero2021infaas} & 21 & Latency, Throughput and Resource Cost & Dynamic Model Variant Selection and Scaling.\\
& Morphling~\cite{wang2021morphling} & 21 & Resource Cost & Model-Agnostic Meta-Learning.\\
& Cocktail~\cite{gunasekaran2022cocktail} & 22 & Accuracy, Latency and Resource Cost & 1. Resource controller. 2. Autoscaler.\\
& Kairos~\cite{li2023kairos} & 23 & Throughput & Query-distribution mechanism.\\
& SHEPHERD~\cite{zhang2023shepherd} & 23 & Throughput and Utilization & HERD: Planner for Resource Provisioning.\\
& SpotServe~\cite{miao2023spotserve} & 23 & Resource Cost & Device Mapper.\\

Edge & Na et al.~\cite{na2018frequency} & 18 & Throughput and Latency & Resource allocation scheme based on Lagrangian and Karush-Kuhn-Tucker conditions.\\
& Avasalcai et al.~\cite{avasalcai2019decentralized} & 19 & Latency & Deployment policy module.\\
& Yang et al.~\cite{yang2019joint} & 19 & Latency & 1. Multi-Dimensional Search and Adjust (MDSA) Algorithm. 2. Cooperative Online Scheduling (COS) Method.\\
& Tong et al.~\cite{tong2020adaptive} & 20 & Latency and Energy consumption & Deep Reinforcement Learning (DRL) Approach.\\
& Xiong et al.~\cite{xiong2020resource} & 20 & Latency and Utilization & Improved deep reinforcement learning (DQN) Algorithm.\\
& CE-IOT~\cite{zhou2020iot} & 20 & Resource Cost & 1. Delay-Aware Lyapunov Optimization Technique 2. Economic-Inspired Greedy Heuristic.\\
& Chang et al.~\cite{chang2020dynamic} & 20 & Latency and Energy consumption & Online Algorithm for Real-Time Decision-Making.\\
& LaSS~\cite{wang2021lass} & 21 & Latency and Utilization & Model-driven approaches.\\
& Ascigil et al.~\cite{ascigil2021resource} & 21 & Latency and Utilization & Decentralized Strategies.\\
& KneeScale~\cite{li2022kneescale} & 22 & Throughput, Latency and Utilization.
& Adaptive Auto-scaling with Knee Detection.\\
& CEC~\cite{hu2022cec} & 22 & Latency, Utilization and Accuracy & Control-Based Resource Pre-Provisioning Algorithm (PRCT).\\
\bottomrule
\end{tabular}\\

\end{table*}
    
\emph{1) Cloud Environments.} The following articles present various advanced solutions to optimize resource allocation in the cloud.
Clipper uses model containers to encapsulate the model inference process in a Docker container~\cite{crankshaw2017clipper}. Clipper supports replicating these model containers across the cluster to increase the system throughput and utilize additional hardware accelerators for serving. 
MArk is a generic inference system built on Amazon Web Services. It utilizes serverless functions to address service delays with horizontal and vertical scaling.
Horizontal scaling expands the system by adding more hardware instances, while vertical scaling expands the system by increasing the resources of a single instance~\cite{zhang2019mark}. 
MArk uses a Long Short-Term Memory (LSTM) network for multi-step workload prediction. Leveraging the workload prediction results, MArk determines the instance type and quantity required to meet the SLOs using a heuristic approach.
Nexus adopts the squishy bin packing method to batch different types of tasks on the same GPU, enhancing resource efficiency by considering the latency requirements and execution costs of each task~\cite{shen2019nexus}. It also merges multiple tasks into the same GPU execution cycle as long as the latency constraints are not violated.
InferLine utilizes a low-frequency planner and a high-frequency tuner to manage the machine learning prediction pipeline effectively~\cite{crankshaw2020inferline}. The low-frequency combinatorial planner finds the cost-optimal pipeline configuration under a given latency SLO. The high-frequency auto-scaling tuner monitors the dynamic request arrival pattern and adjusts the number of replicas for each model. 
Clockwork achieves an adaptive resource management framework through a fine-grained central controller over worker scheduling and resource management~\cite{gujarati2020serving,wang2024fedcda}. DNN workers pre-allocate all GPU memory and divide it into three categories: workspace, I/O cache, and page cache. This can avoid repeated memory allocation calls and improve predictability.
To cope with changes in query load, INFaaS adopts two automatic scaling mechanisms: vertical auto-scaling at the model level and horizontal auto-scaling at the Virtual Machine (VM) level~\cite{romero2021infaas}. The model auto-scaling is handled by the Model-Autoscaler, which decides each model variant's scaling operations (replication, upgrade, or downgrade) by solving an integer linear programming (ILP) problem.
The vertical auto-scaling adds a new VM if the utilization of any hardware resource exceeds a configurable threshold. 
Morphling utilizes model-agnostic meta-learning techniques to effectively navigate the high-dimensional configuration space, such as CPU cores, GPU memory, GPU timeshare, and GPU type~\cite{wang2021morphling}. It can significantly reduce the cost of configuration search and quickly find near-optimal configurations. 
Cocktail designs a resource controller to manage CPU and GPU instances in a cost-optimized manner and a load balancer to allocate queries to appropriate instances~\cite{gunasekaran2022cocktail}. It also proposes an autoscaler that leverages predictive models to predict future request loads and dynamically adjusts the number of instances in each model pool based on the importance weight of the models.
Kairos designs a query distribution mechanism to intelligently allocate queries of different batch sizes to different instances in order to maximize throughput~\cite{li2023kairos}. 
Kairos transforms the query distribution problem into a minimum-cost bipartite matching problem and uses a heterogeneity coefficient to represent the relative importance of different types of instances. This allows for better balancing of the resource usage of different instance types.
SHEPHERD comprises a planner (HERD), a request router, and a scheduler (FLEX) for each serving group~\cite{zhang2023shepherd}. HERD is responsible for partitioning the entire GPU cluster into multiple service groups. Besides, it performs periodic planning and informs each GPU worker of their designated service group and the models it must serve. 
SpotServe is a serverless LLM system that adjusts the GPU instances and updates the parallelism strategy flexibly~\cite{miao2023spotserve}. 
It uses a bipartite graph matching algorithm (Kuhn-Munkres algorithm) to orchestrate model layers to hardware devices, thus maximizing the reusable model parameters and key-value caches.
    
\emph{2) Edge Environments.} Given the limited and heterogeneous nature of edge computing resources, efficient management and scaling of these resources are crucial. The following papers present efficient solutions to improve the performance and efficiency of edge systems.
To optimize resource allocation and interference management, \cite{na2018frequency} proposes a resource allocation scheme with Lagrangian and Karush-Kuhn-Tucker conditions. Based on the number of associated IoT end devices (IDs) and the queue length of the edge gateways (EGs), this scheme calculates the optimal resource allocation parameters and allocates resource blocks to each EG. 
A decentralized resource management technique is proposed in ~\cite{avasalcai2019decentralized} to deploy latency-sensitive IoT applications on edge devices. 
The key design is the deployment policy module, which is responsible for finding a task allocation scheme that meets the application's requirements based on the bids from the bidder nodes and deploying the application to the corresponding edge nodes.
The authors in ~\cite{yang2019joint} introduce a novel method, called MDSA, to address the challenge of joint model partitioning and resource allocation for latency-sensitive applications in mobile edge clouds. 
They employ the task-oriented online scheduling method, COS, to collaboratively balance the workload across computing and network resources, thereby preventing excessive wait times. 
\cite{tong2020adaptive} designs a DRL algorithm to determine whether a task needs to be offloaded and to allocate computing resources efficiently. 
Tasks generated by mobile user equipment (UE) will be submitted to the task queue wait for execution. The algorithm models the task queue as a Poisson distribution. Then it uses the DRL method to select the suitable computing node for each task, learning the optimal strategy during the algorithm training process.
Authors of \cite{xiong2020resource} formulate the resource allocation problem using a Markov decision process model. They enhance the DQN algorithm by incorporating multiple replay memories to refine the training process. This modification involves using different replay memories to store experiences under different circumstances. 
The CE-IoT is an online cloud-edge resource provisioning framework based on delay-aware Lyapunov optimization to minimize operational costs while meeting latency requirements of requests~\cite{zhou2020iot}. The framework allows for online resource allocation decisions without prior knowledge of system statistics. 
The authors of \cite{chang2020dynamic} propose a dynamic optimization scheme that coordinates and allocates resources for multiple mobile devices in fog computing systems. They introduce a dynamic subcarrier allocation, power allocation, and computation offloading scheme to minimize the system execution cost through Lyapunov optimization. 
LaSS is a platform designed to run latency-sensitive serverless computations on edge resources~\cite{wang2021lass}. LaSS employs model-driven strategies for resource allocation, auto-scaling, fair-share allocation, and resource reclamation to efficiently manage serverless functions at the edge. 
In~\cite{ascigil2021resource}, the authors discuss resource provisioning and allocation in FaaS edge-cloud environments, considering decentralized approaches to optimize CPU resource utilization and meet request deadlines. This paper design resource allocation and configuration algorithms with varying degrees of centralization and decentralization.
KneeScale optimizes resource utilization for serverless functions by dynamically adjusting the number of function instances until reaching the knee point, where increasing resource allocation no longer provides significant performance benefits~\cite{li2022kneescale}. KneeScale utilizes the Kneedle algorithm to detect the knee points of functions through online monitoring and dynamic adjustment.
CEC is a containerized edge computing framework that integrates workload prediction and resource pre-provisioning~\cite{hu2022cec}. It adjusts the resource allocation of the container through the PRCT algorithm to achieve zero steady-state error between the actual response time and the ideal response time. 

\vspace{-6pt}
\subsection{Parallelism}

\subsubsection{Parallelism type}

\begin{figure}
    \centering
    \includegraphics[width=1\linewidth]{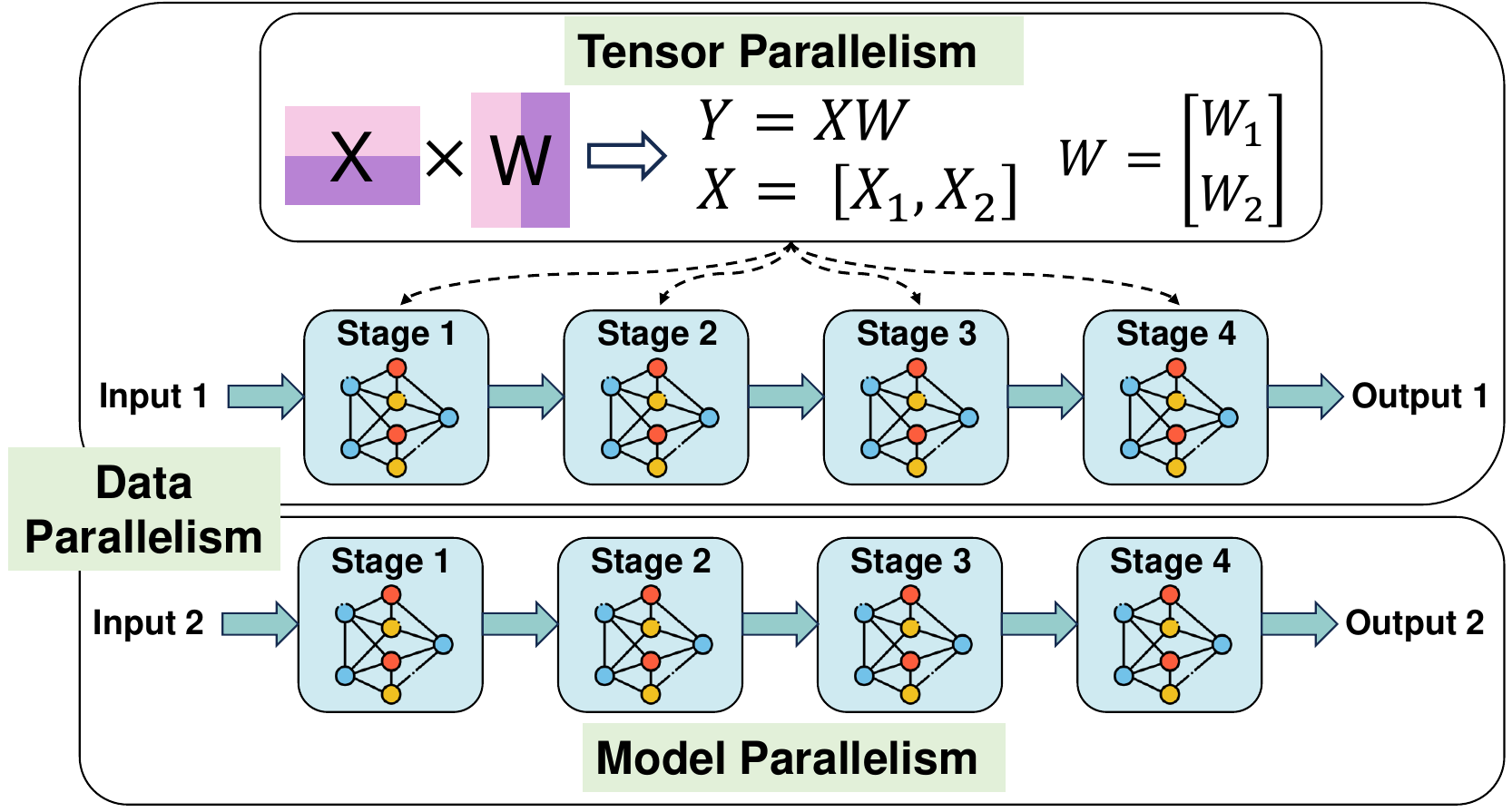}
    \caption{The illustration of different parallelism methods. Data parallelism divides the data into multiple micro-batches for processing. Model parallelism partitions a model into several modules (stages). Tensor parallelism splits a tensor into various segments.}
    \label{fig:parallelism}
\end{figure}

\begin{figure}
    \centering
    \includegraphics[width=0.9\linewidth]{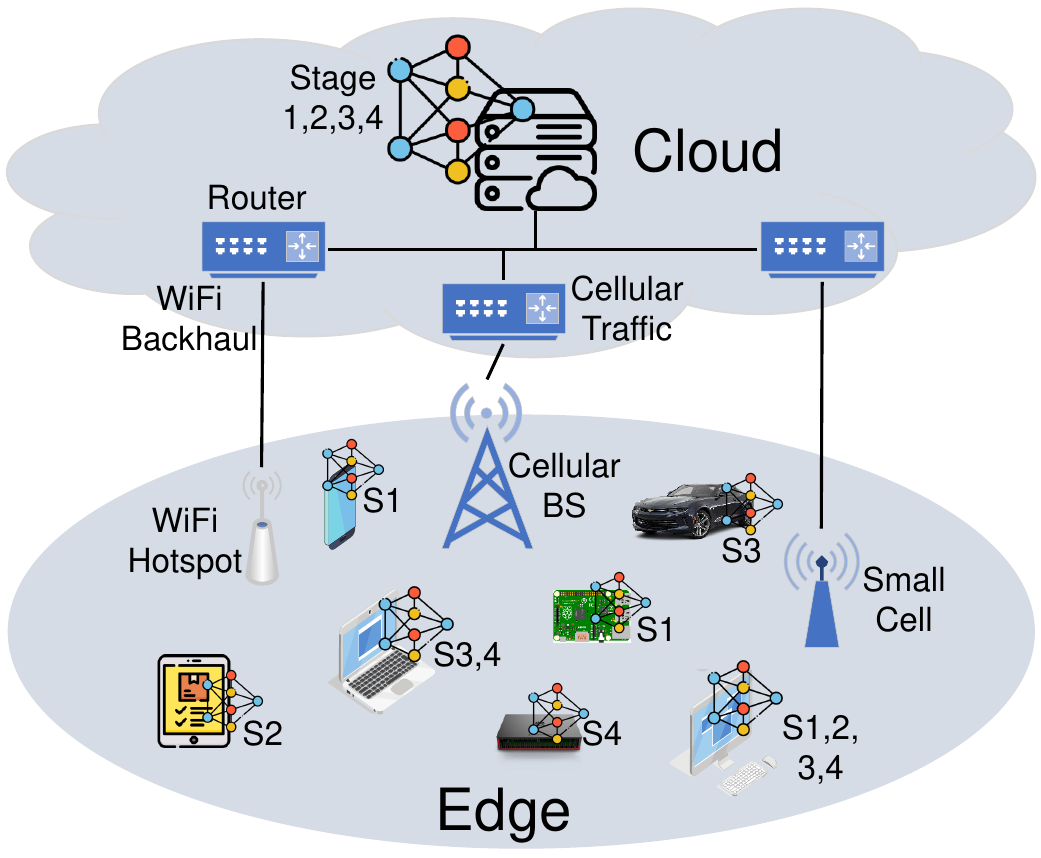}
    \caption{Orchestrate different sub-models to different devices. Lots of research works design auto-parallelism methods to deploy sub-models on different devices according to the execution environment.}
    \label{fig:para_system}
\end{figure}

\begin{table*}
\centering
\caption{Summary of parallel methods.}
\label{tab:parallelism}
\begin{tabular}{@{}p{1cm}llp{1.5cm}p{8cm}@{}}
\toprule
\textbf{Model} & \textbf{Ref.} & \textbf{Type} & \textbf{Target} & \textbf{Design}  \\ 
\midrule
LLM & Megatron-lm~\cite{shoeybi2019megatron}  & TP & Latency & Introduce an intra-layer model parallelism method. \\
 & DeepSpeed-inference~\cite{aminabadi2022deepspeed} & TP, MP, DP & Latency and Throughput & 1.  Leverages heterogeneous memory systems; 2. Custom GEMM operations, kernel fusion, and memory access optimizations. \\
 & Pope et al.~\cite{pope2023efficiently} & TP, MP, DP & Latency and Throughput & Get the best partitioning strategy for a given model size with specific application requirements. \\
 & AlpaServe~\cite{li2023alpaserve} & MP & Latency & Model parallelism \& Statistically multiplexing multiple devices when serving multiple models. \\
 & LightSeq~\cite{li2023lightseq} & Sequence parallelism & Throughput & Partitioning solely the input tokens. \\
 & PETALS~\cite{borzunov2024distributed} & MP, DP & Throughput & Fault-tolerant inference algorithms and load-balancing protocols. \\
  & SpotServe~\cite{miao2023spotserve} & TP, MP, DP & Latency & 1. Dynamic Re-Parallelization. 2. Instance Migration Optimization. 3. Stateful Inference Recovery. \\
  & SARATHI~\cite{agrawal2023sarathi} & MP, DP & Throughput & Constructs a batch using a single prefill chunk and fills the remaining batch with decode requests. \\
  & DistServe~\cite{zhong2024distserve} & MP, DP & Latency & Assigns prefill and decoding computation to different GPUs. \\
 Small Model & Zhou et al.~\cite{zhou2019distributing} & MP & Latency & Utilizes dynamic spatial partitioning and layer fusion techniques to optimally distribute the DNN computation across multiple devices. \\
 & DINA~\cite{mohammed2020distributed} & MP & Latency & A fine-grained, adaptive DNN partitioning and ofﬂoading strategy. \\
  & CoEdge~\cite{zeng2020coedge} & MP & Latency and Resource &  Exploiting model parallelism and adaptive workload partitioning across heterogeneous edge devices. \\
  & Li et al.~\cite{li2021throughput} & MP & Throughput & Partitioning the DNN model on the IOT devices and a cloudlet. \\
  & JellyBean~\cite{wu2022serving} & MP & Throughput & Create optimized execution plans for complex ML workflows on diverse infrastructures. \\
  &  PipeEdge~\cite{9996638} & MP, DP & Throughput & An optimal partition strategy that accounts for the heterogeneity in computing power, memory capacity, and network bandwidth. \\
  & PDD~\cite{wu2023pdd} & MP & Latency & A multipartitioning and offloading approach for streaming tasks with a Directed Acyclic Graph topology. \\
  & B\&B~\cite{feltin2023dnn} & MP & Throughput & 1. Latency Modeling and Prediction. 2. A branch and bound solver for DNN partitioning. \\
  & Li et al.~\cite{li2024distributed} & MP & Latency &  Fine-grained model partitioning mechanism with multi-task learning based A3C approach.\\
  & MoEI~\cite{liu2024moei} & MP & Latency & Optimizes model partition and service migration in mobile systems. \\
 \bottomrule
\end{tabular}\\[5pt]
TP: Tensor Parallelism; MP: Model Parallelism; DP: Data Parallelism.
\end{table*}

The large FMs, such as GPT series and Llama series, and other transformer-based architectures, have significantly advanced the capabilities of AI applications. However, these models come with a substantial increase in computational requirements due to their size and complexity.
Parallelism, including data parallelism (DP), model parallelism (MP), pipeline parallelism (PP), and tensor parallelism (TP), is a critical design aspect that addresses the scalability, efficiency, and performance challenges associated with deploying and serving large-scale machine learning models.
As shown in Figure~\ref{fig:parallelism}, data parallelism is a technique where data is split into smaller batches and distributed across multiple processors. Different machines can execute the inference simultaneously, thus significantly improving the throughput.
Model parallelism aims at splitting the model across different processors, with each processor responsible for a portion of the model's layers or parameters. This approach is useful for very large models that cannot fit into the memory of a single processor. 
Model parallelism can be more challenging to implement than data parallelism because it requires careful partitioning of the model and management of the dependencies between layers.
Pipeline parallelism includes data parallelism and model parallelism, where the data passes through the model in a sequential manner, transitioning from one processor to another as it traverses different layers of the model.
Tensor parallelism is a more fine-grained approach than model parallelism. It splits the individual operations within a layer across multiple processors. For example, if a layer performs a large matrix multiplication, the computation of this matrix can be distributed across multiple processors. 
This approach can be particularly useful for operations that are computationally intensive and can be easily parallelized, albeit at the cost of increased communication volume.
Parallelism enhances performance by enabling simultaneous processing, significantly reducing execution time and increasing the scalability of applications to handle lots of users and more complex models.

There are still some challenges when designing a parallelism strategy for FM in the edge-cloud environment. 
1. Memory Constraints: Large models may not fit into the memory of a single GPU, necessitating strategies to distribute them across multiple processing units.
2. Computational Load: The volume of computations required for a single request can be substantial, requiring efficient distribution of computational tasks.
3. Latency Requirements: Applications often require real-time responses, imposing strict latency constraints on the serving infrastructure.
4. Scalability: The ability to serve a large number of concurrent requests without degradation in performance is crucial for ensuring user's satisfaction.
5. Heterogeneous environment: Edge devices can vary widely in their computation and communication capabilities, operating systems, and available software. A parallelism strategy must be adaptable to different platforms and capable of optimizing execution depending on the specific characteristics of each device.

\vspace{-3pt}
\subsubsection{Auto-parallelism}

Existing research works design different automatic parallelism methods in cloud and edge scenarios.
As shown in Figure~\ref{fig:para_system}, the key problem lies in determining an optimal partitioning scheme of a model and strategically allocating each stage to an appropriate device.
We summarize some automatic parallelism methods in Table~\ref{tab:parallelism}.
NVIDIA Triton and Tensorflow-serving are two famous inference frameworks for machine learning models~\cite{triton2024,tensorflow_serving2024}.
Megatron-lm introduces an efficient intra-layer model parallel approach (i.e., tensor parallelism) that allows for the training of transformer models with billions of parameters without requiring new compilers or significant library changes, fully implementable in PyTorch~\cite{shoeybi2019megatron}. 
DeepSpeed-inference is a comprehensive system designed to address the challenges of efficiently executing transformer model inference at large scales with heterogeneous memory systems, custom GEMM operations, and more~\cite{aminabadi2022deepspeed}.
To understand and optimize the trade-offs (e.g., efficiency and latency) for LLM inference, Pope et al. develop an abstract and powerful partitioning framework designed for model parallelism~\cite{pope2023efficiently}.
It allows for dynamically analyzing the best partitioning strategy based on the specific requirements of a given model size and application scenario.
AlpaServe presents a novel approach that harnesses the power of model parallelism to scale and statistically multiplex multiple devices, enabling efficient serving of multiple models~\cite{li2023alpaserve}.
This approach is particularly beneficial in scenarios where workloads are bursty and the demand fluctuates significantly.
LightSeq designs a sequence parallelism method for long-context transformers that performs partitioning solely the input tokens~\cite{li2023lightseq}.
PETALS explores cost-efficient methods for inference and fine-tuning of LLMs on consumer GPUs that are connected by the Internet~\cite{borzunov2024distributed}.
This approach could enable the pooling of idle compute resources from multiple research groups and volunteers to run LLMs efficiently.
It introduces two main innovations: fault-tolerant inference algorithms and load-balancing protocols.
SpotServe is a novel system designed to serve generative LLMs on preemptible GPU instances in cloud environments~\cite{miao2023spotserve}.
Preemptible instances offer a cost-effective solution for accessing spare GPU resources at significantly reduced prices, although they can be interrupted or terminated at any moment.
SpotServe addresses this challenge with dynamic re-parallelization, instance migration optimization and stateful inference recovery.
SARATHI detects inference bubbles in pipeline parallelism caused by the imbalance (i.e., different execution times) between two distinct phases in the LLM: the prefill phase and the decode phase~\cite{agrawal2023sarathi}.
To tackle this issue, SARATHI addresses it through a decode-maximal batching approach and a chunked-prefills approach. This method divides a prefill request into equal-sized chunks and constructs a batch by utilizing a single chunk from the chunked-prefills and then populates the remaining batch slots with decode requests.
DistServe enhances the performance of serving LLMs by disaggregating the prefill and decoding phases~\cite{zhong2024distserve}.
The disaggregation allows each phase to be assigned to different GPUs, eliminating interferences between prefill and decoding operations and allowing for tailored resource allocation and parallelism strategies for each phase.

Some research works design parallelism methods to deploy models on edge devices~\cite{wang2024fednlr}.
Zhou et al. present a framework for deploying deep neural network (DNN) inference on edge devices using model partitioning with containers~\cite{zhou2019distributing}.
This approach utilizes dynamic model partitioning and layer fusion techniques to optimally distribute the DNN computation across multiple devices, considering the available computational resources and network conditions.
DINA is a system designed to optimize the deployment of DNN across edge devices in fog computing environments~\cite{mohammed2020distributed}.
DINA employs a fine-grained, adaptive DNN partitioning and ofﬂoading strategy.
By leveraging matching theory, DINA dynamically adapts the partitioning and ofﬂoading process based on real-time network conditions and device capabilities, aiming to minimize total latency for DNN inference tasks. 
CoEdge is designed to facilitate cooperative DNN inference across heterogeneous edge devices by dynamically partitioning the DNN inference workload with an adaptive algorithm~\cite{zeng2020coedge}.
Li et al. partition DNN models and utilize parallel processing techniques to meet the real-time requirements of various applications in mobile edge computing (MEC) environments~\cite{li2021throughput}. 
They design an approximation algorithm and an online algorithm to determine the model partitioning and placement on a cloudlet as well as the local device.
JellyBean aims to optimize machine learning inference workflows across heterogeneous computing infrastructures~\cite{wu2022serving}.
JellyBean performs model selection and worker assignment to reduce the costs associated with computing and network resources and meet the constraints of input throughput and accuracy.
Recognizing the computational and memory limitations of edge devices, PipeEdge implements a strategic partitioning approach that considers the diversity in computational power, memory size, and network speed across different devices~\cite{9996638}. 
It designs a dynamic programming algorithm to find the optimal partitioning strategy for distributing the DNN inference tasks.
PDD designs an efficient partitioning and offloading method based on greedy and dichotomy principles for 
 DNNs with directed acyclic graph in streaming tasks~\cite{wu2023pdd}.
B\&B first introduces a prediction model for predicting both inference and transmission latency in distributed DNN deployments. They formulate the optimization problem for DNN partitioning and present a branch and bound solver to tackle this problem~\cite{feltin2023dnn}.
Li et al. investigate a multi-task learning approach with asynchronous advantage actor-critic (A3C) to optimize model partitioning in MEC networks for reducing inference delay~\cite{li2024distributed}.
MoEI is a task scheduling framework for device-edge systems that optimizes model partition and service migration in an MEC environment~\cite{liu2024moei}. 
They develop two algorithms: one based on game theory for offline optimization and another leveraging proximal policy optimization for online, adaptive decision-making processes in a distributed environment.

\section{Foundation Model \& Compression}
\label{sec:model}

\subsection{Current Foundation Model}
\begin{figure}[t]
\centering
    \includegraphics[width=0.45\textwidth]{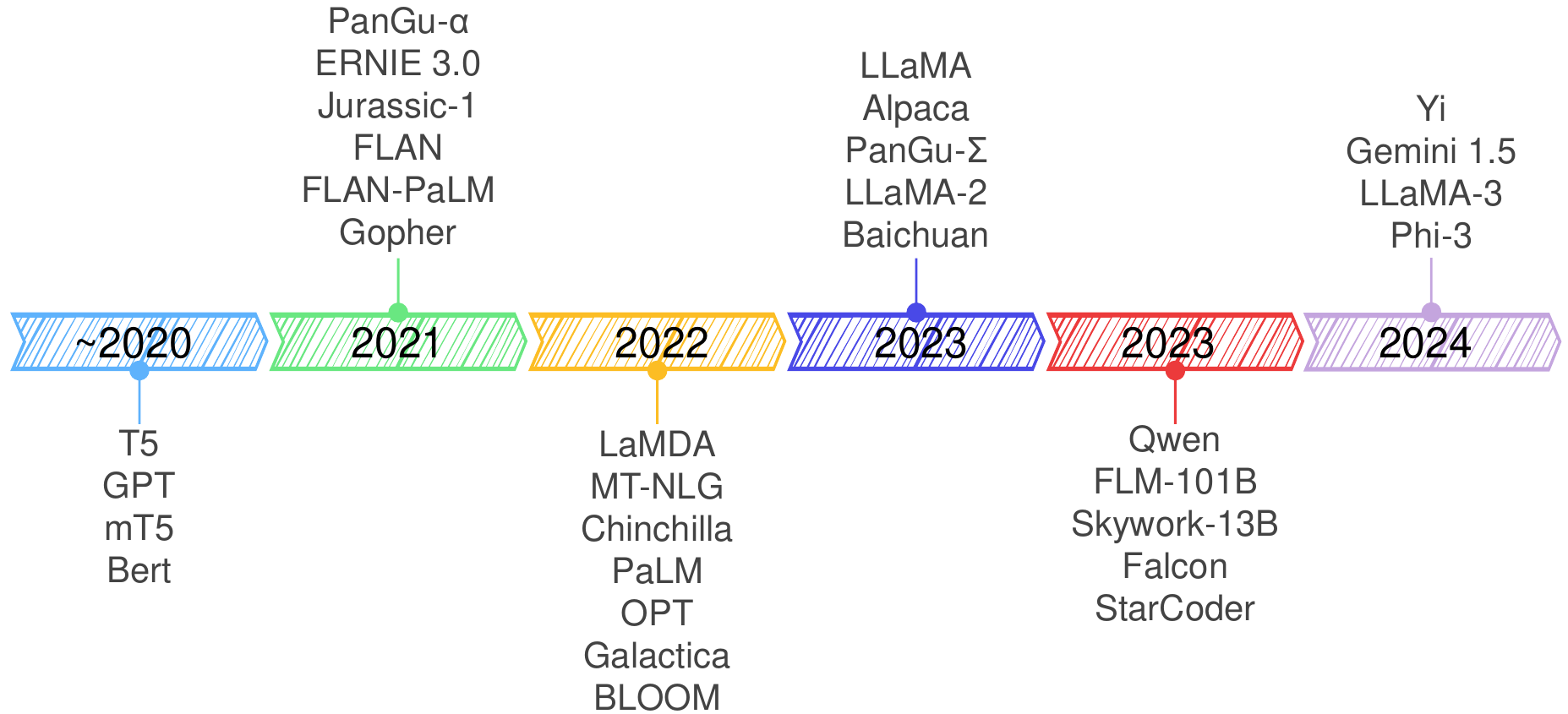}
    \caption{The timeline of some popular LLMs.}
    \label{fig:LLM-Timeline}
\end{figure}

\begin{table*}[]
\centering
\caption{Basic Language Models}
\label{tab:currentfm}
\begin{tabular}{@{}llllllllp{1.5cm}@{}}
\toprule
\textbf{Model} &
  \textbf{Time(Y.M)} &
 \textbf{$n_{para}$} &
  \textbf{$n_{layer}$} &
  \textbf{$n_{head}$} &
  \textbf{$d_{model}$} &
  \textbf{AF} &
  \textbf{Attention Type} &
  \textbf{PE} \\ \midrule
T5           & 19.10 & 60M$\sim$11B       & 6$\sim$24   & 8$\sim$128  & 512$\sim$1024   & ReLU       & Multi-head    & Relative    \\
GPT-3        & 20.05 & 125M$\sim$175B     & 12$\sim$96  & 12$\sim$96  & 768$\sim$12288  & GELU       & Multi-head    & Sinusoidal  \\
PanGu-$\alpha$ &
  21.04 &
  2.6B$\sim$207.0B &
  32$\sim$64 &
  40$\sim$128 &
  2560$\sim$16384 &
  GELU &
  Multi-head &
  Learned \\
ERNIE 3.0   & 21.07 & 10B                & 48, 12      & 64, 12      & 4096, 768       & GELU       & Multi-head    & Relative   \\
Jurassic-1   & 21.08 & 7.5B, 178B         & 32, 76      & 32, 96      & 4096, 13824     & GELU       & Multi-head    & Sinusoidal  \\
Gopher       & 21.12 & 44M$\sim$280B      & 8$\sim$80   & 16$\sim$128 & 512$\sim$16384  & GELU       & Multi-head    & Relative    \\
LaMDA        & 22.01 & 2B$\sim$137B       & 10$\sim$64  & 40$\sim$128 & 2560$\sim$8192  & gated-GELU & Multi-head    & Relative    \\
MT-NLG       & 22.01 & 530B               & 205         & 128         & 20480           & GELU      & Multi-head    & \textbackslash{}           \\
Chinchilla   & 22.04 & 44M$\sim$16.183B    & 8$\sim$47   & 8$\sim$40   & 512$\sim$5120   & \textbackslash{}          & Multi-head    & \textbackslash{}           \\
PaLM         & 22.04 & 8.63B$\sim$540.35B & 32$\sim$118 & 16$\sim$48  & 4096$\sim$18432 & SwiGLU     & Multi-query   & RoPE        \\
OPT*        & 22.05 & 125M$\sim$175B     & 12$\sim$96  & 12$\sim$96  & 768$\sim$12288  & ReLU       & Multi-head?   & RoPE       \\
Galactica*   & 22.11 & 125M$\sim$120.0B   & 12$\sim$96  & 12$\sim$80  & 768$\sim$10240  & GELU       & Multi-head   & Learned     \\
BLOOM*       & 22.11 & 559M$\sim$176.274B  & 24$\sim$70  & 16$\sim$112 & 1024$\sim$14336 & GELU       & Multi-head    & Alibi       \\
LLaMA*       & 23.02 & 6.7B$\sim$65.2B    & 32$\sim$80  & 32$\sim$64  & 4096$\sim$8192  & SwiGLU     & Multi-head    & RoPE        \\
LLaMA 2*     & 23.07 & 7B$\sim$70B        & 32$\sim$80  & 32$\sim$64  & 4096$\sim$8192  & SwiGLU     & Grouped-query & RoPE        \\
Baichuan*    & 23.09 & 7B, 13B            & 32, 40      & 32, 40      & 4096, 5120      & SwiGLU     & Multi-head    & RoPE, AliBi \\
Qwen*        & 23.09 & 1.8B$\sim$14B      & 24$\sim$40  & 16$\sim$40  & 2048$\sim$5120  & SwiGLU     & Multi-head    & RoPE        \\
Skywork-13B* & 23.10 & 13B                & 52          & 36          & 4608            & SwiGLU     & Multi-query   & RoPE        \\
Falcon* &
  23.11 &
  7B$\sim$170B &
  32$\sim$80 &
  64 &
  4544$\sim$14848 &
  GELU &
  Multi-group &
  RoPE\\
StarCoder*   & 23.12 & 15.5B              & 40          & 48          & 2048            & \textbackslash{}          & Multi-query   & Learned     \\
Yi*          & 24.03 & 6B, 34B            & 32, 60      & 32, 56      & 4096, 7168      & SwiGLU     & Grouped-query & RoPE        \\
Llama 3*         & 24.04 & 8B, 70B            & \textbackslash{}      & \textbackslash{}      & \textbackslash{}      & SwiGLU     & Grouped-query & RoPE       \\
\bottomrule
\end{tabular}%
\\[5pt]
* indicates open-source. PE: Positional Embedding. AF: Activation Function
\end{table*}

\begin{table}[t]
\centering
\caption{Some Instruction-Tuned Models}
\label{tab:instruction-tuned}
\begin{tabular}{@{}llll@{}}
\toprule
\textbf{Model} & \textbf{Time(Y.M)} & \textbf{$n_{para}$} & \textbf{Basic LM} \\ \midrule
mT5            & 20.10              & 300M$\sim$13B       & T5                \\
FLAN           & 21.09              & 137B                & LaMDA             \\
Flan-T5        & 21.10              & 80M$\sim$11B        & FLAN, T5          \\
Flan-cont-PaLM & 21.10              & 62B                 & FLAN, PaLM        \\
Flan-U-PaLM    & 21.10              & 540B                & FLAN, U-PaLM      \\
Alpaca         & 23.03              & 7B                  & LLaMA-7B          \\
FLM-101B       & 23.09              & 101B                & FreeLM               \\ \bottomrule
\end{tabular}
\end{table}

\begin{table*}[]
\centering
\caption{Some Popular Multimodal Models}
\label{tab:Multimodal Models}
\begin{tabular}{@{}llllllllll@{}}
\toprule
\textbf{Model} &
  \textbf{Time} &
  \textbf{Language Model} &
  \textbf{Vision Model} &
  \textbf{Vision$\rightarrow$Language} &
  \multicolumn{2}{c}{\textbf{Input}} &
  \multicolumn{3}{l}{\textbf{Output}} \\ \midrule
Flamingo &
  22.04 &
  \begin{tabular}[c]{@{}l@{}}Chichilla 70B \end{tabular} &
  NFNet F6 &
  Perceiver Resampler &
  \multicolumn{2}{l}{text, image, video} &
  \multicolumn{3}{l}{text} \\
MiniGPT-4 &
  23.04 &
  Vicuna &
  EVA-CLIP ViT-g/14 &
  Linear &
  \multicolumn{2}{l}{text, image, box} &
  \multicolumn{3}{l}{text, box} \\
mPLUG-owl &
  23.04 &
  Vicuna &
  CLIP-ViT-L/14 &
  Linear &
  \multicolumn{2}{l}{text, image} &
  \multicolumn{3}{l}{text} \\
PandaGPT&
  23.05 &
  Vicuna &
  ImageBind &
  Linear &
  \multicolumn{2}{l}{\begin{tabular}[c]{@{}l@{}}text, image, audio, video,\\ depth, thermal, IMU\end{tabular}} &
  \multicolumn{3}{l}{text} \\
Shikra &
  23.06 &
  Vicuna &
  CLIP ViT-L/14 &
  Linear &
  \multicolumn{2}{l}{text, box, image, point} &
  \multicolumn{3}{l}{text, point, box} \\
Qwen-VL &
  23.08 &
  Qwen-7B &
  CLIP ViT-G/14 &
  Cross Attention &
  \multicolumn{2}{l}{text, image, box} &
  \multicolumn{3}{l}{text, box} \\
NExT-GPT &
  23.09 &
  Vicuna &
  ImageBind &
  Linear &
  \multicolumn{2}{l}{text. image, audio, video} &
  \multicolumn{3}{l}{text. image, audio, video} \\
CogVLM &
  23.09 &
  \begin{tabular}[c]{@{}l@{}}Vicuna\end{tabular} &
  EVA-CLIP Vit-E/14 &
  MLP &
  \multicolumn{2}{l}{text, image} &
  \multicolumn{3}{l}{text} \\
Ferret&
  23.10 &
  Vicuna &
  CLIP-ViT-L/14 &
  Layer &
  \multicolumn{2}{l}{text, image} &
  \multicolumn{3}{l}{text} \\
OneLLM &
  23.12 &
  LLaMA-2 &
  CLIP-ViT &
  Transformer &
  \multicolumn{2}{l}{\begin{tabular}[c]{@{}l@{}}text, image, video, audio,\\ point, IMU, fMRI\end{tabular}} &
  \multicolumn{3}{l}{text} \\
NExT-Chat &
  23.12 &
  Vicuna &
  CLIP-ViT &
  \textbackslash{} &
  \multicolumn{2}{l}{text, image, box} &
  \multicolumn{3}{l}{text, box, mask} \\
Gemini &
  23.12 &
  \textbackslash{} &
  \textbackslash{} &
  \textbackslash{} &
  \multicolumn{2}{l}{text, image, video, audio} &
  \multicolumn{3}{l}{text} \\
  Gemini 1.5 &
  24.04 &
  \textbackslash{} &
  \textbackslash{} &
  \textbackslash{} &
  \multicolumn{2}{l}{text, image, video, audio} &
  \multicolumn{3}{l}{text} \\
  
  \bottomrule
\end{tabular}
\end{table*}

Since the release of ChatGPT, FMs, especially LLMs, have become increasingly important in daily life.
\autoref{fig:LLM-Timeline} presents a timeline of some popular LLMs. \autoref{tab:currentfm} highlights the structural features of several basic LLMs. \autoref{tab:instruction-tuned} and \autoref{tab:Multimodal Models} list various instruction-tuned models and multimodal models respectively.

\subsubsection{Large Language Models}
Current LLMs are based on Transformer, a neural network based on attention mechanisms\cite{vaswani2017attention}. Most LLMs use the decoder-only architecture, which is beneficial to few-shot capabilities. 
The researchers design different LLMs based on various Transformer architectures with different pretraining datasets.

\emph{T5} is based on the encoder-decoder transformer architecture~\cite{vaswani2017attention}. The authors found that transfer learning can significantly enhance performance, especially when combined with a lot of high-quality data~\cite{raffel2020exploring}, so they design pre-training tasks as text-to-text tasks to pre-train a model. 
\emph{GPT-3} is published by OpenAI, with a major enhancement in in-context learning capabilities compared with GPT-2~\cite{radford2019language} through model scaling~\cite{brown2020language}. 
To improve the multilingual understanding, Zeng et al train \emph{PanGu-$\alpha$} on a 1.1TB high-quality Chinese text corpus, which exhibits decent performance in various Chinese NLP tasks under few-shot or zero-shot conditions~\cite{zeng2021pangu}. 
\emph{ERNIE 3.0} integrates autoregressive and autoencoding networks, which allows easy customization for natural language understanding and generation tasks, solving the disadvantage of downstream language understanding in previous LLMs trained on plain text~\cite{sun2021ernie}.
\emph{Jurassic-1} series includes the J1-Large(7.5B) and the J1-Jumbo(178B). J1-Jumbo's architecture is improved to address the depth-to-width expressivity tradeoff found in self-attention networks~\cite{lieber2021jurassic}.
At a later stage, \emph{Gopher} is introduced and achieves state-of-the-art performance in most of the 152 tasks~\cite{hoffmann2022training}. 

To enhance LLMs' safety and response quality, LaMDA is fine-tuned on labeled data and can reference external knowledge sources. LaMDA generates multiple response candidates in dialogues, filters out those with lower safety scores, and outputs the one with the highest quality score~\cite{thoppilan2022lamda}. 
\emph{MT-NLG}, a 540B model with strong zero, one, and few-shot capabilities, is trained with an efficient and scalable 3D parallel system~\cite{smith2022using}. 
Previous studies focus on increasing the model parameters without enlarging the size of pretraining tokens. 
Hence, DeepMind studied how to balance the number of parameters and tokens within a given computational budget and found that the number of parameters and tokens should scale equally. Based on this, \emph{Chinchilla} is trained with 1.4T tokens and demonstrates superior performance on numerous downstream tasks compared to other LLMs such as Gopher (280B)~\cite{rae2021scaling} and GPT-3 (175B)~\cite{brown2020language}~\cite{hoffmann2022training}. 
\emph{PaLM} uses a decoder-only transformer architecture instead of an encoder-decoder architecture~\cite{radford2019language} to enhance few-shot capability. 
\emph{OPT}, comparable to GPT-3, is developed to address the problem that most LLMs are not open-sourced and have limited access to other developers~\cite{zhang2022opt}. 
To help researchers find useful information, Galactica is trained on a vast amount of scientific corpora, reference materials, and other academic databases and outperforms other LLMs on various scientific tasks~\cite{taylor2022galactica}. 
To promote the transparency of LLM research, \emph{BLOOM} is published and open-sourced. It is trained on a dataset with hundreds of sources in 46 natural languages and 13 programming languages. Benchmarks suggest that fine-tuning BLOOM with multitask prompts can improve its performance~\cite{le2023bloom}.

Based on Chinchilla's contribution, which indicates the number of tokens and parameters should scale equally~\cite{hoffmann2022training}, \emph{LLaMA} focuses on using more training tokens to achieve optimal performance. Although Hoffman et al. suggest training a 10B model on 200B tokens, LLaMA-7B is trained on 4T tokens, indicating that increasing the number of tokens can still enhance the model's performance~\cite{touvron2023llama}. 
Building on this observation, MetaAI publishes and open-sources \emph{LLaMA 2}. It utilizes a larger corpus, longer context lengths, and grouped-query attention to train the model~\cite{touvron2023llama2}. 
Following the LLaMA series, \emph{Baichuan} is released to enhance the performance of Chinese NLP tasks. 
To enhance the compression rate for Chinese, Byte-Pair Encoding is adopted as the tokenization algorithm, and the tokenization model is trained on 20M multilingual corpora. It separates all numbers into individual digits to enhance the model's mathematical capabilities and integrates various optimizations for Chinese support, including operator, tensor partitioning, mixed-precision, training recovery, and communication technologies~\cite{baichuan7B}~\cite{baichuan13B}. 
Bai et al. publish the \emph{QWEN} series of language models, which includes QWEN, Qwen-Chat, CODE-QWEN, CODE-QWEN-CHAT, and MATH-QWEN-CHAT. QWEN series shows strong performance compared to other open-source models, while a little inferior compared to the proprietary models~\cite{bai2023qwen}.
Wei et al. develop \emph{Skywork-13B}, which is trained on a corpus of over 3.2T tokens extracted from English and Chinese texts~\cite{wei2023skywork}. 
\emph{Falcon} series is trained on high-quality corpora primarily assembled from web data. 
Almazrouei et al. release a custom distributed training codebase that allows efficient pretraining of these models on up to 4,096 A100s on cloud AWS infrastructure~\cite{almazrouei2023falcon}. 
Given the widespread use of code-generating LLMs, \emph{StarCoder}, trained in over 80 programming languages with multi-query attention, is released and open-sourced for public use~\cite{li2023starcoder}.

The team of \emph{Yi} series designs a model of 34B parameters to retain complex reasoning and emergent capabilities while enabling inference on consumer-grade hardware, such as the RTX 4090. The model is trained on a high-quality dataset with 3.1T tokens~\cite{young2024yi}. 

\subsubsection{Instruction-tuned Models}
Instruction-tuning is an approach that fine-tunes pretrained LLMs on a formatted language dataset~\cite{wei2022finetuned}. 
This technique improves the model's understanding and response to inputs with specific instructions by training them on instructional task datasets.
An instance of the dataset usually consists of instruction, input, and output. For example, the instruction is ``What is the answer to the formula?", the input is ``7+3", and the output is ``10".

In this subsection, several instruction-tuned models are introduced to demonstrate the development of instruction-tuning.
To improve T5's multilingual capabilities~\cite{raffel2020exploring}, \emph{mT5} is trained on a dataset that includes 101 languages~\cite{xue2020mt5}. 
FLAN is based on LaMDA 137B~\cite{thoppilan2022lamda} and instruction-tuned on over 60 datasets with natural language instruction templates, significantly improving its performance. Ablation studies reveal that the number of fine-tuning datasets, model scales, and the utilization of natural language instructions are crucial to the success of instruction tuning~\cite{shen2023flan}. 
The process of fine-tuning \emph{Flan-T5, Flan-PaLM, Flan-cont-PaLM}, and \emph{Flan-U-PaLM} is expanded to include datasets from Muffin, T0-SF, NoV2, and CoT. Notably, adding nine CoT datasets greatly enhances the models' reasoning capabilities~\cite{chung2022scaling}.

To promote academic research of LLMs, Taori et al. train \emph{Alpaca} based on LLaMA-7B, using 52k instruction-following demonstrations generated by OpenAI’s text-davinci-003. Alpaca's performance is very similar to that of text-davinci-003, yet surprisingly, it is smaller in scale and also inexpensive to train($<600\$$)~\cite{taori2023alpaca}. 
Based on FreeLM, different strategies are adapted to train \emph{FLM-101B} on 0.31T tokens, significantly reducing training costs. With just a \$100k training budget, FLM-101B is comparable to GPT-3~\cite{brown2020language} and GLM-130B~\cite{zeng2023glm130b}~\cite{li2023flm}.

\subsubsection{Multimodal Models}
LLMs are designed to process and generate text-based information, whereas humans interact with the world through multiple sensors, such as visual and auditory modalities. To bridge this gap, multimodal LLMs (MLLMs) are designed to handle text, images, videos, audio, points, boxes, inertial measurement unit, functional magnetic resonance imaging, etc, enabling them to process diverse tasks. Currently, the training of MLLMs typically involves three stages: pre-training, fine-tuning, and prompting. 
Training an MLLM from scratch is very costly. Most prior works on MLLMs have focused on aligning existing VMs with pre-trained LLMs and fine-tuning the alignment module to improve performance. 
We provide some popular MLLMs in \autoref{tab:Multimodal Models}.

\vspace{-3pt}
Many research works design different methods to align a VM to an LLM.
\emph{Flamingo} designs new benchmarks for few-shot visual and language tasks. It uses the Perceiver Resampler and Gated Xattn-Dense to align an LLM and a VM, allowing it to process and integrate visual and textual data sequences. It demonstrates strong performance with few-shot capabilities in visual question answering and close-ended tasks~\cite{alayrac2022flamingo}. 
Different from Flamingo, miniGPT-4, mPLUG-owl, and PandaGPT use a linear layer to align. \emph{miniGPT-4} demonstrates that correctly aligning visual features with an LLM can unlock the advanced multimodal capabilities of GPT-4~\cite{achiam2023gpt}. Training solely on a large-scale image-text paired dataset might lead to unnatural language outputs like repetition and fragmentation. To overcome this, MiniGPT-4 is fine-tuned on a small but higher-quality dataset with more detailed textual descriptions~\cite{zhu2023minigpt}. Meanwhile, \emph{mPLUG-owl} also demonstrates impressive command and visual comprehension abilities, multi-turn dialogue, and knowledge reasoning~\cite{ye2023mplug}. Previous works align LLM and VM by training them on joint image-text tasks, which could compromise LLM's language capabilities. 
\emph{CogVLM} innovatively incorporates trainable visual expert modules between the attention and feed-forward neural network layers to align LLM and VM. This approach allows for deep integration of visual and textual elements without compromising language capabilities because all parameters of the LLM are fixed~\cite{wang2023cogvlm}.

Lots of research works focus on enhancing the MLLMs to process different input and output modalities because previous models can only deal with a small set of modalities. 
PandaGPT utilizes ImageBind as an encoder that can embed data from different modalities into the same feature space~\cite{girdhar2023imagebind}. 
Therefore, PandaGPT has strong cross-modal zero-shot capabilities, allowing it to naturally integrate multimodal inputs and perform complex multimodal tasks efficiently~\cite{su2023pandagpt}. \emph{NExT-GPT} is an end-to-end, general-purpose any-to-any MLLM. It connects the Vicuna with multimodal adapters and various diffusion decoders, enabling it to perceive different inputs and generate outputs in arbitrary combinations of text, images, videos, and audio. It leverages pre-trained encoders and decoders and requires only a few parameters to be tuned, thus reducing training costs and facilitating expansion to more modalities~\cite{wu2023next}. \emph{OneLLM} aligns eight modalities through a universal encoder and projection modules, along with a step-by-step multimodal alignment process, pioneering a universal MLLM architecture. Initially, it connects the LLM and visual encoder via an image projection module. It then expands to additional modalities using a universal projection module and dynamic routing, demonstrating its scalability and generality~\cite{han2023onellm}. \emph{Gemini} is an MLLM pre-trained on a large multimodal dataset, capable of handling a variety of text inputs interleaved with audio, video, and images.
Gemini 1.5 utilizes an MoE architecture, enabling it to process multimodal inputs up to 10M tokens in length, thus exhibiting exceptional performance across modalities~\cite{team2023gemini,reid2024gemini}.

A few research works enhance the MLLM to understand fine-grained visual information in an image.
Based on QWEN~\cite{bai2023qwen}, \emph{Qwen-VL} uses a high-quality dataset with fine-grained vision-language annotation and a larger image input resolution, achieving remarkable performance on fine-grained visual understanding capabilities. It outperforms other models in various vision-centered benchmarks such as image description, question answering, and visual localization~\cite{bai2023qwen}. \emph{Shikra} is the first MLLM capable of detecting specific areas in images. It takes natural language as input and outputs the coordinates. This feature supports visual question answering, image description, and more specialized spatial tasks without additional complex setups or external modules~\cite{chen2023shikra}. 
To address spatial understanding issues, especially in referring and grounding tasks, \emph{Ferret} designs a hybrid regional representation that integrates discrete coordinates and continuous features to represent an area in an image jointly. 
Considering that grid-based processing (e.g., convolution, patch attention) has difficulty in handling regions with an irregular shape, a spatial-aware visual sampler is adopted, allowing it to handle various regional inputs~\cite{you2023ferret}.  
\emph{NExT-Chat} adopts the innovative pix2emb approach instead of pix2seq\cite{chen2022pix2seq}, enabling it to process and output different positional formats, which allows for more flexible and precise object localization and representation, such as bounding boxes and segmentation masks~\cite{zhang2023next}. 


\subsubsection{MoE Models}

The MoE model is an advanced sparse model based on the Transformer architecture, where parameters are divided into groups, each representing an ``expert" with unique weights. Only a subset of these experts are activated and contribute to the computations during inference. This is achieved through a dynamic routing mechanism. It enables MoE models to have higher computational efficiency, easier scalability, and faster pretraining and inference speeds.

Based on PanGu-$\alpha$'s work~\cite{zeng2021pangu}, \emph{PanGu-$\Sigma$} is trained using the MindSpore framework on the Ascend 910 AI processor. Random Routed Experts is utilized to expand the dense transformer model into a sparse one. They implement expert computation and storage separation for efficient training on 329 billion tokens, which increased the training throughput by 6.3 times~\cite{ren2023pangu}.
\emph{Mixtral 8$\times$7B} is the most famous MoE model. Its architecture is similar to Mistral 7B~\cite{jiang2023mistral}, but each layer consists of eight feed-forward blocks (experts) and is pretrained on a multilingual dataset. Two experts assigned by the routing network process each token, and their outputs are merged. Therefore, although it has 47B parameters, only 13B are active in computations. The performance of Mixtral 8$\times$7B matches or exceeds that of LLaMA 70B~\cite{touvron2023llama} or GPT-3~\cite{brown2020language}, showing its broad application potential~\cite{jiang2024mixtral}.
\emph{FLAN-MoE} series is also a well-known MoE model. Shen et al. discovered that after instruction tuning, MoE models demonstrate significant efficiency and performance compared to dense models with equivalent computational power, indicating that the MoE structure can enhance the computational efficiency and performance of models~\cite{shen2023flan}.

Based on Gemini~\cite{team2023gemini}, \emph{Gemini 1.5 Pro} leverages the advantages of MoE, scaling the model size and enabling it to process multimodal data of up to 100k tokens. This model achieves desirable performance in cross-modal long-context retrieval tasks~\cite{reid2024gemini}.
\emph{Switch Transformer} extends T5~\cite{raffel2020exploring} to a MoE model to scale the parameters up to 1.6T while maintaining constant computational costs, showing remarkable scalability and effectiveness across various NLP tasks~\cite{fedus2022switch}.

\subsubsection{Tiny Models}
The LLMs mentioned above typically require high-capacity memory and computational resources, preventing their deployment on resource-constrained edge devices. Tiny models fill this gap by employing techniques such as model pruning, quantization, tensor decomposition, and knowledge distillation to compress neural networks, allowing these models to be executed on edge devices like smartphones. 

\emph{TinyBERT} is an optimized version of BERT designed for deployment on resource-constrained devices. The team proposes a knowledge distillation method for Transformer models, transferring BERT's knowledge to TinyBERT. TinyBERT significantly reduces model size, increases inference speed, and exhibits comparable performance to BERT~\cite{devlin2019bert}~\cite{jiao2019tinybert}. 
PanGu-$\pi$ 1B Pro and PanGu-$\pi$ 1.5B Pro optimize the tokenizer by removing low-frequency vocabularies to enhance the models' representational efficiency. Tiny models have severe data forgetting issues, which can be improved by multiple-round training. A sample selection strategy is employed to decrease the training cost. 
Model configurations (e.g., depth, width, FFN expansion rate, etc) are changed to enhance the models' performance. Initial parameters are inherited from an LLM to improve the tiny models' generalization capabilities. The optimized models achieved significant improvements in benchmark evaluations~\cite{tang2024rethinking}. 
Based on the LLaMA series, \emph{TinyLlama} has 1.1B parameters and is trained on 3T tokens, showing competitive performance in various downstream tasks compared to existing similar open-source LMs.~\cite{zhang2024tinyllama}.
Phi-3-mini, a 3.8B-parameter model, is small enough
to be deployed on mobile devices. 
3.3T high-quality training data are used to enhance the performance of small models. 
Despite its small size, Phi-3-mini’s performance is comparable to larger models like Mixtral 8$\times$7B or GPT3.5~\cite{abdin2024phi}.



\vspace{-6pt}
\subsection{Model Compression}

\subsubsection{Pruning}

\begin{table*}[]
\caption{Pruning Methods}
\label{tab:Model Pruning}
\begin{tabular}{@{}p{1cm}p{2cm}p{6.8cm}p{7cm}@{}}
\toprule
Type &
  Ref. &
  Challenge &
  Method \\ \midrule
LLMs &
  Wanda~\cite{sun2023simple} &
  Pruning LLMs requires costly retraining.&
  Prune weights with the smallest magnitudes multiplied by the corresponding input activations.\\
 &
  Llm-pruner~\cite{ma2024llm} &
  The training corpus of LLMs is enormous.&
  Remove non-critical coupled structures selectively based on gradient information.\\
 &
  LoRAPRUNE~\cite{zhang2023loraprune} &
  Unstructured pruning cannot work with LoRA weights.&
  Use the weights and gradients of LoRA for importance estimation.\\
 &
  LoRAShear~\cite{chen2023lorashear} &
  The enormous size of LLM leads to computational costs.&
  A dynamic fine-tuning scheme with dynamic data adaptors.\\
 &
  An et al.~\cite{an2023fluctuation} &
  Existing retraining-free pruning approaches require hardware support for acceleration.&
  FLAP:fluctuation-based adaptive structured pruning.\\
 &
  Shao et al.~\cite{shao2023one} &
  Existing pruning methods require extensive retraining of pruned models.&
  Allocate sparsity adaptively based on sensitivity.\\
 &
  Compresso~\cite{guo2023compresso} &
  One-shot structured pruning leads to performance decline.&
  Learn optimal pruning decisions during the training process.\\
 &
  SHEARED LLAMA~\cite{xia2023sheared} &
  Training smaller yet powerful LLMs from scratch is costly.&
  Targeted structured pruning and dynamic batch loading.\\
 &
  Ji et al.~\cite{ji2023pruning} &
  The manual design of pruning features leads to a complex optimization pipeline.&
  Train a non-neural model as an accuracy predictor.\\
 &
  GBLM-Pruner~\cite{das2023beyond} &
  Prior approaches overlooked informative gradients derived from pretrained LLMs.&
  Harness normalized gradients from calibration samples to determine pruning metric.\\
 &
  Anagnostidis et al.~\cite{anagnostidis2024dynamic} &
  Autoregressive transformers are hard to scale to long sequences.&
  Employ a learnable mechanism to determine when to drop uninformative tokens from the context.\\
 &
  ZipLM~\cite{kurtic2024ziplm} &
  Generalize to different pruning settings.&
  Identify and remove components with the worst loss-runtime trade-off iteratively.\\
Audio &
  DPHuBERT~\cite{peng2023dphubert} &
  High inference cost hinders speech model deployment. &
  Joint distillation and pruning for speech model compression.\\
 &
  Peng et al. ~\cite{peng2023structured} &
  The frontend network of speech models has a large computational cost.&
  Design three task-specific structured pruning methods for heterogeneous networks.\\
Vision \&  &
  MoPE-CLIP~\cite{lin2024mope} &
  Uni-modal compression metrics lead to limited performance and costly mask-search processes.&
  Evaluate the importance of modules by performance decline on cross-modal tasks.\\
Multi-modal &
  X-pruner~\cite{yu2023x} &
  Overlooking the relationship between network units and target classes leads to inferior model performance. &
  An explainable pruning framework considering the explainability of pruning criterion.\\
 &
  Fang et al.~\cite{fang2024structural} &
  Generative models have large computational overheads.&
  Disregard non-contributory diffusion steps and ensemble informative gradients to identify important weights.\\
 &
  UP-ViT~\cite{yu2023unified} &
  Limited research on model pruning for ViTs.&
  Prune the channels in ViTs in a unified manner.\\
 &
  CAP~\cite{kuznedelev2024cap} &
  It is hard to prune highly accurate ViTs.&
  A theoretically-justified pruner and an efficient finetuning procedure.\\ \bottomrule
\end{tabular}
\end{table*}

LLMs demonstrate remarkable capabilities in various tasks, including text generation, translation, and sentiment analysis. However, the deployment of them in real-world applications is often hindered by their huge size and computational requirements. Pruning, a technique aimed at lowering the computational cost and maintaining the performance of LLMs, has attracted significant attention from researchers. Pruning selectively removes weights or neurons from a pre-trained model to reduce its size and computational requirements. However, pruning LLMs poses unique challenges due to their complex architectures and the need to preserve specific language patterns during compression. 
Firstly, the size of LLMs is particularly large, which often consists of countless parameters, leading to significant computational expenses during both training and inference stages. Secondly, existing pruning methods often necessitate retraining or fine-tuning, resulting in additional computational overhead. Moreover, pruning techniques should retain performance across various NLP tasks, such as text modeling, text classification, and machine translation. Achieving explainability in pruned models is also essential for understanding their decision-making processes. 
Furthermore, the pruning techniques must adapt to different tasks with various language data. Language models trained on diverse datasets may exhibit significant differences and biases, requiring pruning methods to handle such variations while maintaining model performance.

Recent advancements in pruning techniques have addressed several challenges associated with LLMs. They typically design their pruning methods by simplifying dependency on pruning techniques or enhancing compatibility. For instance, they aim to eliminate the need for retraining or weight updates, avoid task-specific compression, and reduce dependence on the original training corpus. Alternatively, they combine the strengths of various pruning methods or focus on making the pruning techniques compatible with hardware platforms. To improve the effectiveness and flexibility of these methods, new criteria are also introduced for pruning, such as sensitivity-based sparsity allocation.

Dynamic context~\cite{anagnostidis2024dynamic} selectively removes contextual information in autoregressive transformers during inference to improve the scalability. This approach reduces memory and computational requirements without significantly sacrificing model performance. Additionally, dynamic context pruning enhances interpretability by studying the model's decision-making process, as uninformative tokens are dynamically pruned based on learnable mechanisms. Joint distillation and pruning methods, such as DPHuBERT~\cite{peng2023dphubert}, achieve task-agnostic compression by training a student model to emulate a teacher model, integrating structured pruning techniques. The method exhibits versatility and applicability to various speech SSL models. By leveraging sensitivity analysis, these methods identify and prune redundant weights or neurons to obtain compressed models with reduced computational costs and memory requirements. 

Wanda~\cite{sun2023simple} aims to induce sparsity for LLMs without requiring retraining or extensive computational resources. Wanda utilizes a unique pruning metric that considers both the weights and the relevant norm input activation. LLM-Pruner~\cite{ma2024llm} removes non-critical coupled structures according to gradient data. LLM-Pruner reduces reliance on the original training data and is an automatic structural pruning framework. Loraprune~\cite{zhang2023loraprune} considers neural network pruning with LoRA, introducing a criterion for weight importance estimation and integrating parameter-efficient fine-tuning, demonstrating superior compression rates and reduced memory usage over existing methods on LLaMA models. 
An et al.~\cite{an2023fluctuation} propose FLAP, which introduces structured importance metrics and compensation mechanisms to mitigate performance loss. Shao et al.~\cite{shao2023one} introduces a pruning method based on mixed Hessian-sensitive sparsity to achieve at least 50\% sparsity in LLMs without retraining, which reduces pruning errors and maintains overall sparsity levels by adaptively allocating sparsity based on sensitivity. Compresso~\cite{guo2023compresso} incorporates techniques such as LoRA and L0 regularization to mitigate training expenses and data collection challenges and introduces a collaborative prompt to enhance interaction between the LLM and the pruning algorithm, leading to notable performance improvements. Sheared llama~\cite{xia2023sheared} prunes LLaMA2 from 7B to 1.3B and 2.7B parameters, outperforming equivalent-sized models while needing only 3\% of the compute for training. It effectively addresses challenges in optimizing pruned architectures and continuing pre-training, showing superior performance. Ji et al.~\cite{ji2023pruning} employ gradient boosting decision trees (GBDT) as an accuracy predictor to guide pruning process based on specific performance requirements. This predictor is then utilized to optimize the search space and select the most suitable pruned model. Das et al.~\cite{das2023beyond} introduce GBLM-Pruner, a sparsity-centric pruning method for billion-parameter LLMs, operating without retraining and surpassing competitors like SparseGPT and Wanda across benchmarks. It unveils structural patterns in unstructured pruning within LLMs, maintaining simplicity and efficiency. Anagnostidis et al.~\cite{anagnostidis2024dynamic} employ an adaptable mechanism to identify and eliminate irrelevant tokens from context, thereby reducing memory and computational demands during inference.
ZipLM~\cite{kurtic2024ziplm} achieves advanced performance in different compression settings and model categories. Structured pruning algorithm in ZipLM considers both local and global correlations, ensuring precise pruning. Additionally, the algorithm is enhanced with a layer-wise token-level distillation technique.

The above research works describe pruning methods for LLMs. Next, we enumerate some pruning methods aimed at audio, visual, and multi-modal models. 
Pada~\cite{lodagala2023pada} introduces cross-domain task-aware pruning, a novel pruning paradigm that utilizes fine-tuned out-of-domain models to enhance adaptation to the target domain. Peng et al.~\cite{peng2023structured} propose heterogeneous joint pruning, which prunes both the CNN and Transformer components.
MoPE-CLIP~\cite{lin2024mope} accurately evaluates the importance of CLIP modules by assessing the performance decline on cross-modal tasks with model pruning. MoPE-CLIP effectively harnesses knowledge from the teacher model, substantially reducing pre-training costs and generating competitive task-specific models. X-pruner~\cite{yu2023x} proposes an explainable pruning framework that quantifies each unit's contribution to class prediction using explainability-aware masks. Applied to representative transformer models such as DeiT and Swin Transformer, X-Pruner demonstrates superior performance with reduced computational costs and minimal performance degradation. Fang et al.~\cite{fang2024structural} present Diff-Pruning, which utilizes a Taylor expansion over pruned timesteps to identify crucial weights and alleviate the computational burden. This approach achieves a notable 50\% reduction in FLOPs with a mere 10\% to 20\% of the original training resources. Yu et al.~\cite{yu2023unified} introduce UP-ViTs, a unified pruning framework tailored for ViTs, which focuses on structurally pruning ViTs and their variants while ensuring consistency in model structure. UP-ViTs achieves high accuracy in compressed models, surpassing previous ViTs and their variants.  CAP~\cite{kuznedelev2024cap} effectively and efficiently manages weight correlations throughout the pruning process. It is compatible with structured pruning and quantization, facilitating practical speedups without compromising accuracy. CAP achieves high sparsity levels with minimal impact on accuracy.

\subsubsection{Quantization}

\begin{table*}[]
\caption{Quantization Methods}
\label{tab:Model Quantization}
\begin{tabular}{@{}p{1.5cm}p{1.8cm}p{6.5cm}p{6.7cm}@{}}
\toprule
Type &
  Ref. &
  Challenge &
  Method \\ \midrule
W8A8 &
  Smoothquant~\cite{xiao2023smoothquant} &
  The presence of outliers in activation.&
  Outlier smoothing and per-channel scaling transformation.\\
 &
  RPTQ~\cite{yuan2023rptq} &
  Varying ranges across channels.&
  Rearrange channels and quantize them in clusters.\\
 &
  LoftQ~\cite{li2023loftq} &
  Performance differences between 2 fine-tuning method.&
  Find a proper low-rank initialization for LoRA fine-tuning.\\
 &
  Outlier suppression+~\cite{wei2023outlier} &
  Existence of detrimental outliers in activations.&
  Channel-wise shifting for asymmetry and channel-wise scaling for concentration.\\
&
  FPTQ~\cite{li2023fptq} &
  The W4A8 faces notorious performance degradation.&
  Layerwise activation quantization strategies.\\
Low-bit weight-only &
  OWQ~\cite{lee2023owq} &
  The presence of activation outliers.&
  Prioritize structured weights sensitive to quantization in high-precision.\\
 &
  AWQ~\cite{lin2023awq} &
  The significant model sizes of modern LLMs.&
  Search for optimal per-channel scaling by observing the activation.\\
 &
  Zhang et al.~\cite{zhang2023integer} &
  The superiority of low-bit Integer versus Floating Point formats is unclear.&
  Select the optimal format on a layer-wise basis.\\
 &
  Omniquant~\cite{shao2023omniquant} &
  Hand-craft quantization parameters lead to low performance.&
  Learnable Weight Clipping (LWC) and Learnable Equivalent Transformation (LET).\\
 &
  IntactKV~\cite{liu2024intactkv} &
  Previous quantization methods compromise LLM performance.&
  Construct KV cache of pivot tokens from full-precision model.\\
 &
  Kim~\cite{kim2024memory} &
  Previous quantization methods are designed for inference.&
  Update solely the quantization scales during fine-tuning.\\
 &
  QLLM~\cite{liu2023qllm} &
  Activation outliers in particular channels.&
  Reallocate the magnitude of outliers to other channels.\\
QAT &
  LLM-QAT~\cite{liu2023llm} &
  Post-training quantization methods cannot perform well at lower bit precision.&
  A data-free distillation method leveraging outputs generated by a pre-trained model.\\
&
  QuIP~\cite{chee2024quip} &
  High memory usage of LLMs.&
  Use random orthogonal matrices to guarantee weight and Hessian incoherence.\\
 &
  Norm tweaking~\cite{li2023norm} &
  Lower-bit quantization leads to severe performance degradation.&
  Update normalization weights with calibration data generation and channel-wise constraints.\\
 &
  Zeroquant-v2~\cite{yao2023zeroquant} &
  Lack of a systematic examination of various quantization schemes.&
   An evaluation and comparison of existing quantization methods.\\
 &
  QA-LORA~\cite{xu2023qa} &
  The imbalance of quantization and adaptation during fine-tuning.&
  Use group-wise operators.\\
 &
  Int2.1~\cite{chai2023int2} &
  Errors induced by the quantization process.&
  Add LoRA layers to bring quantized model close to its float point counterpart.\\
\bottomrule
\end{tabular}\\[5pt]
W8A8: 8-bit weights and activation. QAT: Quantization-Aware Training.
\end{table*}

LLMs have brought NLP to a new era, showing great performance across a mass of missions ranging from text translation to generation. However, these models' widespread deployment presents severe challenges, such as the massive memory requirements, computational overhead, and huge model sizes. To alleviate these issues, researchers have developed different quantization techniques tailored specifically for LLMs. Quantization reduces the bit-width of models, thereby diminishing memory footprint and enhancing inference speed. Researchers need to strike a balance between compression ratios and task-specific performance metrics. Key concepts in quantization include post-training quantization (PTQ), quantization-aware training (QAT), and methodologies for managing weight and activation quantization.

Despite the promise of quantization, many challenges persist in its application to LLMs. 
Achieving high compression rates while ensuring task performance is difficult. It is necessary to balance quantization ratios and task-specific accuracy.
Quantization introduces quantization errors, particularly at lower bit precisions, which can significantly impair model accuracy. 
Moreover, quantization methods should be compatible with a variety of hardware, such as different edge devices.

In recent years, the research community has made significant efforts in addressing the challenges associated with quantizing LLMs.
The quantization methods can be divided into three categories: 8-bit weight and 8-bit activation (W8A8) quantization, low-bit weight-only quantization and quantization-aware training. 
The W8A8 method employs a quantization process that reduces both weights and activation to 8-bit formats.
Smoothquant~\cite{xiao2023smoothquant} achieves a balance between accuracy and hardware efficiency by considering activation outliers and simplifying quantization complexities. Enabling W8A8 quantization, it delivers significant performance enhancement with up to $1.56\times$ speedup and $2\times$ memory reduction without sacrificing accuracy. 
RPTQ~\cite{yuan2023rptq} saves up to 80\% memory with high accuracy levels when quantizing OPT-175b. 
LoftQ~\cite{li2023loftq} quantizes LLMs while identifying an appropriate low-rank initialization for LoRA fine-tuning, thereby enhancing generalization in downstream tasks. 
Outlier suppression+~\cite{wei2023outlier} presents efficient techniques for determining optimal shifting and scaling values. Its performance is comparable to floating-point precision and establishes new benchmark criteria for 4-bit BERT models.
FPTQ~\cite{li2023fptq} introduces layerwise activation quantization strategies, including a novel logarithmic equalization technique, to enhance performance. 
FPTQ achieves outstanding W4A8 quantized performance without the need for additional fine-tuning, thereby simplifying the production of LLMs. 
QLLM~\cite{liu2023qllm} is an efficient low-bitwidth quantization method with a channel reassembly technique for handling activation outliers.
It also includes an adaptive strategy to determine the optimal number of disassembled channels and an efficient error correction mechanism with low-rank parameters.

The low-bit weight-only quantization methods quantize LLM weights into low-bit integers, usually 4 bits or fewer. 
OWQ~\cite{lee2023owq} prioritizes critical weights for high-precision storage while applying quantization to the remaining dense weights, achieving desirable performance by reducing the quantization error significantly.
AWQ~\cite{lin2023awq} utilizes activation information to identify significant weights and optimizes per-channel scaling to preserve these important weights during quantization. AWQ surpasses existing methods on various language modeling and domain-specific benchmarks, exhibiting outstanding quantization performance for instruction-tuned and multi-modal LLMs. The Mixture of Formats Quantization (MoFQ)~\cite{zhang2023integer} selects the optimal format on a layer-wise basis, performing well in weight-only and weight-activation post-training quantization scenarios. MoFQ incurs no hardware overhead compared to INT/FP-only quantization. Omniquant~\cite{shao2023omniquant} efficiently optimizes quantization parameters and preserves original full-precision weights with limited learnable parameters, performing well at low-bit scenarios. IntactKV~\cite{liu2024intactkv} uses full-precision model to generate KV cache of pivot tokens, thereby effectively reducing the quantization error and achieving lossless weight-only INT4 quantization. 
PEQA~\cite{kim2024memory} combines parameter-efficient fine-tuning with quantized LLMs. It updates only the quantization scales, minimizing memory overhead and model sizes. PEQA-tuned LLMs exhibit competitive performance in language modeling, few-shot learning, and comprehension, even at sub-4-bit precision.

Some research works propose quantization-aware training (QAT) methods to simulate quantization in training, making the model adapt to lower bits without decreasing accuracy.
LLM-QAT~\cite{liu2023llm} preserves the original output distribution by distilling data freely, facilitating the quantization of any generative model. This method could quantize LLMs to 4 bits for weights and 6 bits for activation.
Norm tweaking~\cite{li2023norm} redistributes the quantized activation to match its floating-point value. Norm tweaking achieves high-performance quantization for general LLMs. Zeroquant-v2~\cite{yao2023zeroquant} explores PTQ to reduce memory and computational costs in LLMs, introducing Low-Rank Compensation (LoRC) to maintain model quality in low-bit settings. 
QA-LORA~\cite{xu2023qa} algorithm employs group-wise operators to increase quantization flexibility while reducing adaptation complexity. This enables efficient fine-tuning by quantizing the weights of LLMs and integrating them into a quantized model without sacrificing accuracy. Int2.1~\cite{chai2023int2} incorporates an Extremely Memory-Efficient Fine-Tuning (EMEF) framework utilizing LoRA alongside an Error Correction Framework (LREC) to minimize quantization errors. Memory requirements are reduced by up to $5.6\times$, enabling fine-tuning on consumer laptops. Michaud et al.~\cite{michaud2024quantization} elucidates the power law drop-off of loss with model and data size and the emergence of new capabilities with scale. They propose a method for automatically discovering quanta in language models and find that the frequency at which these quanta are used follows a power law.  MobileBERT~\cite{rahman2023quantized} assesses the performance of converted and quantized models on edge devices for English tweet reputation analysis, reducing accuracy loss with smaller footprints. 

These approaches represent tremendous advances in mitigating the challenges associated with quantizing LLMs, preparing for future more efficient and scalable language models that can be seamlessly deployed across diverse applications and platforms.

\subsubsection{Knowledge Distillation}

\begin{table*}[]
\caption{Distillation Methods}
\label{tab:Model Distillation}
\begin{tabular}{@{}p{1cm}p{2.5cm}p{6.8cm}p{7cm}@{}}
\toprule
Type &
  Ref. &
  Challenge &
  Method \\ \midrule
Language Models &
  Hsieh et al.~\cite{hsieh2023distilling} &
  Finetuning and distillation require large amounts of training data.&
  Train small models in a multi-task system by extracting LLM rationales as extra supervision.\\
 &
  ZEPHYR~\cite{tunstall2023zephyr} &
  Models from distilled supervised fine-tuning do not respond well to natural prompts.&
  Apply distilled direct preference optimization (dDPO) to learn a chat model with significantly improved intent alignment.\\
 &
  Lion~\cite{jiang2023lion} &
  Overlook of incorporating reciprocal feedback.&
  An adversarial cycle including imitation, discrimination, and generation.\\
 &
  PaD~\cite{zhu2023pad} &
  Synthetic Chain-of-Thought (CoT) data often contains faulty reasoning.&
  Utilize the reasoning program to substitute the CoT, allowing automated error checking of synthetic data.\\
 &
  DISTILLSPEC~\cite{zhou2023distillspec} &
  Identify a well-aligned compact draft model with a target model.&
  Use knowledge distillation to better align the draft model with the target model for speculative inference.\\
 &
  Latif et al.~\cite{latif2023knowledge} &
  Deploy large models on constrained devices.&
  Use prediction probabilities of LLM as soft labels to train smaller student models.\\
 &
  Less is more~\cite{liang2023less} &
  Student models are often under-fitted.&
  Match hidden features of student and teacher by task-aware filters for every layer.\\
 &
  HOMODISTIL~\cite{liang2023homodistil} &
  Student models cannot produce predictions that match the results of teacher models over massive training data.&
  A task-agnostic distillation approach equipped with iterative pruning.\\
 &
  SCoTD~\cite{li2023symbolic} &
  Only large models (beyond 50B parameters) can gain from the chain of thought.&
  Sample in a larger teacher model to generate a smaller student model.\\
 &
  EvoKD~\cite{liu2024evolving} &
  Lack of exploring LLMs' potential to comprehend the target task and acquire valuable knowledge.&
  Interactively enhance data generation using LLMs with active learning.\\
 &
  MiniDisc~\cite{zhang2024minimal} &
  Teacher assistant-based distillation requires numerous trials to find the optimal teacher assistant.&
  Introduce a new $\lambda$-tradeoff metric that quantifies the optimality of the teacher assistant.\\
 &
  SLaM~\cite{kontonis2024slam} &
  The noise of teacher's pseudo-labels leads to students' ineffectiveness.&
  Student-Label Mixing: Knowledge distillation with unlabeled examples.\\
 &
  UniversalNER~\cite{zhou2023universalner} &
  Student models trail original LLMs by large margins in downstream applications.&
  Targeted distillation with mission-focused instruction tuning to train student models.\\
 &
  SCOTT~\cite{wang2023scott} &
  Generated rationales of LLMs are seldom consistent with the predictions or faithfully justify the decisions.&
  Train counterfactual inference student model by teacher-provided concepts.\\
Visual Models &
  DETRDistill~\cite{chang2023detrdistill} &
  Knowledge distillation methods designed for convolution-based detectors may not be directly applicable to Transformers.&
  A Hungarian-matching logits distillation, a target-aware feature distillation, and a query-prior assignment distillation.\\
 &
  AdaAD~\cite{huang2023boosting} &
  Student models are more likely to encounter adversarial attacks at the edge.&
  Adaptively searches for optimal match points in the inner optimization.\\
 &
  DIME-FM~\cite{sun2023dime} &
  The difficulty of training a small custom vision-language FM for resource-limited applications.&
  Transfer knowledge from large vision-language FMs to compact, personalized foundation models.\\ \bottomrule
\end{tabular}
\end{table*}

LLMs have revolutionized NLP and other AI domains, yet their deployment poses challenges because of immense calculating requirements and storage inefficiency. To resolve these difficulties, researchers have explored knowledge distillation techniques to compress LLMs into smaller, more deployable models while retaining their performance. These techniques leverage teacher-student paradigms, where a large teacher model transfers its knowledge to a smaller student model. 
A previously trained model often distills its knowledge into a smaller model using task-specific data.

However, the optimization of large model distillation faces some challenges. One major challenge is the significant capacity gap between teacher and student models, leading to suboptimal distillation performance. Additionally, noisy pseudo-labels generated by the teacher model may negatively influence the distillation process. 
LLMs possess the capability for chain-of-thought reasoning, and it is crucial to transfer these reasoning abilities to smaller models through distillation.

Researchers have put forward innovative methods to improve knowledge distillation with large FMs. 
Distilling step-by-step~\cite{hsieh2023distilling} is a novel method with a multi-task framework that utilizes LLM rationales for additional supervision.
It performs better with fewer training examples and enhances performance with smaller model sizes.
Zephyr~\cite{tunstall2023zephyr} employs distilled direct preference optimization (dDPO) to enhance user intent alignment in chat models, leveraging preference data from AI feedback. 
It achieves significant enhancements to align 7B models with user intent in chat tasks. 
Lion~\cite{jiang2023lion} has three-stage adversarial loops, including imitation, discrimination, and generation, shifting knowledge from a sophisticated large LLM to a compact, open-source one.
Program-aided Distillation (PaD) ~\cite{zhu2023pad} utilizes reasoning programs to check errors in synthetic data for distillation, enabling small models to outperform large LLMs with significantly fewer parameters and training data. 
Distillspec~\cite{zhou2023distillspec} uses KD to align a compact draft model with a larger target model in speculative decoding.
This method has a substantial reduction in latency with minimal performance drop.
TED~\cite{liang2023less} aligns hidden representations and selects pertinent knowledge by task-aware filters, achieving notable advancements in compressing language models.
Homodistil~\cite{liang2023homodistil} mitigates large prediction discrepancies between teacher and student models. This approach initializes the student model from the teacher and gradually prunes neurons until reaching the desired width, ensuring consistent knowledge transfer with minimal prediction discrepancies throughout the distillation process. 
Li et al.~\cite{li2023symbolic} introduce Symbolic Chain-of-Thought Distillation (SCoTD) techniques, which equip smaller language models with the capability for chain-of-thought prompting, thus performing better in supervised and few-shot settings.
Liu et al.~\cite{liu2024evolving} presents EvoKD, which leverages active learning to enhance data generation with LLMs, thereby improving the capabilities of smaller domain models (student models). EvoKD integrates evolving knowledge distillation and active learning to optimize model training and distill informative knowledge effectively. 
Zhang et al.~\cite{zhang2024minimal} introduce Minimal Distillation Schedule (MINIDISC), which aims to identify an optimal teacher assistant in a single trial for extreme compression scenarios, such as compressing to 5\% scale. 
Slam~\cite{kontonis2024slam} introduces the teacher model's noise to modify the student's loss function and improve the performance.
Universalner~\cite{zhou2023universalner} aims to train student models that excel in open named entity recognition, thus creating more cost-efficient student models.
Scott~\cite{wang2023scott} utilizes contrastive decoding to extract rationales that support gold answers from the teacher model. It also employs counterfactual reasoning to ensure faithful distillation in the student model. Scott generates more faithful CoT rationales compared to baselines while maintaining comparable end-task performance. 
Li et al.~\cite{li2024contextualization} instructs LLMs to transform structural triplets into context-rich segments and introduces auxiliary tasks for smaller knowledge graph completion (KGC) models, enhancing KGC models by leveraging LLMs.
Hu et al.~\cite{hu2024large} propose Linguistic Graph Knowledge Distillation (LinguGKD), which enhances the predictive accuracy and convergence rate of GNNs by distilling knowledge from LLMs without requiring additional data or model parameters. Marrie et al.~\cite{marrie2024good} propose leveraging Mixup based on stable diffusion as a data augmentation strategy to enhance distillation. Their findings demonstrate the effectiveness of linear probing, task-specific distillation, and the successful use of diffusion models for data augmentation without class information, offering insights for improved distillation techniques.

Some research works aim to design knowledge distillation methods for visual tasks, such as object detection~\cite{wang2023dafkd}. 
Detrdistill~\cite{chang2023detrdistill} includes Hungarian-matching logits distillation, target-aware feature distillation, and query-prior assignment distillation. DETRDistill enhances various methods by more than 2.0 mAP, often surpassing their teacher models. 
Huang et al.~\cite{huang2023boosting} address the vulnerability of student models to edge adversarial attacks by introducing AdaAD. This approach incorporates the teacher model in knowledge optimization, significantly enhancing the student model's performance in terms of both accuracy and adversarial robustness.
Dime-fm~\cite{sun2023dime} transfers knowledge from large Vision-Language Foundation Models (VLFMs) to smaller models with minimal data requirements and employs a novel distillation mechanism that matches the similarity of images to sentences, ensuring transferability and robustness. It selects visually-grounded sentences efficiently to construct a distillation text corpus.


\subsection{Model adaptation}

\begin{figure}[t]
    \centering
	\subfloat[Model Selection.]
        {\includegraphics[width = 0.48\textwidth]{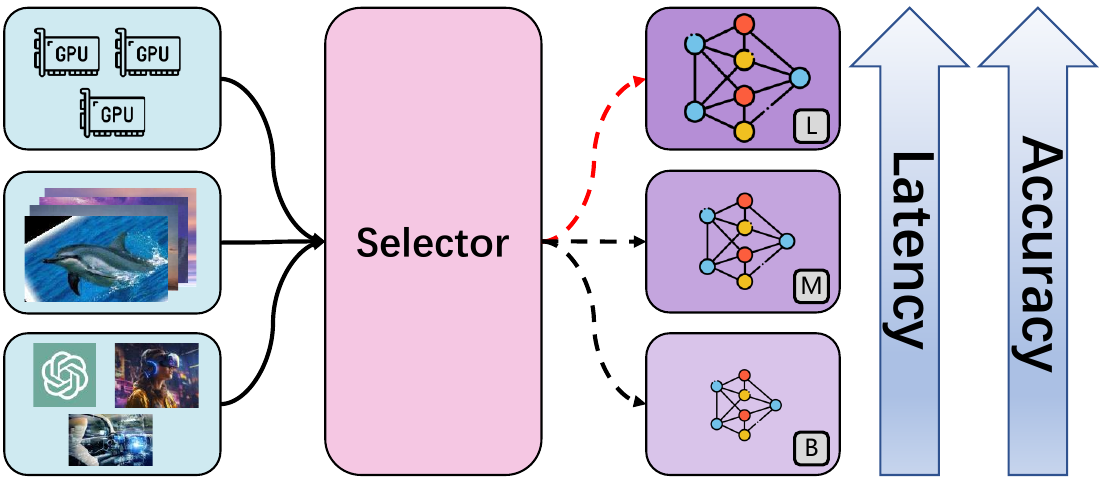}\label{fig:model_select}}
	\hfill
	\subfloat[Model Iteration.]
        {\includegraphics[width = 0.48\textwidth]{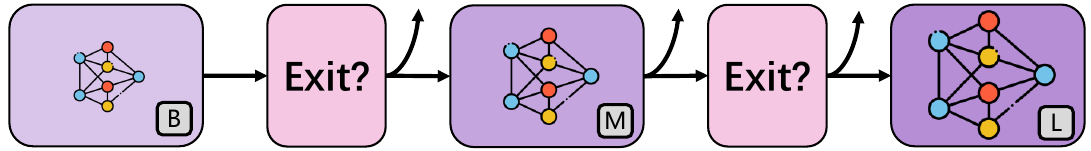}\label{fig:model_iterative}}
        \hfill
	\subfloat[Speculative Decoding.]
        {\includegraphics[width = 0.48\textwidth]{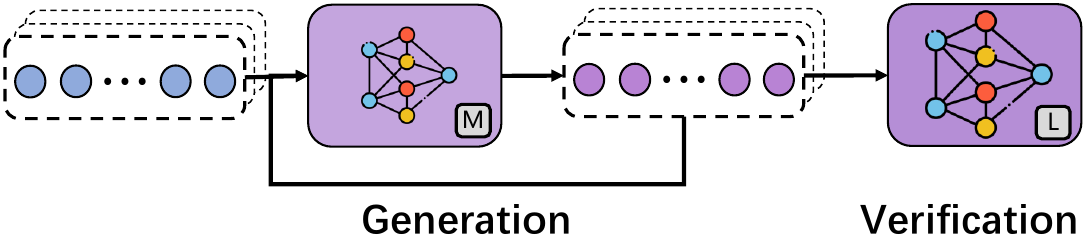}\label{fig:model_spec}}
    \caption{The illustration of model adaptation methods. Model selection dynamically selects a suitable model for inference. Model iteration sequentially runs models and decides when to return results. Speculative decoding uses a small model for text generation and a large model for verification.}
    \label{fig:model_adap}
    \end{figure}

\begin{table*}
\centering
\caption{The model adaptation methods.}
\label{tab:model_adapt}
\begin{tabular}{@{}lllll@{}}
\toprule
\textbf{Type} & \textbf{Ref.} & \textbf{Selector/Exit} & \textbf{Target} & \textbf{Scenario}   \\ 
\midrule
Model Selection & INFaaS~\cite{romero2021infaas} & A Greedy Heuristic & Accuracy \& Latency \& Recourse cost & Cloud \\
& Edgeadaptor~\cite{zhao2022edgeadaptor} & Online optimization and approximate optimization & Accuracy \& Latency \& Recourse cost &  Edge\\
 & JellyBean~\cite{wu2022serving} & Profile and a beam-search. & Accuracy \& Throughput & Edge/Cloud \\
  &STI~\cite{guo2023sti} & A greedy method & Accuracy \& Latency & Edge \\

Model Iteration& Tabi~\cite{wang2023tabi} & Confidence of logits & Accuracy \& Latency & Cloud \\
& CATs~\cite{schuster2021consistent} & A meta consistency classifier & Accuracy \& Latency & Edge/Cloud \\
 \bottomrule
\end{tabular}
\end{table*}

Model adaptation in edge-cloud systems aims to dynamically select and possibly adapt ML models for inference tasks based on the current execution context, such as available computational resources, network conditions, and specific requirements of the task. 
Dynamic model selection and adaptation enable elastic acceleration in edge-cloud systems. 
By exploring the trade-off between performance, efficiency, privacy, and cost, this approach significantly enhances the feasibility and effectiveness of AI applications across different edge-cloud scenarios.
While model adaptation for inference in edge-cloud systems brings substantial benefits, it also introduces several challenges.
First, edge devices, such as IoT sensors or smartphones, often have limited computational power, memory, and energy resources. Adapting and running complex FMs within these constraints without compromising performance or accuracy is a significant challenge.
Second, edge environments can be highly variable, with changes in network conditions, device capabilities, and application contexts. Dynamically adapting models to these changing conditions without human intervention requires real-time monitoring and long-term decision-making algorithms that can accurately assess the current environment and predict the best model or configuration.
Third, quickly finding a sweet spot between inference speed (latency) and model accuracy is a challenge in different applications. 

As shown in Figure~\ref{fig:model_adap}, we categorize research works of model adaptation into three types.
\emph{1) Model Selection.} The most prevalent method is model selection, which dynamically selects a suitable model for inference in a serving system.
Research works design a model selector by considering various system characteristics. 
The first factor is hardware resources, which fundamentally determine the computational capacity of the system.
Generally, lightweight models are deployed on edge devices, and stronger models are deployed on cloud centers.
The second factor is the input. If the images are complex and belong to different domains, or if the language task is challenging to process, it is necessary to deploy a large model to enhance accuracy.
The third factor is the service demands of users in different applications.
Users' requirements for accuracy and latency can vary significantly based on their use cases. For instance, those using entertainment devices like virtual reality may prioritize low-latency AI services, even if it means accepting a slight decrease in accuracy~\cite{ribeiro2023virtual}. Conversely, users in the medical industry require impeccable accuracy, even if it results in longer waiting time for results~\cite{esmaeilzadeh2020use}.
The selector takes these factors as input and outputs the model selection strategy based on an optimization algorithm or a heuristic greedy method.
A larger model can deliver more accurate results at the expense of increased latency, while a smaller model may provide faster responses but with potentially reduced accuracy.
The model selector should find a sweet spot between latency and accuracy under different settings of a serving system.
\emph{2) Model Iteration.}
The model selection process should estimate both accuracy and latency in advance and then execute only one model per request.
However, directly selecting a small model may lead to an inaccurate result.
A few works execute model inference iteratively, starting from a small model and progressively moving to a larger model, and decide when to return results based on the prediction probability.
The existing methods are designed based on the entropy of the probability, where a higher entropy value indicates increased uncertainty and necessitates forwarding the request to a larger model.
\emph{3) Speculative Decoding.} 
Speculative decoding has emerged as an efficient and widely adopted technique in LLM inference to address the limitations of multi-step decoding. It utilizes a smaller model to generate a sequence of candidate words, which are then simultaneously verified by a larger model, resulting in improved performance.
The benefits of speculative decoding stem from two aspects: (1) Small models can quickly generate text sequences, and (2) large models can group generated sequences into a batch to improve throughput by leveraging hardware resources more effectively.

We summarize recent works of model adaptation in Table~\ref{tab:model_adapt}.
INFaaS is an automated model-less serving system, which generates model variants optimized along different dimensions and automatically selects the most appropriate variant for each query based on performance, cost, and accuracy objectives~\cite{romero2021infaas}.
EdgeAdaptor is designed to efficiently manage the trade-offs between inference accuracy, latency, and resource costs for edge-based DNN inference services. This framework addresses the problem by jointly optimizing application configuration, DNN model selection, and edge resource provisioning dynamically, in response to fluctuating demand and system conditions~\cite{zhao2022edgeadaptor}.
JellyBean is a system designed for optimizing and serving machine learning inference workflows across heterogeneous computing infrastructures~\cite{wu2022serving}. For each ML operator within a workflow, JellyBean selects a model that meets the accuracy requirements at the lowest possible cost. This process considers the interaction between models to estimate the overall workflow's accuracy and utilizes a beam search algorithm to explore the space of possible model configurations efficiently.
STI is an on-device inference system with two novel techniques: model sharding and elastic pipeline planning with a preload buffer~\cite{guo2023sti}. STI manages model parameters as independently tunable shards, profiling their importance to accuracy and managing them on disk. Moreover, the elastic pipeline planning module utilizes a small preload buffer to initiate execution without delays, selecting and assembling shards according to their importance to maximize inference accuracy within resource constraints.
Tabi is an inference system that employs a multi-level inference engine to serve queries using smaller models by default and only switches to more computationally expensive LLMs for more complex applications~\cite{wang2023tabi}. Tabi uses a calibrated confidence score to decide whether the results from smaller models are accurate or if a query should be rerouted to a larger model for processing.
CATs trains additional prediction heads on intermediate layers of a transformer model and uses a meta consistency classifier to dynamically decide when to stop computation for each input, based on a unique extension of conformal prediction~\cite{schuster2021consistent}.
Leviathan et al. first introduce speculative decoding. Let $p(x_t|x_{<t})$ be the distribution of a target model $M_p$ and $q(x_t|x_{<t})$ be the distribution of a draft model $M_q$, outputs can be sampled from these two distributions.
Then it samples from $q(x)$ and keeps it if $q(x) \leq p(x)$. If $q(x)>p(x)$, it rejects the sample with probability $1-\frac{p(x)}{q(x)}$ and samples $x$ again from an adjusted distribution $p'(x)=norm(\max(0,p(x)-q(x)))$.
LLMCad designs an on-device system with three novel techniques: constructing a token tree for broader candidate token pathways, a self-adjusting fallback strategy for error correction, and a speculative token generation during the verification process to maintain efficiency~\cite{xu2023llmcad}.
SpecInfer introduces a novel approach where small speculative models predict the LLM's outputs, organizing these predictions into a token tree, with each node representing a candidate token sequence~\cite{miao2023specinfer}. The system then verifies the correctness of all candidate sequences in parallel against the LLM, significantly reducing end-to-end latency and computational requirements while maintaining model quality. 
Sequoia employs a dynamic programming algorithm to determine the optimal speculative token tree structure, enabling it to scale with the size of the speculation budget~\cite{chen2024sequoia}.
Sequoia also features a hardware-aware tree optimizer that selects the optimal token tree size and depth based on the available resources to maximize speculative decoding performance.
Minions employs multiple small speculative models to predict the output of an LLM, using a majority-voted approach to improve inference performance~\cite{wang2024minions}. It also dynamically adjusts the speculation length of small models, optimizing the trade-off between the number of tokens speculated by small models and the verification cost by the LLM.

\vspace{-6pt}
\subsection{Token adaptation}

\begin{table*}
\centering
\caption{The token reduction methods.}
\label{tab:token_reduction}
\begin{tabular}{@{}p{1cm}p{2cm}lp{6.5cm}p{3cm}@{}}
\toprule
\textbf{Type} & \textbf{Reduction Method} & \textbf{Ref.} & \textbf{Highlight}  &\textbf{Model} \\ 
\midrule
Language & Token Pruning  & FastGen~\cite{ge2023model} &  Pruning KV cache based on special tokens, punctuation, locality and frequency. & LLaMa \\
&  & StreamingLLM~\cite{xiao2023efficient} &  1. Attention sink mechanism: A small set of initial tokens is important. 2. Placeholder token as dedicated attention sink. & Llama-2, MPT, PyThia and Falcon \\
& & H2O~\cite{zhang2024h2o} & KV cache pruning based on the accumulated attention scores &  OPT, LLaMA, and GPT-NeoX \\
& & ToP~\cite{li2023constraint} &  1. Ranking-distilled token distillation. 2. A two-tier binary masking system.  &  Bert \\
& & LLMLingua~\cite{jiang2023llmlingua} & 1. Budget controller: Allocate different compression ratios. 2. Token-level Iterative compression algorithm: Evaluating the importance of each token and iteratively removing tokens. & GPT-3.5, Claude-v1.3\\
& & Longllmlingua~\cite{jiang2023longllmlingua} & A post-compression subsequence recovery strategy. & Llama-2 \\
& & Selective Context~\cite{li2023compressing} &  Pruning tokens on lexical units based on self-information. & GPT-3.5, GPT-4, LLaMA, Vicuna\\
& & LTP~\cite{kim2022learned} & 1. Learnable threshold-based token pruning. 2. Differentiable soft binarized mask.  & Bert \\
& & AdapLeR~\cite{modarressi2022adapler} &1. Contribution predictor. 2. Gradient-based saliency method to train the predictor. 3. A soft-removal function for masking.  & Bert  \\
& & LengthDrop~\cite{kim2021length}  & 1. Randomly generates a length configuration during training. 2. Multi-objective evolutionary search to find an optimal length. 3. Drop-and-restore Process.  & Bert\\
& & Power-bert~\cite{goyal2020power} &   1. Determining word-vector significance with attention. 2. Learning how many word vectors to eliminate. & Bert \\

& Token Summary & ICAE~\cite{ge2023context}  &  The encoder is adapted from an LLM with LoRA for encoding a long context into a small number of memory slots. & LlaMa \\
&  & Gist~\cite{mu2024learning}  & Compress prompts into smaller sets of ``gist" tokens for language models. & LLaMA, FLAN-T5\\
& & AutoCompressor~\cite{chevalier2023adapting} & Adapting language models to process long text documents by compressing them into compact summary vectors. & Llama-2 \\

Vision & Token Pruning & STAR~\cite{zhangsynergistic} & 1. Dynamic evaluation of intra-Layer patch importance. 2. Offline evaluation of inter-Layer patch importance. & DeiT \\
& & METR~\cite{liu2023simple} & METR integrates a multi-exit architecture into ViTs to encourage the model to prioritize task-relevant information from the initial stages. & ViT, DeiT \\
& & HeatViT~\cite{dong2023heatvit}  & 1. An attention-based selector for pruning less informative tokens. 2. Hardware optimization strategies such as 8-bit quantization. & DeiT, LV-ViT \\
& & Slimming~\cite{tang2022patch} & Dynamically prunes less informative patches with the guidance of the last layer. & ViT \\
& & Dynamicvit~\cite{rao2021dynamicvit} &  The framework includes several trainable prediction modules to determine the tokens to be pruned.  & DeiT, LV-ViT \\

& Token Merging & ToMe~\cite{bolya2022token} &  ToMe employs a simple, lightweight matching algorithm to gradually combine similar tokens within a transformer. & ViT, DeiT\\
& & EViT~\cite{liang2021evit}  & The approach involves computing token attentiveness between image tokens and the class token, preserving attentive tokens, and fusing inattentive tokens.  & DeiT, LV-ViT \\

& Token Pruning \& Token Merging  & DiffRate~\cite{chen2023diffrate}   & Automatic learning of different compression rates for different layers. & ViT, DeiT \\
& & Beyond~\cite{long2023beyond} &  An efficient token decoupling and merging method that considers both token importance and diversity for token pruning. & DeiT, LV-ViT \\
& & TPS~\cite{wei2023joint}  & 1. Split tokens into reserved and pruned subsets. 2. Merge pruned tokens into the reserved tokens. &  DeiT, ViTs\\

Multi-modal \& Video & Token Pruning \& Token Merging & PuMer~\cite{cao2023pumer} &  PuMer employs a token reduction strategy that combines text-informed pruning and modality-aware merging techniques to selectively reduce the number of tokens from both input images and text. & ViLT, METER \\
& Token Merging & TESTA~\cite{ren2023testa} & 1. Sampling input frames from the video. 2. Adaptively aggregating similar frames and patches within frames. & BLIP \\

 & Token Pruning & STA~\cite{ding2023prune}  &  Evaluates tokens based on temporal redundancy and semantic importance. & ViT, VideoSwin \\
& & STTS~\cite{wang2022efficient}  & Dynamically selects informative tokens in both temporal and spatial dimensions. & MViT-B16 \\
 
 \bottomrule
\end{tabular}
\end{table*}

    \begin{figure}[t]
    \centering
	\subfloat[Token pruning.]
        {\includegraphics[width = 0.24\textwidth]{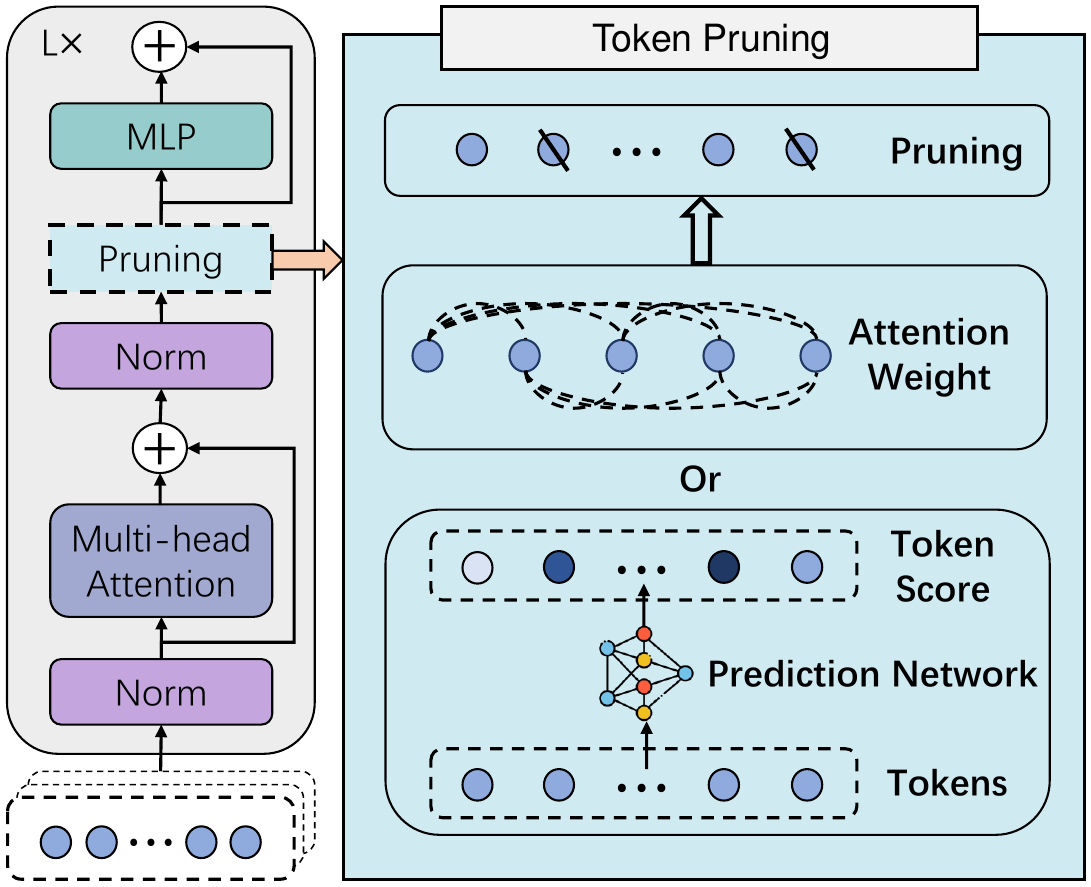}\label{fig:token_pruning}}
	\hfill
	\subfloat[Token merging.]
        {\includegraphics[width = 0.24\textwidth]{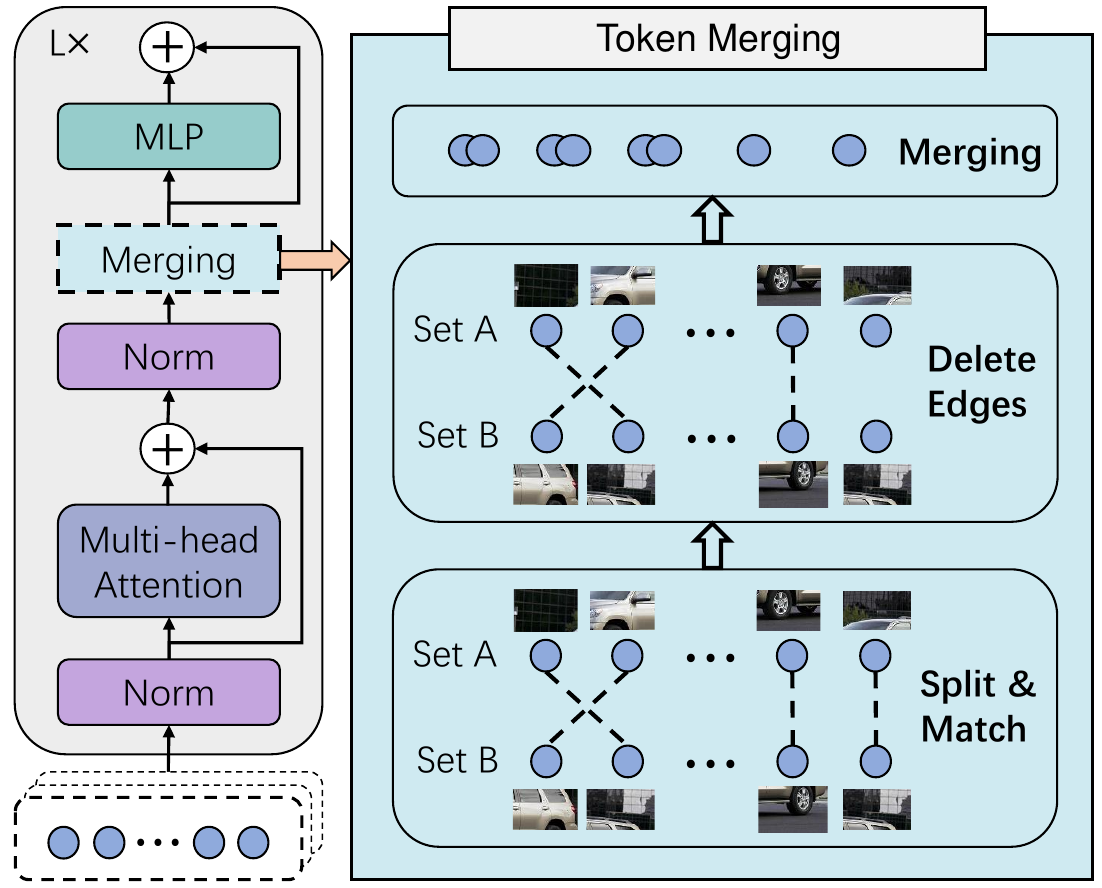}\label{fig:token_merging}}
        \hfill
	\subfloat[Token summary.]
        {\includegraphics[width = 0.48\textwidth]{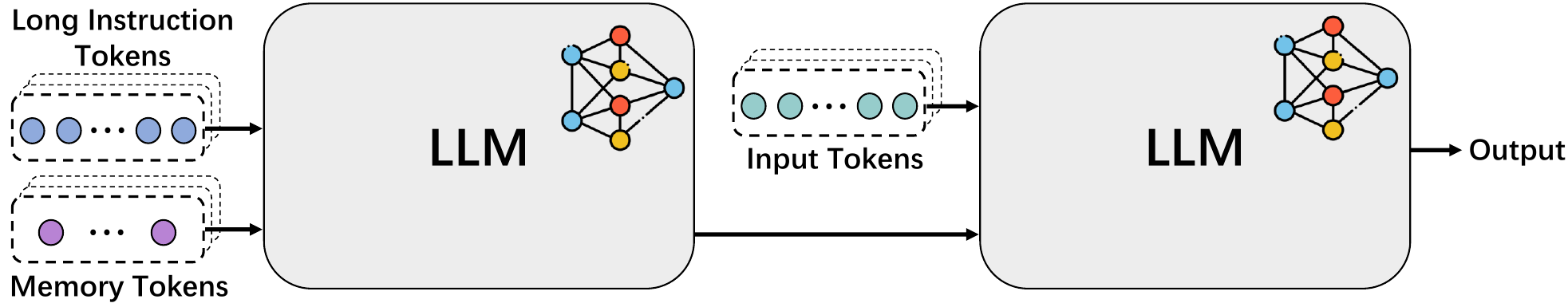}\label{fig:token_summary}}
    \caption{The illustration of token reduction methods. Token pruning removes useless tokens. Token merging combines similar tokens. Token summary summarizes long tokens into memory tokens.}
    \label{fig:p_m}
    \end{figure}

Although model adaptation can accelerate the inference of a transformer model, it may lead to large I/O latency because it needs to switch different versions of models between GPU memory and disk, which becomes the bottleneck during inference.  
By analyzing the inference process, researchers find that the significant computational cost of transformer models primarily comes at the self-attention mechanism~\cite{10.1145/3530811}. 
It calculates the matrix multiplication between the key and query and scales quadratically with the sequence length. 
Processing long sequences or large batches of data can become computationally intensive because the number of tokens increases dramatically.
Token reduction, encompassing token pruning, token merging and token summary, is an advanced technique in the field of artificial intelligence aimed at enhancing the efficiency and performance of large transformer models~\cite{haurum2023tokens}. 
The core idea behind token reduction is to shorten the input data that a model processes, thereby reducing computational overhead and potentially improving the model's ability to focus on the most important information. 
A key research topic in token reduction is to identify redundant or similar tokens in the input and remove or merge them without negatively affecting model performance.

Designing token reduction methods faces several challenges. 
1) Information loss due to token reduction. Although token reduction can decrease computation costs, it often leads to significant information loss. The errors introduced by pruning strategies can adversely affect the model performance, as essential context and details necessary for accurate predictions may be discarded.
2) Trade-off between performance and efficiency. Achieving a good balance between model performance and computational efficiency is a major challenge. Aggressive token pruning can lead to large accuracy drops due to the loss of essential information.
3) Robustness to various reduction ratios. A large reduction ratio may inevitably remove crucial tokens, resulting in incomplete inputs. 
A robust token reduction method should maintain optimal performance across different reduction ratios.
4) Adaptability to different transformer architecture. The diversity of transformer architectures, including vanilla transformers and hybrid transformers, poses a challenge for developing a compression method that is both effective and flexible across different models.
5) Maintaining hardware-friendly inference. Another challenge lies in ensuring that reduction methods facilitate hardware-compatible inference, particularly when running on certain edge devices. This is crucial for the practical deployment of compression methods in real-world applications.

To solve the above challenges, a number of token reduction methods have been developed to accelerate the inference of the transformer.
\emph{ 1) Token pruning.}
There are only a limited number of words or image patches that contribute to the prediction of final results, and a lot of redundant tokens can be regarded as noisy information.
Therefore, token pruning selectively removes tokens from the input sequence during the inference process of a transformer model. 
The core idea is to identify and retain only the most informative tokens for the task, thereby reducing the sequence length and the computational load.
As shown in Figure~\ref{fig:token_pruning}, there are two common approaches to indicate the significance of tokens that can guide the pruning process.
The first method is \emph{training a prediction module}, which takes tokens as input and outputs the importance score for each token. 
This module is constructed with a two-layer linear network and trained with soft masking of tokens.
During inference, the $k$ tokens with the lowest importance scores will be removed to reduce the token number.
The second method is directly \emph{collecting the attention weight} to prioritize tokens. 
The attention mechanism calculates the interdependence among tokens, and the attention weight reflects the relationship between them.
Tokens with lower attention values contribute less to the output of other tokens, therefore indicating their smaller significance.
Therefore, the attention weight serves as a valuable metric for evaluating the token importance.
\emph{ 2) Token Merging.}
Token merging, on the other hand, combines adjacent or similar tokens into single tokens, effectively condensing the content and further reducing the sequence length without a substantial loss of information. 
This technique is particularly useful for large transformer models, where processing extensive sequences of tokens can be computationally intensive. 
The motivation behind token merging is the presence of numerous similar patches in images and videos, which exhibit similar functionalities within the transformer model.
One of the most well-known methods is TOME, which demonstrates the effectiveness of token merging~\cite{bolya2022token}. 
As shown in Figure~\ref{fig:token_merging}, the input tokens are initially divided into two sets, with tokens in set A selecting the most similar token in set B using cosine similarity of features. 
The top $k$ edges, representing the highest similarity of tokens, are retained. 
Finally, these tokens are merged using a weighted average approach to reduce the length of the input.
\emph{ 3) Token Summary.}
Token summary is a crucial technique for optimizing LLMs by condensing lengthy instructions and external knowledge sources.
Instructions typically contain detailed task information and example references, while knowledge pools provide additional external information to assist the LLM's inference process.
However, these sources can be voluminous, potentially exceeding the model's window size and causing latency issues.
To address this, token summary expedites inference by summarizing the content into concise forms.
It distills the content into smaller memory tokens using the LLM and concatenates them with user input tokens for inference.
By leveraging token summaries, LLMs can effectively manage the overloaded information and enhance their ability to generate accurate responses.

Based on the application scenarios, we can classify existing token reduction methods into three types. 
We summarize all token reduction methods and list their contributions in Table~\ref{tab:token_reduction}.

\emph{1) Language: } Token reduction is initially applied to the BERT model. 
Power-bert determines the pruned tokens based on the attention value and learns a configuration of how many tokens should be eliminated~\cite{goyal2020power}.
To improve the robustness, LengthDrop randomly generates a token length configuration during training and designs a multi-objective evolutionary search method to find an optimal token length during inference. Besides, a drop-and-restore process is employed to recover certain pruned tokens that might be important for the deeper layers~\cite{kim2021length}.
AdapLeR trains a contribution predictor to evaluate tokens and design a soft-removal function to mask tokens for gradient propagation~\cite{modarressi2022adapler}.
LTP assesses token importance with attention weight and learns a threshold to dynamically remove tokens~\cite{kim2022learned}.
To improve the accuracy of token importance evaluation, ToP transfers the token importance rankings derived from the final layer of unpruned models to the initial layers of pruned models~\cite{li2023constraint}.
They also introduce a coarse-to-fine pruning approach to dynamically select pruning layers.
The above methods are designed for discriminative language models, such as Bert.
In recent years, there has been a significant advancement in generative language models, such as GPT-3 and Llama. As a result, several studies propose token reduction techniques specifically tailored for these models.
The self-information, such as entropy and perplexity, is used to measure the token importance for pruning~\cite{li2023compressing}.
LLMLingua uses a small model to compress prompts, sets different compression ratios for instructions, questions and demonstrations, and iteratively prunes fine-grained tokens to improve the accuracy~\cite{jiang2023llmlingua}.
A few methods aim to reduce the size of the Key-Value (KV) caches to expedite the decoding process.
The authors of Heavy Hitter Oracle (H20) observe that a small subset of tokens, termed Heavy Hitters (H2), contribute most of the value when computing attention scores~\cite{zhang2024h2o}. 
Based on this, they propose the H2O, a KV cache eviction policy based on the accumulated attention scores that dynamically retains a balance of recent and H2 tokens, effectively reducing the cache size without sacrificing performance.
This method can improve the throughput by up to $29\times$ on OPT-6.7B and OPT-30B compared to other leading inference systems.
FastGen incorporates four policies to dynamically adjust the KV cache~\cite{ge2023model}. These policies are as follows: 1. Retaining special tokens like $<$s$>$ and $<$INST$>$. 2. Preserving punctuation tokens like ``.'' and ``?''. 3. Evicting tokens that are distantly located from the current token. 4. Retaining tokens with high attention scores.
StreamingLLM is an efficient method for deploying LLMs in streaming applications, such as multi-round dialogue~\cite{xiao2023efficient}. 
StreamingLLM leverages an observed phenomenon called ``attention sink'', where maintaining the KV states of initial tokens significantly improves performance.
This method can accelerate the inference of Lllama-2 by up to $22.2\times$ compared to other methods.
For token summary, AutoCompressor and ICAE adopt compact summary vectors to distill information from the original contexts~\cite{chevalier2023adapting, ge2023context}.
Gist compresses instruction prompts into ``gist'' tokens that represent a specific task~\cite{mu2024learning}.

\emph{2) Vision.}
There are a lot of redundant tokens in an image, such as the background. 
DynamicViT introduces a trainable prediction module to determine the token scores. 
Paper~\cite{tang2022patch} dynamically prunes less informative patches from the input image in a top-down manner by formulating token pruning as an optimization problem.
HeatViT deploys attention-based token pruning methods on FPGA devices and also incorporates hardware optimization strategies such as 8-bit quantization~\cite{dong2023heatvit}.
Authors of METR find that the attention weight with class token ([CLS]) fails to gather task-specific information because the attention mechanism tends to concentrate on more general tokens at the initial layers~\cite{liu2023simple}.
This phenomenon may degrade the performance of attention-based token reduction because of the inaccurate token importance estimation.
This paper proposes Multi-Exit Token Reduction (METR) that integrates a multi-exit architecture into ViTs to encourage the model to prioritize task-relevant information from the initial stages~\cite{liu2023simple}.
The motivation of STAR is to evaluate and prune patches based on their importance within and across layers~\cite{zhangsynergistic}. 
It combines online intra-layer importance assessment with offline inter-layer importance analysis, using a fusion mechanism to selectively prune patches, aiming to retain those most critical for model performance.
The idea of token merging originated from EViT~\cite{liang2021evit}, which divides tokens into attentive and inattentive tokens based on the attention weight and fuses inattention tokens into one token.
TOME is a pure token merging method to gradually combine similar tokens within a transformer, achieving a balance of speed and accuracy that rivals pruning methods~\cite{bolya2022token}.
Recent works focus on combining token pruning and token merging to improve accuracy.
Long et al. delve into the importance and diversity among image patches to preserve discriminative local tokens while maximizing global token diversity~\cite{long2023beyond}.
First, tokens are decoupled into two separate groups by leveraging class token attention. Inattentive tokens are then clustered using a density peak clustering algorithm, and attentive tokens are merged with a gentle matching method.
TPS is also a token pruning and squeezing method~\cite{wei2023joint}. First, it splits tokens into reserved and pruned subsets. Then, it employs unidirectional nearest-neighbor matching and similarity-based fusing steps to combine the information from pruned tokens into the reserved tokens.
DiffRate enables automatic learning of different compression rates for different layers, thereby enhancing the optimization of the pruning and merging configuration~\cite{chen2023diffrate}.

\emph{3) Multi-modal \& Video.} 
When designing token reduction methods for multi-modal models, it is important to take into account the interactions between different modalities and adjust the strategy for evaluating token importance accordingly.
In order to speed up the inference of video transformers, it is crucial to leverage both the redundancy within individual frames (intra-frame redundancy) and the redundancy between consecutive frames (inter-frame redundancy).
TESTA extends the idea of token merging to video-language applications~\cite{ren2023testa}.  
It begins by densely sampling input frames from the video. Subsequently, the framework incorporates two distinct merging schemes: frame merging, which combines similar frames, and token merging, which merges similar tokens. By employing these merging strategies, TESTA enhances the efficiency and effectiveness of video processing and analysis.
PuMer employs a novel token reduction strategy that combines text-informed pruning and modality-aware merging techniques to selectively reduce the number of tokens from both input images and text~\cite{cao2023pumer}. 
It selectively removes image tokens that are irrelevant to the accompanying text and are deemed unimportant for the vision-language task predictions.
STTS dynamically selects informative tokens in both temporal and spatial dimensions to improve the efficiency of video transformers~\cite{wang2022efficient}. By formulating token selection as a ranking problem, a lightweight scorer network estimates the importance of each token, and only those with top scores are used for downstream evaluation.
Ding et al. introduce a method to optimize the speed-accuracy trade-off in video recognition tasks by pruning spatio-temporal tokens using a Semantic-aware Temporal Accumulation (STA) score. This score evaluates tokens based on temporal redundancy and semantic importance, allowing for the pruning of tokens that are temporally redundant or semantically less significant without the need for additional parameters or re-training.

\emph{Adaptive token reduction.}
Token reduction techniques demonstrate remarkable advantages in expediting transformer inference while maintaining prediction performance. Furthermore, it is crucial to develop adaptation methods for token reduction to cater to various applications across diverse execution environments. By doing so, we can fully leverage the potential of token reduction and optimize its effectiveness.
OTAS incorporates token prompting and token reduction in an elastic serving system, enabling dynamic selection of an optimal token number~\cite{chen2024otas}.
This selection process takes into account factors such as fluctuating query load, diverse service targets, and limited hardware resources.
It designs an optimization method to investigate the balance between accuracy and throughput across various token reduction ratios.
AdaTape enhances the flexibility and performance of a transformer model by dynamically adjusting the computation based on the input's complexity~\cite{xue2023adaptive}. It employs an elastic input sequence mechanism through adaptive tape tokens (i.e., prompt), which are generated from a tape bank and appended to the input sequences, allowing for dynamic read-and-write operations. This method enables the model to adaptively control both the content and the number of tape tokens used for each input, thereby adjusting the computational budget and potentially improving efficiency and performance on tasks.

\section{AI Agent}
\label{sec:agent}

As early as the 1950s, Alan Turing had already expanded the concept of ``intelligence" to artificial entities and proposed the famous Turing Test to assess whether machines could exhibit intelligence similar to that of humans. This test determines the level of machine intelligence based on whether it can make its text-based communication indistinguishable from that of humans. Such evaluated AI entities are commonly referred to as Agents. The concept of an agent originally stemmed from philosophy and is used to describe an entity with autonomy, not only capable of action but also possessing the desire, belief, and intent to decide when and how to act. This concept has been further developed in AI. Agents are not just entities executing programmed instructions. They also make decisions, solve problems, and navigate complex environments.

Over time, the application of intelligent agents has expanded from simple automation tasks to scenarios requiring complex decision-making and adaptability. For example, intelligent agents are now widely used in areas such as autonomous vehicles, high-frequency trading, personalized medicine, and smart home systems. These systems can respond to external inputs, learn users' behavior patterns, and adjust their strategies according to environmental changes. This evolution from philosophy to technology marks technical advancements in AI and triggers meaningful sociological and ethical discussions about machine ethics and the control of intelligent systems. As AI technology develops and becomes more widespread, intelligent agents will keep shaping our ways of working and living while challenging our traditional understanding of intelligence, autonomy, and control.

The development from NLP to AGI can be divided into five levels: corpora, internet, perception, embodiment, and social attributes\cite{morris2023levels}. LLMs have reached the second level, where they can handle text input and output on the internet scale. If perceptual and action spaces are introduced to LLM-based agents, these agents could advance to the third and fourth stages of development. This means that these intelligent agents would not only be able to understand vast amounts of textual data but also understand and influence their physical or virtual environments through perception and interaction.

In an autonomous agent system powered by LLM, the LLM serves as the agent's central nervous system or cognitive core, playing a pivotal role in processing information, making decisions, and generating actions. The agent is augmented by other essential elements that ensure the agent's robustness, adaptability, and efficiency: Multi-agent framework, Planning, Memory, and Tool use.

\vspace{-6pt}
\subsection{Multi-agent Framework}
An LLM multi-agent framework is a system that leverages multiple LLMs as independent agents working collaboratively to handle complex tasks. Each agent may specialize in different areas or subtasks, and they communicate and cooperate to share information, verify outputs, and enhance overall performance. This approach improves robustness, efficiency, and scalability by enabling parallel processing and modular updates.
For example, A multi-agent framework in software development organizes specialized roles such as architects, programmers, and testers into a cohesive, collaborative system. Each agent has distinct responsibilities and communicates effectively through defined protocols. This multi-agent framework enhances efficiency, reduces errors, and ensures consistency by leveraging task management systems, collaboration tools, and shared knowledge bases.  
\begin{figure}
    \centering
    \includegraphics[width=1.0\linewidth]{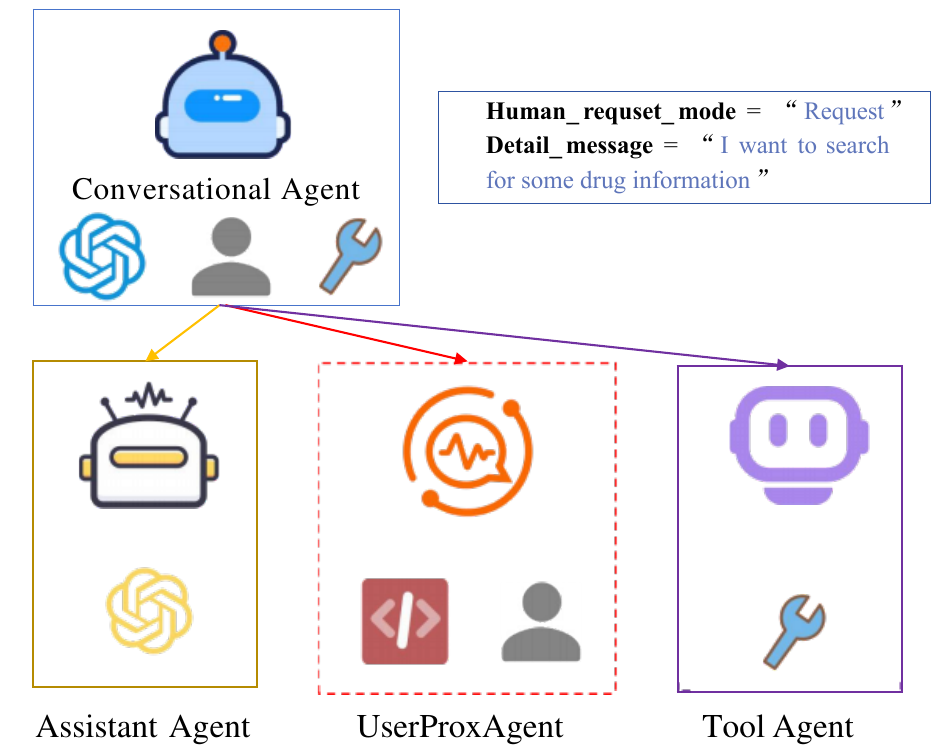}
    \caption{LLM Agent framework. The LLM serves as the central nervous system or cognitive core of the agent and plays a pivotal role in processing information, making decisions, and generating actions}
    \label{fig:agent framework}
\end{figure}

In multi-agent systems, communication and coordination are the foundations of achieving cooperation.  It is necessary to design practical communication protocols to ensure that agents exchange information efficiently and accurately.  Agents must maintain consistent semantic understanding and objectives to avoid misunderstandings and conflicts.  The reasonable allocation and division of tasks are central to multi-agent cooperation.  Tasks should be distributed among agents reasonably to avoid overloading some agents while leaving others idle.  Knowledge sharing is the basis for achieving collaborative work in multi-agent systems.  Different agents may have different backgrounds, and efficiently integrating this knowledge is challenging.

Multi-agent systems integrate the advantages of multiple LLMs with distinct functionalities. Each agent within these systems possesses expertise in specific domains, and this broad spectrum of specialized knowledge ensures that the generated results are comprehensive and accurate. Solving complex problems requires a multifaceted approach, and by leveraging the strengths of multiple agents, these systems can offer more refined and practical solutions.
AgentVerse\cite{chen2023agentverse} is a multi-agent framework designed to facilitate collaborative problem-solving among autonomous agents powered by LLMs. AgentVerse emulates human group dynamics to improve task performance and explores emergent behaviors within agent collaborations.  Agentbench\cite{liu2023agentbench} presents a comprehensive benchmark designed to evaluate the capabilities of LLM when acting as autonomous agents across a variety of real-world challenges. It consists of eight distinct environments categorized into code-grounded, game-grounded, and web-grounded scenarios to test the LLMs' reasoning, decision-making abilities, and their performance in multi-turn, open-ended generation settings. CoELA\cite{zhang2023building} presents a novel framework for building cooperative embodied agents using LLM, focusing on multi-agent cooperation without requiring fine-tuning or few-shot prompting. This framework is evaluated in various embodied environments, demonstrating that agents can effectively plan, communicate, and cooperate on long-horizon tasks. The approach surpasses traditional planning methods and shows that natural language communication enhances cooperation with humans. LLM-Co explores the potential of LLM for multi-agent coordination, focusing on their ability to collaborate in various scenarios effectively\cite{agashe2023evaluating}. CLIP\cite{Huang2023Grounded} explores a novel approach to enhance the ability of LLM to execute tasks in embodied settings, such as with robots, where understanding the physical world is crucial. Semantic ICL offers a promising approach to improve conversational agents by leveraging both semantic search and LLM\cite{Omidvar2023Empowering}. Yashar\cite{talebirad2023multiagent} introduces a novel framework to enhance the capabilities of LLM by leveraging multi-agent systems, allowing for collaborative environments where intelligent agents work to handle complex tasks efficiently. The authors demonstrate the framework's superior performance by conducting case studies on models such as Auto-GPT, BabyAGI, and Gorilla, which incorporate external APIs into the LLM. DyLAN\cite{liu2023dynamic} aims to optimize the collaboration of LLM agents for complex tasks such as reasoning and code generation. DyLAN is introduced to enhance the performance and efficiency of LLM-agent collaboration on complex tasks by enabling dynamic interaction architecture and inference-time agent selection. Traditional methods employ a static ensemble of agents, limiting adaptability and requiring extensive human input for agent design. DyLAN overcomes these limitations through a framework that dynamically constructs agent teams based on task requirements, supports multi-round interactions among agents, and incorporates an automatic agent team optimization algorithm. This optimization relies on an unsupervised metric, the Agent Importance Score, to assess and select the most contributory agents. Empirical evaluations demonstrate DyLAN's superiority in reasoning and code generation tasks, showcasing significant improvements over single-agent performances and traditional static approaches.

Multi-agent systems can fully leverage the expertise and skills of multiple independent agents to achieve solutions for complex tasks through collaborative cooperation. Each agent focuses on in-depth research in a specific domain, ensuring high professionalism, while the overall system integrates this information to provide comprehensive and accurate results. This architecture enhances the ability to handle complex problems. It possesses high flexibility and scalability, allowing adjustments and optimizations based on different needs, ultimately offering a more robust solution than a single LLM.

\begin{figure}
    \centering
    \includegraphics[width=1.0\linewidth]{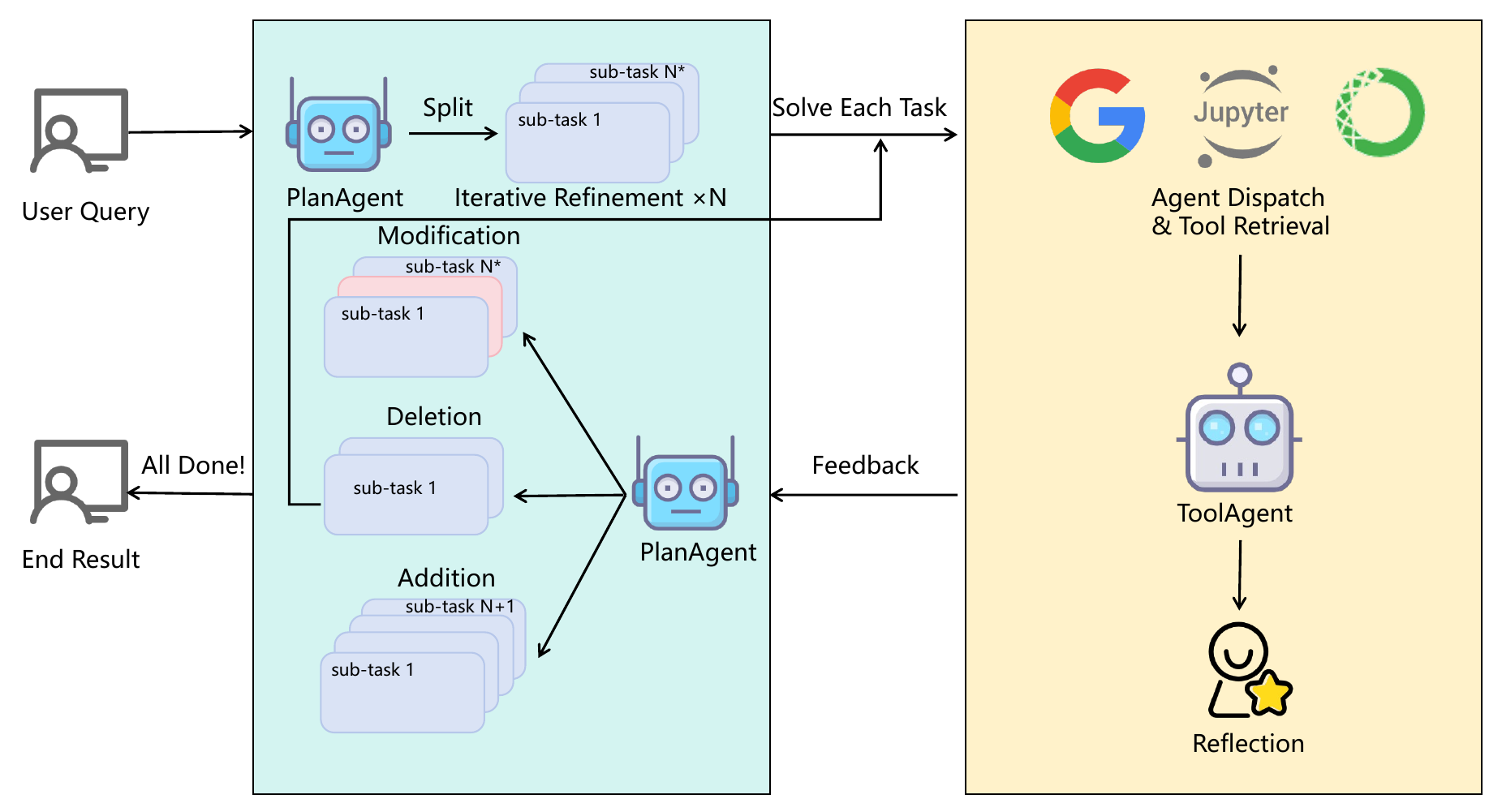}
    \caption{LLM for planning. Intelligent agents enhance task-handling capabilities in complex systems by breaking down large tasks into smaller, more specific sub-goals}
    \label{fig:planning}
\end{figure}

\vspace{-6pt}
\subsection{Planning}
Intelligent agents enhance task-handling capabilities in complex systems by breaking down large tasks into smaller, more specific sub-goals. This decomposition strategy makes task management more feasible and improves overall efficiency and effectiveness. The  planning has the following characteristics:
\begin{itemize}
    \item Hierarchical approach: Agents solve problems layer by layer by creating a hierarchical structure of tasks. From high-level strategic decisions to low-level specific operations, let LLM think step-by-step.
    \item Parallel processing: After task decomposition, agents can process multiple sub-goals in parallel, utilizing resources effectively. For example, in multi-agent systems, different agents can simultaneously deal with various tasks.
    \item Dynamic adjustment: Agents adjust the priority and resource allocation of sub-goals based on real-time feedback to adapt to environmental changes and unforeseen circumstances, ensuring the optimal task execution strategy.
\end{itemize}

Intelligent agents can enhance their ability to handle complex tasks through effective sub-goal setting, task decomposition, and continuous reflection and improvement. This strengthens the agents' ability to solve problems independently and optimizes the entire system's performance, contributing to efficient, adaptive, intelligent system design.

MindAgent\cite{gong2023mindagent} is designed to evaluate planning and coordination capabilities in gaming interactions.  
DEPS \cite{wang2024describe}is an innovative approach leveraging LLM to plan tasks in multi-task agents in open-world environments. DEPS uniquely integrates interactive planning with LLMs, focusing on error correction and efficiency improvement in plan execution through a goal selector module. It significantly improves task success rates in Minecraft and other tasks like ALFWorld and tabletop manipulation. MCTS\cite{zhao2024large} explores the integration of LLM with Monte Carlo Tree Search for task planning. This combination, termed LLM-MCTS, leverages the commonsense knowledge of LLMs to enhance the efficiency of planning algorithms. The key findings indicate that LLM-MCTS significantly outperforms traditional MCTS and LLM-induced policies in complex task scenarios, demonstrating the advantage of using LLMs as both a world model and a heuristic policy guide. ICPI\cite{brooks2024large} introduces a method for implementing policy iteration using LLM. ICPI  updates the prompt content to derive policies through trial-and-error interaction with an RL environment. The experiments demonstrate the efficacy of this method across six RL tasks, employing a range of LLMs such as Codex and GPT variants. Code-LLMs\cite{dinh2024large} investigate the limitations of current code-oriented LLM in handling code completion tasks when the provided code contains potential bugs. 

There are limitations to the direct use of LLMs as planners, such as their limited reasoning and planning capabilities and the inefficiency of the utilization of human feedback. PDDL\cite{guan2024leveraging} proposes a new method for planning tasks using pre-trained LLM.  The method uses LLMs to construct PDDL models, employs PDDL validation and human feedback to correct initial errors in these models, and generates plans using the corrected PDDL models.\cite{guan2024leveraging}. ReCon\cite{wang2023avalon} incorporates two processes: formulation contemplation for generating initial thoughts and speech and refinement contemplation for polishing these thoughts. MPC\cite{Lee2023Prompted} proposes a novel approach for creating high-quality conversational agents without fine-tuning. 
This approach harnesses the power of pre-trained LLMs as discrete modules, thereby ensuring sustained coherence and adaptability in open-domain conversational contexts.
The author demonstrates that MPC performs comparably with fine-tuned chatbot models in open-domain settings, offering an effective solution for generating consistent and engaging chatbots through human evaluations.

In an LLM-powered autonomous agent system, LLM functions as the agent’s brain. The agent breaks down large tasks into smaller, manageable subgoals, enabling efficient handling of complex tasks. 
The agent is capable of engaging in autonomous introspection and critiquing its past actions, extracting lessons from its errors, and subsequently refining its strategies for subsequent endeavors, culminating in an enhancement of the ultimate outcome's caliber.

\vspace{-6pt}
\subsection{Memory}
\begin{figure}[t]
    \centering
    \includegraphics[width=1.0\linewidth]{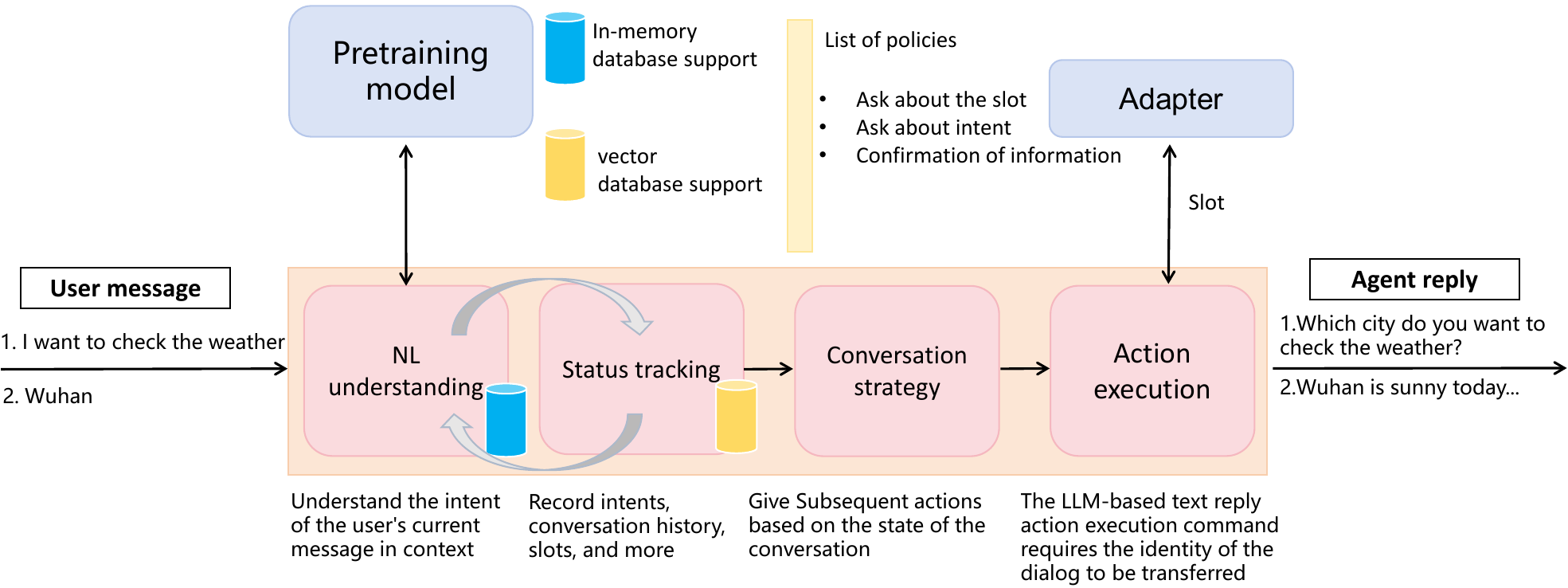}
    \caption{LLM Agent of Memory. The memory in the historical sequence is dynamically maintained, and it is extracted through a retrieval method.}
    \label{fig:LLM agent of Memory}
\end{figure}

Based on the LLM, memory of the historical chat sequence is dynamically maintained and extracted through a retrieval method. The historical chatbot manages and updates a memory system to ensure relevant information is retained and can be accessed efficiently. This rethinking method helps refine, organize, and integrate this information for future use, enhancing the model's ability to learn from past interactions and apply this knowledge to new situations. This methodology effectively improves the model's decision-making capabilities and adapts to evolving contexts.

Many scholars are currently studying long-term memory in agents. Agents are capable of storing and retrieving information over extended periods, allowing them to maintain context and consistency across multiple interactions.
REMEMBERER\cite{zhang2024large} introduces a novel framework for LLM that integrates a long-term experience memory, allowing the model to leverage past interaction experiences for decision-making across different tasks.  This approach, termed reinforcement learning (RL) with experience memory, enables the LLM to evolve its capabilities without fine-tuning parameters, positioning it as a semi-parametric reinforcement learning agent. The methodology updates the experience memory through RL processes, and experiments on two RL task sets demonstrate REMEMBERER's superiority over state-of-the-art methods. LONGMEM\cite{wang2024augmenting} enhances LLM by incorporating long-term memory capabilities.  This framework addresses the limitations of current LLMs, which are constrained by a fixed-sized input, preventing them from leveraging extensive long-context information from past interactions. LONGMEM can handle unlimited-length context, significantly expanding its use cases, particularly in scenarios requiring understanding and generation based on extensive historical data. The model demonstrates superior performance on long-context modeling benchmarks such as ChapterBreak and shows remarkable improvements in memory-augmented in-context learning tasks over traditional LLMs. MaLP \cite{zhang2023memoryaugmented}is a novel framework for personalizing LLM like GPT-3.5 without fully retraining them, which is resource-intensive. The authors introduce a dual-process enhanced memory mechanism and a parameter-efficient fine-tuning schema, aiming to make LLMs more user-oriented by leveraging both short-term and long-term memory components in coordination. Hatalis\cite{hatalis2023memory} focuses on the agents' memory management, particularly long-term memory, which is implemented using vector databases for storing and retrieving information. This aspect is crucial for maintaining context-specific knowledge and recalling past experiences, enabling coherent and efficient interactions. 
However, these methods require a lot of extra space to store long-term memory. To alleviate this issue, some research investigates the combination of self-reflection and memory mechanisms to achieve self-improvement without the need for additional data. 
MoT\cite{li2023mot} aimed at enhancing the capabilities of LLM like ChatGPT without the need for additional annotated datasets or computationally expensive fine-tuning. This approach draws inspiration from the human ability to self-improve through self-reflection and memory. It proposes a mechanism for LLMs to self-improve by leveraging their internal generation processes and an external memory component.
Liu\cite{liu2024llm} introduces Reasoning and Acting through Scratchpad and Examples (RAISE), a sophisticated architecture designed to improve the integration of LLM like GPT-4 into conversational agents. The architecture uses a memory system to maintain the continuity of contextual conversations better. It outlines a comprehensive method for the construction of agents, including historical conversation selection, scene extraction, chain of thought completion, and scene augmentation, leading to the LLMs training phase. The approach aims to enhance agent controllability and adaptability in complex, multi-turn dialogues.
Maharana\cite{maharana2024evaluating} introduces a machine-human pipeline to generate high-quality, very long-term dialogues leveraging LLM-based agent architectures. These dialogues are grounded on personas and temporal event graphs, and agents are equipped with the capability to share and react to images. Human annotators verify and edit the generated conversations for consistency and grounding to the event graphs. 
Wang\cite{wang2015largercontext} presents a novel approach to language modeling that incorporates corpus-level discourse information, termed a larger-context language model. This model introduces a late fusion technique to a recurrent language model based on LSTM units, which enables the model to distinguish between intra-sentence dependencies and inter-sentence dependencies more effectively.
To improve the efficiency of retrieval-based generation, RaLMSpec and Speculative RAG are proposed to use a small knowledge pool to approximate a large knowledge base with a small knowledge base~\cite{wang2024speculative,zhang2024accelerating}.

Traditional LLMs remember patterns, facts, and information across the data they were trained on, allowing them to generate knowledgeable responses. 
With the support of the memory system, LLMs can quickly retrieve information from memory, making them highly effective for tasks that require rapid access to a broad range of information, such as question-answering systems or recommendation engines.

\vspace{-6pt}
\subsection{Tool Use}
\begin{figure}[t]
    \centering
    \includegraphics[width=1.0\linewidth]{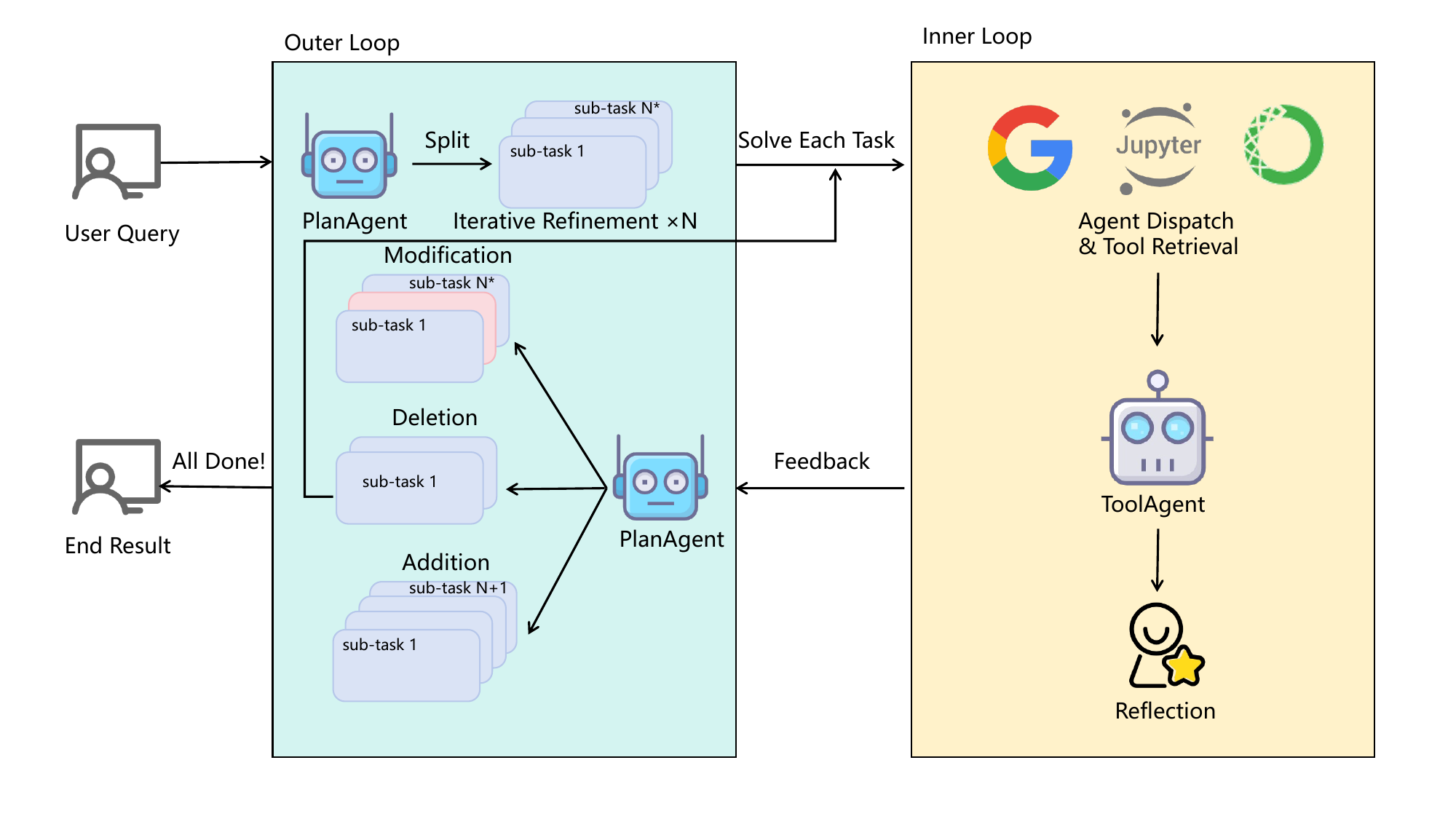}
    \caption{LLM agent of using tools. External APIs and tools can extend the abilities of an LLM beyond its original training.}
    \label{fig:LLM agent of using tools}
\end{figure}
In the context of multi-agent systems, particularly those involving LLM, the capability for agents to call external APIs represents a significant enhancement to their functionality and adaptability. This approach addresses several limitations of pre-trained models, offering the following benefits:
\begin{itemize}
    \item Access to current information: LLMs, once trained, do not inherently update their knowledge unless retrained with new data. By enabling agents to call external APIs, they can access the most current information available, which is crucial for tasks that depend on up-to-date data such as news updates, stock prices, or weather forecasts.
    \item Enhanced Functional Capabilities: External APIs can extend the abilities of an LLM beyond its original training. For instance, agents can perform calculations, access mapping services, or retrieve specialized data that is not stored within their pre-trained weights.
    \item Interactivity with Proprietary Systems: Agents can interact with proprietary information sources that are not publicly available or part of their training data. This is particularly useful in enterprise settings where access to internal databases or specific industry-related systems is required.
    \item Dynamic Adaptation to New Domains: The ability to call external APIs allows agents to dynamically adapt to new domains or changes in their operating environment without the need for retraining.  This makes the system more flexible and responsive to user needs.
\end{itemize}

Many researchers design different automation and standardization methods for integrating diverse tools into the execution of LLMs, enabling LLMs to seamlessly invoke various APIs and external resources. They are dedicated to developing context-aware tool selection mechanisms that allow LLMs to intelligently choose the most suitable tools based on the dialogue content and achieve multi-step decision-making in complex tasks.
Toolformer\cite{schick2024toolformer} is specifically designed to make informed decisions regarding the selection and invocation of appropriate Application Programming Interfaces (APIs).  It determines not only the optimal timing for API calls but also the most suitable arguments to be passed to these interfaces.  
They integrate a diverse array of utilities, encompassing tools such as a computational calculator, a question-and-answer system, a robust search engine, a multilingual translation service, and a scheduling calendar. Remarkably, Toolformer demonstrates significantly enhanced performance in zero-shot scenarios across a wide array of downstream tasks, frequently matching or even outperforming much larger models.
Gpt4tools\cite{yang2024gpt4tools} introduces an innovative methodology aimed at empowering LLMs, such as LLaMA and OPT, with the capacity to leverage tools. This is achieved by crafting an instruction adherence dataset through the strategic prompting of a sophisticated teacher model across diverse multimodal environments. Low-Rank Adaptation optimization techniques are then employed during the fine-tuning process, enabling these LLMs to undertake visual understanding tasks and generate images. The resultant system showcases marked enhancements in accuracy for both familiar and novel tool applications.
ToolEmu\cite{ruan2023identifying} is a framework utilizing LLMs to emulate tool execution for testing LM agents against diverse tools and scenarios, identifying risks such as data leakage or financial losses.  It features an LM-based automatic safety evaluator to quantify risks from agent failures.
LLMs have shown remarkable abilities in various NLP tasks but struggle with issues like hallucination and numerical reasoning\cite{NEURIPS2023_9cb2a749}. To enhance their capabilities, the integration of external tools has been explored. However, existing evaluation methods fail to clearly distinguish between LLMs leveraging their internal knowledge and those effectively using external tools. The ToolQA\cite{NEURIPS2023_9cb2a749} dataset is introduced to address this gap, consisting of data from 8 domains and incorporating 13 types of tools for accessing external information. This dataset is unique in requiring the use of external tools for question answering, minimizing reliance on LLMs' internal knowledge.
Ruan\cite{ruan2023tptu} focuses on task planning and tool usage. They introduce two types of agents, one-step and sequential, to execute tasks through planning and using external tools.
ToolkenGPT\cite{NEURIPS2023_8fd1a81c} enhances LLMs with external tools without the need for fine-tuning. This method aims to solve complex problems by incorporating a vast array of tools, such as calculators or databases. The key innovation of ToolkenGPT is representing each tool as a token and learning an embedding for it. This enables the LLM to call tools as easily as generating text, thereby combining the strengths of fine-tuning while overcoming their limitations.
Automatic Reasoning and Tool-use\cite{paranjape2023art} enhances the capabilities of LLMs by automatically generating multi-step reasoning chains and integrating external tool use. It generates reasoning programs for new tasks by retrieving related demonstrations and then uses external tools as needed within those programs. This approach allows for a structured and extensible way to incorporate complex reasoning and external knowledge into LLMs' responses.
INFERCEPT is a novel framework designed to optimize inference for augmented LLMs that interact with external tools or environments~\cite{abhyankarinfercept}. Current LLM inference systems handle external interactions by discarding the context and re-initiating a new request after receiving the response, leading to significant GPU resource wastage due to the recomputation of context. INFERCEPT addresses this by minimizing GPU memory waste through improved handling of intercepted LLM generations, employing techniques like swap pipelining and recomputation chunking. It dynamically selects a strategy from discarding, serving, and swapping. The framework significantly improves throughput and request completion rates.

Using APIs, agents can leverage external computing resources, which can be more efficient and scalable than processing everything locally. This is especially valuable for resource-intensive tasks. This approach transforms the agent from a static information processor to a dynamic information seeker, greatly enhancing its utility and ensuring its relevance regardless of when it was last trained. This capability is central to developing intelligent systems that are expected to operate in real-world, ever-changing environments.

\section{Application Layer}
\label{sec:app}

In this section, we first illustrate batching, an effective strategy that groups requests for execution in an application to improve throughput. Next, we describe some representative applications of foundation models and AI agents.

\vspace{-8pt}
\subsection{Batching}

\begin{figure}[t]
      \centering
      \includegraphics[width=\linewidth]{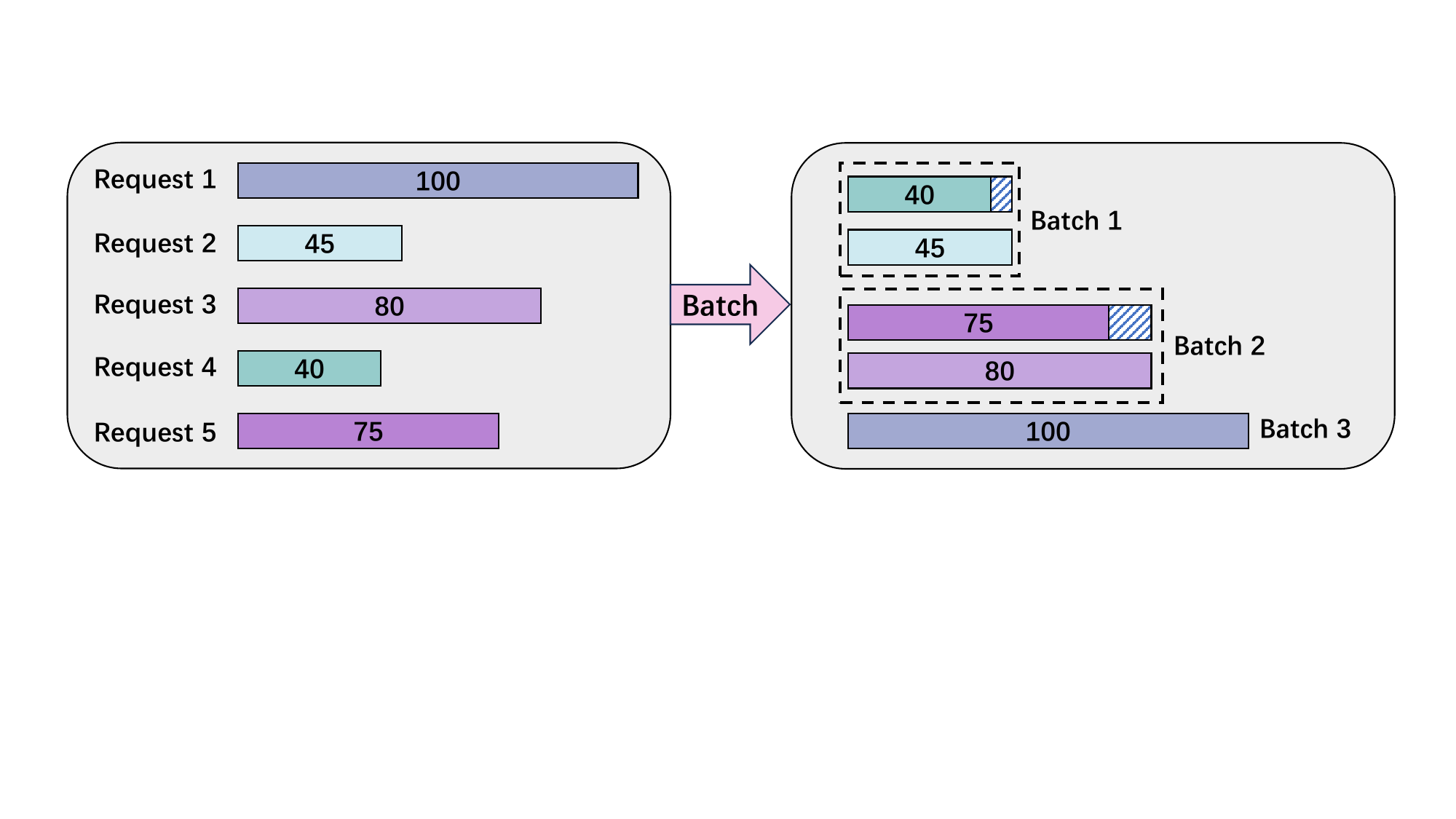}
      \caption{The illustration of batching. Requests with similar lengths are grouped together to reduce zero padding.}
      \label{fig:batch}
    \end{figure}

\begin{table*}
\centering
\caption{The summary of batching methods.}
\label{tab:batch}
\begin{tabular}{@{}llp{7cm}l@{}}
\toprule
\textbf{Issue} & \textbf{Ref.} & \textbf{Contribution} & \textbf{Model}   \\ 
\midrule
Different lengths & ORCA~\cite{yu2022orca} & Iteration-level batching instead of request-level batching. & GPT \\
 & Dvabatch~\cite{cui2022dvabatch} &  A multi-entry multi-exit batching scheme. & CNN, Bert \\
  & PiA~\cite{zheng2024response} & Accurately perceive and predict the response length. & GPT-2, Llama, Vicuna \\
  & TurboTransformers~\cite{fang2021turbotransformers} & Use dynamic programming for batching to maximize the response throughput. & Bert series \\
Different service features & Clipper~\cite{crankshaw2017clipper} & Dynamic batch size and delayed batching. & CNN  \\
 & OTAS~\cite{chen2024otas} & Batch queries with similar service characteristics. & ViT \\

Different weights & FLORA~\cite{wen2023batched} & Introduction of example-specific adapter.  & StarCoder \\
 & S-LoRA~\cite{sheng2023slora} & Highly optimized custom CUDA kernels for heterogeneous batching of LoRA computation. & Llama \\
 & Punica~\cite{chen2023punica} & A new CUDA kernel for the batching of GPU operations for different LoRA models & Llama-2 \\
 \bottomrule
\end{tabular}
\end{table*}

Batching is a widely adopted and beneficial strategy in serving systems, as it involves grouping incoming queries into batches and executing inferences for them together. This approach offers several advantages, including improved throughput and reduced resource consumption.
The advantages of batching can be attributed to two primary factors. Firstly, it helps mitigate the costs associated with RPC (Remote Procedure Call) calls and minimizes overhead in internal frameworks, such as the need to copy inputs to GPU memory.
Secondly, batching enables machine learning frameworks to leverage data-parallel optimizations more effectively. By performing simultaneous batch inference on multiple inputs, the framework can exploit parallel processing capabilities, such as those offered by GPUs, to enhance overall inference performance.

However, there are certain challenges associated with implementing batching in a serving system.
We summarize the issues and their corresponding methods in Table~\ref{tab:batch}.
First, when processing text queries, requests usually have different lengths.
To handle this, padding with zeros is often used for shorter requests.
However, this approach can result in computational inefficiency when there are significant differences in request lengths. 
Additionally, requests can also have different output lengths, ranging from just a few words to long articles. In such cases, shorter requests may need to wait for longer requests before returning results.
TurboTransformer addresses this issue by employing a dynamic programming algorithm that groups requests with similar input lengths~\cite{fang2021turbotransformers}. By doing so, it maximizes the overall throughput and reduces the inference latency.
PiA takes into account both the input length and the output length of requests~\cite{zheng2024response}. It introduces a predictor that accurately perceives and predicts the response length.
PiA groups similar requests together, allowing for better resource allocation and improved efficiency. 
Dvabatch proposes a multi-entry multi-exit batching scheme, utilizing three meta operations—new, stretch, and split to dynamically adjust ongoing batches for different types of diversities. This approach allows for more flexible and efficient processing, such as enabling short queries to exit early, splitting batches to match an operator's preferred batch size, and allowing later queries to join ongoing batches.~\cite{cui2022dvabatch}.
ORCA introduces a novel scheduling mechanism called iteration-level scheduling, which allows for more efficient processing of inference requests by scheduling execution at the granularity of iteration rather than the entire request~\cite{yu2022orca}. 

The second challenge involves addressing various service features of requests, including arrival time, latency requirements, and utility. For instance, one common dilemma is balancing throughput and latency, as prioritizing high throughput may lead to longer wait times for individual requests, potentially violating latency requirements.
Clipper tackles this issue by employing dynamic adjustments to the batch size and modify the delayed batching~\cite{crankshaw2017clipper}.
This approach allows more queries to be batched together, enhancing throughput, while still considering the latency requirements of individual queries.
Similarly, OTAS designs its batching algorithm by taking into account similar service characteristics~\cite{chen2024otas}. This ensures that requests with comparable attributes are grouped together, optimizing the overall system performance.
The premise of batching is that requests are processed by the same model. 
However, pre-trained models are fine-tuned with LORA for different tasks, resulting in distinct model parameters tailored for specific queries~\cite{hu2021lora}.
To deal with this issue, FLORA introduces an example-specific adapter~\cite{wen2023batched}. This approach is computationally efficient, as it leverages matrix multiplications and element-wise operations that are inherently batch-friendly on modern GPUs.
S-LoRA and Punica design some highly optimized CUDA kernels for heterogeneous batching of LoRA computation~\cite{sheng2023slora,chen2023punica}.

\vspace{-8pt}
\subsection{Foundation Model Applications}
Generative AI has demonstrated robust performance and vast potential in commercial applications, playing a significant role across various industries. Commercial generative AI can be categorized into four types: (1) text-generating AI (including some multimodal AI), (2) code-generating AI, (3) image-generating AI, and (4) video-generating AI. Among these, text-generating, code-generating AI, and image-generating AI have already seen widespread application.
Video-generating AI slightly lags behind the former three, but it exhibits considerable potential for future application.

\begin{itemize}
    \item Text-generating and multimodal AI. ChatGPT is the most famous chatbot developed by OpenAI, and performs well in various NLP tasks\cite{OpenAI_Chat}. ERNIE Bot (Wenxin Yiyan) is similar to ChatGPT, supporting input and output of images and text, and outperforms ChatGPT 4.0 in several Chinese tasks\cite{wenxin_yiyan}.
    \item Code-generating AI. Github Copilot is the most widely adopted code-generating tool in the world. It can automatically generate code and comments, and provide code explanations, allowing developers to focus on problem-solving and collaboration\cite{copilot}. Codex, similar to GitHub Copilot, is a product based on GPT-3 and is proficient in lots of programming languages. OpenAI claims that Codex is more powerful than GitHub Copilot\cite{openai_codex}.
    \item Photo and Video-generating AI. Midjourney can generate high-quality images according to natural language. It entered public beta on July 12, 2022, and is continuously evolving\cite{midjourney}. Sora, based on diffusion models, GPT and DALL·E, can generate videos from images and language, or expand on already generated videos. OpenAI claims that it is an important step towards AGI\cite{sora}. Runway Gen-2 is a high-performance video-generation AI developed by Runway AI, which can take text, image, or video input and convert it into new videos\cite{gen2}.
\end{itemize}

\subsection{AI Agent Applications}
The rise of LLMs has significantly accelerated research related to AI agents, which are now considered a principal avenue toward achieving AGI. Andrej Karpathy, co-founder of OpenAI, has stated that AI agents represent a significant future direction for AI. In the development of digital entities across various industries, these agents are expected to play a pivotal role in applying AGI. AI agents, as products, are anticipated to conduct business operations.
Since March 2023, the field of AI Agents has witnessed several groundbreaking research developments. 
AutoGPT, a project developed by Significant Gravitas, automatically optimizes GPT's hyperparameters by searching for algorithms that enhance its performance across diverse tasks~\cite{autogpt2023}. Users simply set one or more objectives, and AutoGPT can autonomously generate and execute tasks, continually analyzing and refining its strategies throughout the process. 
Researchers at Stanford and Google have developed a simulated environment named Smallville, where 25 AI agents exhibit complex behaviors and interactions~\cite{park2023generative}. Each agent possesses a unique seed memory, which enables them to form and recall relationships through social interactions with other agents. 
VOYAGER, designed with the long-term goal of `discovering as many different things as possible', is the first AI agent in Minecraft to continuously explore the world, acquiring a diverse set of skills and making new discoveries~\cite{wang2023voyager}. It introduces an expanding arsenal of skills for storing and retrieving complex behavioral executables, alongside a novel iterative cueing mechanism that enables the agent to autonomously explore and adapt to unknown environments without human intervention, demonstrating superior proficiency. 
ChatDev (Chat-powered Software Development) is proposed as a virtual software company operated by multiple intelligent entities~\cite{qian2023communicative}. After a user specifies a task, ChatDev utilizes the software engineering waterfall model. Through a sequence known as the Chat Chain, consisting of atomic tasks, it facilitates automated interactions, collaboration, and decision-making among diverse AI agents to produce comprehensive software, including source code, environment dependency specifications, and user manuals.
Research on AI Agents is currently led primarily by the academic community and developers. The outstanding performance of AI Agents in independent operation, collective collaboration, and human-machine interaction is increasingly recognized as a potent tool for enhancing productivity across various industries in the digital economy. However, the widespread application of LLM-based agents faces significant challenges due to the costs associated with token interactions. Additionally, the absence of a profit-sharing mechanism indirectly hampers the development of AI Agents. The main strategies for reducing costs include combining different model sizes and optimizing inference infrastructure. Furthermore, rapid advancements in hardware and models are expected to resolve cost issues in the near future. OpenAI predicts that AI will surpass human intelligence levels within the next decade, achieving what is termed superintelligence. AI Agents are expected to become mainstream in future products, with numerous agent-centered products likely to emerge and be implemented across various fields in the coming years.

\section{Summary and Future Work}
\label{sec:conclusion}

\subsection{Lesson Learned}
\subsubsection{Heterogeneous edge computing for FM inference}
Our survey of various edge computing devices for FM inference has yielded several important insights. Edge computing devices each have unique characteristics, designed to meet specific requirements. 
\begin{itemize}
    \item ASICs, while optimal for fixed model architectures in FMs, are immutable post-production, limiting their adaptability to new models. FPGAs offer greater flexibility through programmability, supporting various model sizes. However, they are primarily suited for linear computations, necessitating additional design considerations for non-linear operators prevalent in transformer architectures.
    \item Among IMCs, TPUs are representatives that have limited memory capacity for FMs. Another drawback is their support for integer operations, which constrains precision and performance, making them suboptimal for direct FM inference.
    \item  CPUs, while significantly less powerful than other accelerators, offer larger memory. This characteristic enables offloading some computation workload and part of model caches to CPUs, facilitating collaborative inference with other accelerators.
    \item  GPUs benefit from a mature CUDA ecosystem, making them suitable for out-of-the-box FM inference. Consequently, numerous research efforts focus on accelerating GPU inference in data centers. However, their high cost and energy consumption hinder widespread adoption in edge scenarios. This limitation has spurred interest in edge GPU devices, exemplified by NVIDIA Jetson, which has limited computing capacity and enough memory.
\end{itemize}

In conclusion, heterogeneous edge computing for different kinds of FMs is an under-explored field because these designs only focused on LLM.  
Researchers tended to assume that the transformer architecture of FMs was fixed and did not focus on hybrid architecture, such as ViT plus LLM, which are pretty popular in multi-modal FMs. 
Thus, these approaches were limited to specific models and sub-optimal to others.

On the other hand, most works do not consider inter-accelerator optimization (e.g., parallelism strategy and communication optimization between heterogeneous devices) and focus on designing hardware-specific optimization methods. However, these accelerators can cooperate to form a better FMs serving system if designed approaches consider the complementary advantages and disadvantages of the heterogeneous devices.



\subsubsection{Trade-off on the FMs}
While foundation models exhibit stronger reasoning and generation capabilities as their scale increases, this improvement is not linear and comes with significant trade-offs in terms of resource consumption and complexity. 
Larger models require exponentially more computation, memory, and energy, which can lead to bottlenecks in real-world deployments, especially when there are scarce resources at the edge, but specialized tasks and private data require fine-tuning locally. Tuning larger models is more challenging because they tend to overfit by relying too much on patterns within massive pre-trained datasets, potentially limiting their broader reasoning abilities. Therefore, practical applications require balancing model size with inference efficiency, using domain-specific optimizations to maximize performance while controlling computational costs.

Another key lesson is the challenge of effectively integrating and balancing multiple data modalities (e.g., text, image, audio) within a single model. While multi-modal FMs have shown the potential to understand and generate rich, cross-modal content, they often struggle with maintaining performance consistency across all modalities. This is largely due to discrepancies in data availability, quality, and structure for different modalities, which can lead to imbalanced learning. Moreover, aligning diverse modalities in a common representational space requires sophisticated cross-attention mechanisms or specialized fusion techniques, which are still areas of active research. A major takeaway is that while multimodal FMs can offer robust flexibility and broad applicability, ensuring seamless integration and coherence across modalities remains a significant hurdle.

\subsubsection{Elastic agent serving system}
A key lesson learned from examining the current serving systems for FM agents is the significant gap in elasticity at the agent layer. While substantial progress has been made to introduce elasticity at the levels of the resources, models, or execution tokens, there remains a notable lack of dynamic adaptability within the agent itself. The flexibility at the agent layer is essential for unleashing the full potential of FM-based systems, particularly in handling complex, real-world tasks. 
The agent layer should be more flexible by incorporating adaptivity in API calls, external knowledge retrieval, reasoning capabilities, and multi-agent collaboration. 
\begin{itemize}
    \item Firstly, an adaptive agent could decide when it needs to access external databases or APIs for more specialized information, or when it should rely on internal reasoning capabilities to make decisions autonomously. 
    \item Additionally, instead of hardcoded processes for knowledge integration, agents could be equipped with mechanisms to interact with evolving knowledge repositories and search from different levels of knowledge databases. 
    \item Besides, agents could be adaptive in reasoning and multi-agent collaboration. Current systems tend to employ static reasoning approaches, where the agent either follows a predefined logic or relies entirely on a single model’s output. In contrast, a more flexible agent could dynamically choose between different reasoning strategies depending on the task complexities, and potentially break down a problem into subtasks to be handled by specialized agents within a collaborative network. This multi-agent coordination could allow the system to generate more scalable and robust solutions.
\end{itemize}

Without this level of flexibility, LLM agents remain limited in their ability to handle a broad spectrum of tasks. They miss opportunities to leverage external knowledge, adapt their problem-solving strategies, or collaborate effectively in multi-agent environments. As more complex and specialized applications arise, the lack of agent-layer flexibility becomes a bottleneck, preventing these systems from fully realizing their potential. Thus, future work in this field should prioritize developing more flexible, adaptable agents capable of handling diverse, evolving tasks in real-world environments.

\vspace{-12pt}
\subsection{Future Directions}
\vspace{-3pt}

\subsubsection{Efficiently deploying large FMs on heterogeneous devices}

Efficiently deploying FM on heterogeneous devices at the edge-cloud involves leveraging a combination of edge computing and cloud resources to optimize performance, scalability, and cost-effectiveness.
Existing research works primarily emphasize optimizing the deployment of small models at the edge or large models in the cloud, leaving effectively deploying large FMs on heterogeneous devices at the edge cloud as an unresolved issue.
This paradigm can make use of the idle computational resources at the edge, enabling low-latency and low-cost inference while also ensuring privacy preservation.
There are still some challenges when deploying FMs at the edge.
First, large FMs have substantial computational requirements, while edge devices have limited processing powers and vary significantly in their computational capabilities. 
Second, FMs typically require significant memory to store model parameters, especially for large-scale datasets. Memory constraints on heterogeneous devices, such as edge devices with limited RAM or storage, can hinder the deployment of large FMs.
Third, in distributed deployments at the edge, the communication ability is weak. Transferring data between devices and coordinating computation introduces communication overhead and greatly increases inference latency. 
Additionally, energy consumption is also a critical consideration, especially for edge devices powered by batteries or with limited power budgets. Deploying large FMs on such devices requires energy-efficient algorithms and optimizations to prolong battery life and reduce operating costs.
In summary, advanced model compression methods and resource scaling methods need to be designed in this scenario.

\subsubsection{Deploying multi-modal models and MoE models on edge-cloud devices}
Traditional serving systems have mainly been developed to accommodate vision or language models that exclusively handle visual or textual inputs. Recently, advanced multi-modal models are developed to support diverse types of inputs, thereby expanding the range of interaction possibilities for users.
Besides, different model architectures (e.g., MoE) have been designed to enlarge the parameter space and enhance the model intelligence.
To effectively support these new models, it is necessary to enhance the design of the serving system. 
First, multi-modal models consist of multiple modules, such as modality-aware encoders and decoders.
These modules have different dependencies in different tasks. Consequently, the serving system necessitates new resource allocation and inference acceleration methods tailored to these models.
Second, multi-modal and MoE models have particularly large parameter spaces, and the inference process relies on the received input. For example, different inputs activate distinct experts. Thus, during online inference, it becomes crucial to dynamically schedule available resources to accommodate these varying demands.

\subsubsection{Specific serving system for agents}
Currently, the serving systems are designed for the inference of FMs.
With the increasing prevalence of agent services, it needs to specifically optimize the serving system by considering the specific characteristics of agents.
For example, different agent services employ various execution strategies, such as different planning methods, memory retrieval schemes, and multi-agent frameworks in different applications. 
These variations offer opportunities to optimize the inference process and resource scheduling approaches for agent systems. For instance, we can strategically allocate simple agents to weaker devices and complex agents to more powerful devices. 
By considering the computation and communication capacities, we can optimize their collaboration frequency and method effectively.

\vspace{-6pt}
\subsection{Conclusion}
FM-powered agent services are widely recognized as a crucial way of achieving AGI and are expected to spearhead development in this field over the next decade. Constructing a robust infrastructure for these agent services within an edge-cloud environment is of great importance and has a substantial impact on the user experience. In this survey, we have presented a unified framework for a thorough literature review on diverse techniques for deploying FM-powered agent services. We have introduced a series of low-level optimization methods for the model execution, including computation, communication, and I/O optimization. Subsequently, we have discussed the parallelism methods and resource allocation schemes designed to optimize resource utilization. We've also highlighted several popular FMs and introduced two lightweight methods, namely model compression and token reduction, to expedite the inference processes. Furthermore, we have also reviewed the latest research endeavors focusing on the development of an effective multi-agent framework and explored several recent applications. Finally, the future research directions for serving multi-modal agents on heterogeneous devices are outlined.

This survey is anticipated to significantly advance the development of large-scale model applications in both academia and industry.
The proposed framework, along with the referenced research works, provides a comprehensive perspective on deploying FM-powered agent services. It showcases the latest advancements and encourages more researchers to contribute to this practical and compelling field. 
The technologies discussed in the paper collectively contribute to the development of a reliable and flexible serving system, enabling low-latency agent services that enhance users' daily experiences with increased intelligence.
The integrated optimization of system architecture and AI algorithms will expedite the deployment of large-scale model applications for a diverse range of users, thereby promoting societal progress.
Looking ahead, we are dedicated to advancing research in large model applications and aim to expand our investigation to include a broader range of models, execution environments, and practical applications.



\bibliographystyle{IEEEtran}
\bibliography{main}

\newpage

 




\vfill

\end{document}